\documentclass[a4paper,fleqn,usenatbib]{mnras}
\usepackage{newtxtext,newtxmath}

\usepackage[T1]{fontenc}
\usepackage{ae,aecompl}

\usepackage{graphicx}	
\usepackage{amsmath}	
\usepackage{amssymb}	
\usepackage{enumerate}

\title[Multifractal Behaviour of the 3C 273 Light-curves]{Multifractality Signatures in Quasars Time Series. I. 3C 273}

\author[A. Bewketu et al.]{
A. Bewketu Belete,$^{1}$\thanks{E-mail: asnakew@fisica.ufrn.br}
J. P. Bravo,$^{1,2}$
B. L. Canto Martins,$^{1,3}$
I. C. Le\~ao,$^{1}$
\and J. M. De Araujo,$^{1}$
 J. R. De Medeiros$^{1}$
\\
$^{1}$Departamento de F\'isica Te\'orica e Experimental, Universidade Federal do Rio Grande do Norte, Natal, RN 59078-970, Brazil\\
$^{2}$Instituto Federal de Educa\c{c}\~ao, Ci\^encia e Tecnologia do Rio Grande do Norte, Natal, RN 59015-000, Brazil\\
$^{3}$Observatoire de Gen{\`e}ve, Universit\'e de Gen\`eve, Chemin des Maillettes 51, Sauverny, CH-1290, Switzerland\\
}

\date{Accepted 2018 May 15. Received 2018 May 15; in original form 2017 December 21}

\pubyear{2017}

\begin{document}
\label{firstpage}
\pagerange{\pageref{firstpage}--\pageref{lastpage}}
\maketitle

\begin{abstract}
The presence of multifractality in a time series shows different correlations for different time scales as well as intermittent behaviour that cannot be captured by a single scaling exponent. The identification of a multifractal nature allows for a characterization of the dynamics and of the intermittency of the fluctuations in non-linear and complex systems. In this study, we search for a possible multifractal structure (multifractality signature) of the flux variability in the quasar 3C 273 time series for all electromagnetic wavebands at different observation points, and the origins for the observed multifractality. This study is intended to highlight how the scaling behaves across the different bands of the selected candidate which can be used as an additional new technique to group quasars based on the fractal signature observed in their time series and determine whether quasars are non-linear physical systems or not. The Multifractal Detrended Moving Average algorithm (MFDMA) has been used to study the scaling in non-linear, complex and dynamic systems. To achieve this goal, we applied the backward ($\theta=0$) MFDMA method for one-dimensional signals. We observe weak multifractal (close to monofractal) behaviour in some of the time series of our candidate except in the mm, UV and X-ray bands. The non-linear temporal correlation is the main source of the observed multifractality in the time series whereas the heaviness of the distribution contributes less.
\end{abstract}




\section{Introduction}

Active galactic nucleus (AGN) is a very small fraction of a galaxy whose luminosity outshines the entire galaxy - indicating that the excess luminosity is not produced by stars. The luminosity observed almost in all the electromagnetic (EM) spectrum further ensures that the energy production mechanism in AGNs is completely different from the one in stars. The exciting modern era of AGN studies begins with the identification of the first high-luminosity AGNs \citep{1963Natur.197.1040S}. Even though the history of AGN research begun many years before, the science still remains with a lot of open questions, such as understanding the energy production mechanisms, determining the size of the broad line region, and many others. As might be expected AGNs are among the most energetic and complex objects in the universe.
Variability is one of the main observational features of AGNs noticed at the whole electromagnetic wavebands. \citet{1990A&A...234...73C} affirm that flux variation is a well-known property of AGNs. Quasars are a very luminous type of AGNs which are roughly classified as radio loud and radio quiet. The variability observed in quasars light-curves at short and long time scale, from less than one hour up to several decades, is one of the most important observed characteristics of quasars. The physical features of quasars depend on several properties such as the mass of the central black hole, the rate of gas accretion onto the black hole, the orientation of the accretion disk, the degree of obscuration of the nucleus by dust, and the presence or absence of jets. 

The quasar 3C 273 is a non-extreme, but a very bright type of AGN \citep{2007MNRAS.375.1521M}. The quasar 3C 273 has been extensively studied on diverse time scales across the entire EM spectrum, with measurements made in single bands and often as well as in multi-bands \citep{2014ApJS..213...26F,2009AJ....138.1428F,2010ApJ...714L..73A,2009A&A...494...49P,2009MNRAS.392.1181D,2008A&A...486..411S,2007MNRAS.375.1521M,2007A&A...465..147C,2006ApJ...648..900J,2006A&A...446...71S,2005ApJ...633L..85A,2002MNRAS.336..932K,2002A&A...390L..19G,2001ApJ...549L.161S,2000A&A...354..513C,2000A&A...354..497M}. Blazars show rapid flux variability across the complete EM spectrum \citep{2015MNRAS.451.1356K}. Fluctuations ranging from a few tens of minutes (and sometimes even more rapid) to less than a day is often known as intra-day variability (IDV) \citep{1995ARA&A..33..163W}. It has been reported by \citet{1990A&A...234...73C} that the flux variation across the spectra of 3C 273 shows no simple correlation.
It was also noticed that, at higher frequencies, the strength of the flux variations in 3C 273 are higher than in the lower frequencies \citep[e.g.][]{2009AJ....138.1428F}. Fluctuation in the flux of astronomical objects is an important and widespread phenomenon which provides relevant information on the dynamics of the system that drives the fluctuation. The IDV in blazars is the least well-understood type of variations but it can provide an important tool for learning about structures on small spatial scales and it also provides us with a better understanding of the different radiation mechanisms that are important in the emitting regions \citep[e.g.][]{1995ARA&A..33..163W}. Any characteristic timescale of variability can effectively provide us with information about the physical structure of the central region and complex phenomena such as hot spots on accretion discs \citep[e.g.][]{1993ApJ...406..420M}. It has been known that quasars are non-linear physical systems whose temporal evolution is described by non-linear dynamical equations and their light-curves cannot be analysed only by means of the classical method (e.g., PS, SF, and covariance analyses) which are suitable only for dealing the signals of a linear system \citep{1991ApJ...380..351V}. The presence of non-linearity in a time series leaves open the possibility that the time series is produced by a chaotic system. However, as indicated by \citet{1992ApJ...391..518V}, non-linearity is a necessary, but not sufficient, condition for a light curve to be produced by a chaotic system. There are a number of real-world signals from complex systems that exhibit self-similarity and non-linear power-law behaviour that depends on higher-order moments and scale, which is the signature of a fractal signature in the system \citep{1988Feder}. 

To study the variability seen in quasars light-curves, astronomers have been working on the time-frequency analysis using different approaches to unveil the physics behind the observed physical characteristics. These days, fractals are widely used in the modelling and interpretation of many different natural phenomena \citep[e.g.][]{PhysRevLett.57.1098,mandelbrot1983fractal}, including astrophysical phenomena \citep[e.g.][]{1991HeckandPerdang}, due to their capability of compacting the information on the scaling and clustering behaviour of the system-fractal nature. Fractal and multifractal behaviour is common in natural and social sciences \citep{mandelbrot1983fractal}. Fractals can be classified into two categories: monofractals and multifractals. If this scaling behaviour is characterized by a single scaling exponent, or equivalently is a linear function of the moments, the process is monofractal. A multifractal system is a generalization of a fractal system in which a single exponent is not enough to describe its dynamics; instead, a continuous spectrum of exponents (the so-called singularity spectrum) is needed. The multifractal analysis consists of determining whether some type of power-law scaling exists for various statistical moments at different scales. The main feature of multifractals is that the fractal dimension is not the same on all scales. In the case of a one-dimensional signal, the fractal dimension can vary from unity to a dot. Thus, a fundamental characteristic of the multifractal structure is that the scaling properties may be different in different segments of the system requiring more than one scaling exponent to be completely described. The multifractal spectrum effectively shows the distribution of scaling exponents for a signal and provides a measure of how much the local regularity of a signal varies in time. The presence of multifractality in a physical system indicates the non-linearity, inhomogeneity, complexity of the system dynamics \citep{2002PhyA..316...87K}, the presence of intermittency \citep{1991ApJ...380..351V}, and provides a way to describe signals from a complex, non-linear and dynamic systems. A multifractal analysis performed by \citet{1996A&A...312..424L} clearly indicated non-linear intermittent behaviour in the long term (1910 - 1991) B-band light curve of NGC 4151.

Several techniques have been proposed in the literature to study the scaling and self-similarity or fractality properties in the time series, such as Autocorrelation Function (ACF), detrended fluctuation analysis (DFA), the multifractal detrended fluctuation analysis (MFDFA), rescaled range statistical (R/S) analysis that provide type of self-affinity for stationary time series \citep{2008PhyA..387.5080B}, the periodogram regression (GPH) method, the (m, k)-Zipf method, and the detrended moving average (DMA) analysis \citep{2012NatSR...2E.835S,2004PhRvE..69b6105C}. The simplest type of multifractal analysis is based upon the standard partition function multifractal formalism, which has been developed for the multifractal characterization of normalized, stationary measurements \citep{2004Peitgen,2001PhRvE..64b6103B,1991PhRvA..44.2730B,1988Feder}. However, this standard formalism does not give correct results for non-stationary time series that are affected by trends or that cannot be normalized. Therefore, an improved multifractal formalism was developed known as wavelet transform modulus maxima (WTMM) method \citep{1991PhRvL..67.3515M}, which is based on the wavelet analysis and involves tracing the maxima lines in the continuous wavelet transform over all scales. This method has been widely used in diverse fields of solar activities to the earth science (e.g. Geology, DNA sequences, neuron spiking, heart rate dynamics, economic time series and also weather related and earthquake signals) \citep{2005NPGeo..12..157I,2004GeoRL..3110206V,2003PhRvE..67b1109V,2002PhRvE..66a1902V,2001PhyA..295..441K,2001GeoRL..28.4323T,1995PhRvL..74.3293A,1995GeoRL..22.3091M}. Readers are referred to \citet{2011Ouahabi} to better understand WTMM based multifractality analysis. By generalizing this standard DFA method, \citet{2002PhyA..316...87K} has been introduced the MFDFA, which allows the global detection of multifractal behaviour. By applying this MFDFA method to the sunspot number time series, \citet{2006JSMTE..02..003S} and \citet{2009JSMTE..02..066H} have found that the presence of multifractality/complexity in the sunspot number fluctuations is almost due to long-range correlation.
  
In this work, we apply the extended version of DMA, called multifractal detrended moving algorithm (MFDMA) for one-dimensional time series. This technique has been applied by \citet{0004-637X-843-2-103} in the search for multifractality traces in the magnetic activity of stars. To our knowledge, this is the first time that this technique is applied to quasar time series analysis. We hope that it will give insights into the different degrees of non-linearity of the system across its electromagnetic spectra.

This paper is structured as follows. In Section \ref{data}, we discuss the data, method and procedures used, and in Section \ref{res}, we present our results and discuss the multifractal nature of the quasar 3C 273. Summary and conclusions are given in Section \ref{concl}.

\section{Observational data, Method and Procedures}\label{data}

\subsection{Observational data: Light-curves for 3C 273}

\begin{figure}
\centering
\includegraphics[scale=0.253]{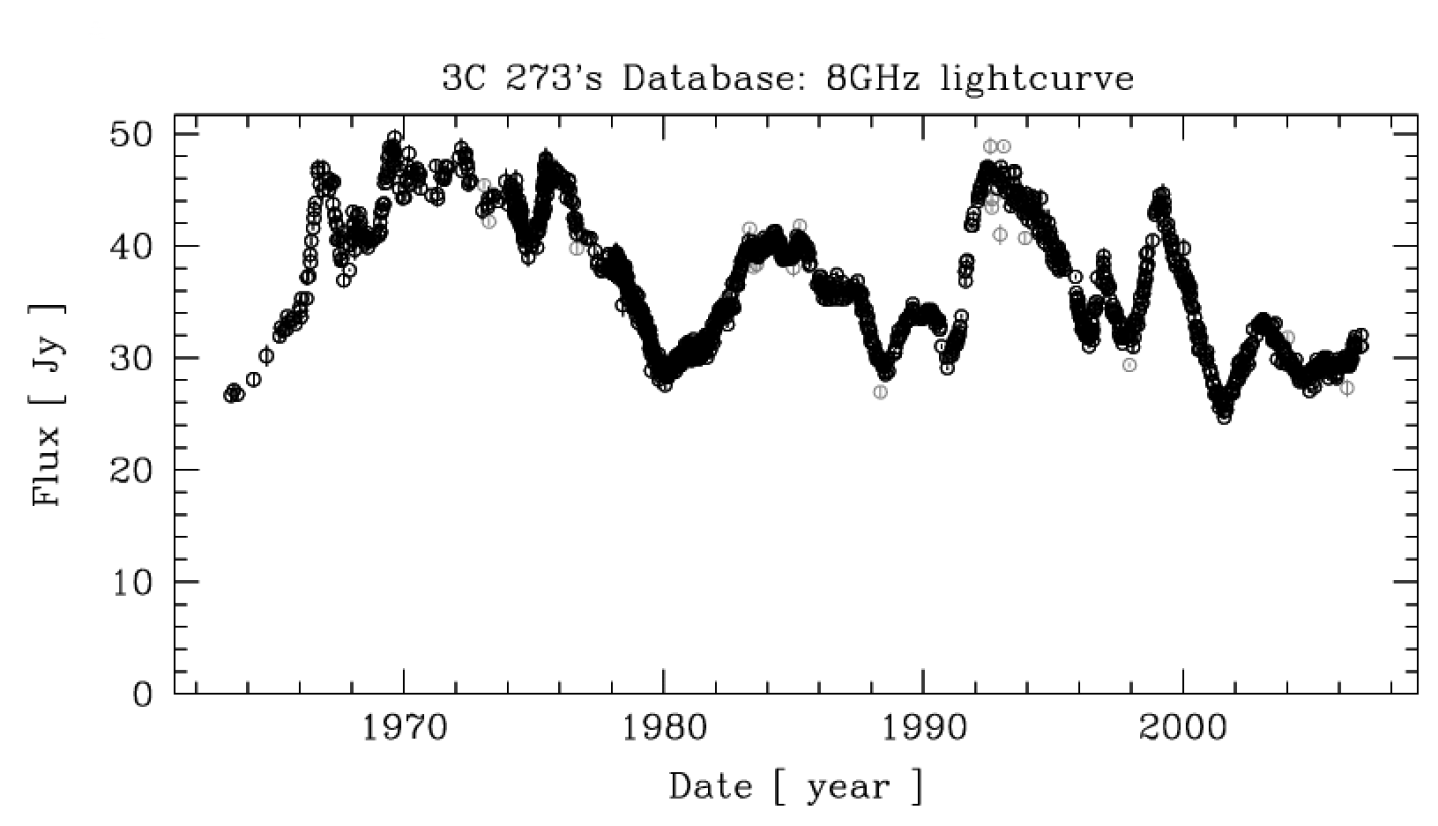}
\includegraphics[scale=0.253]{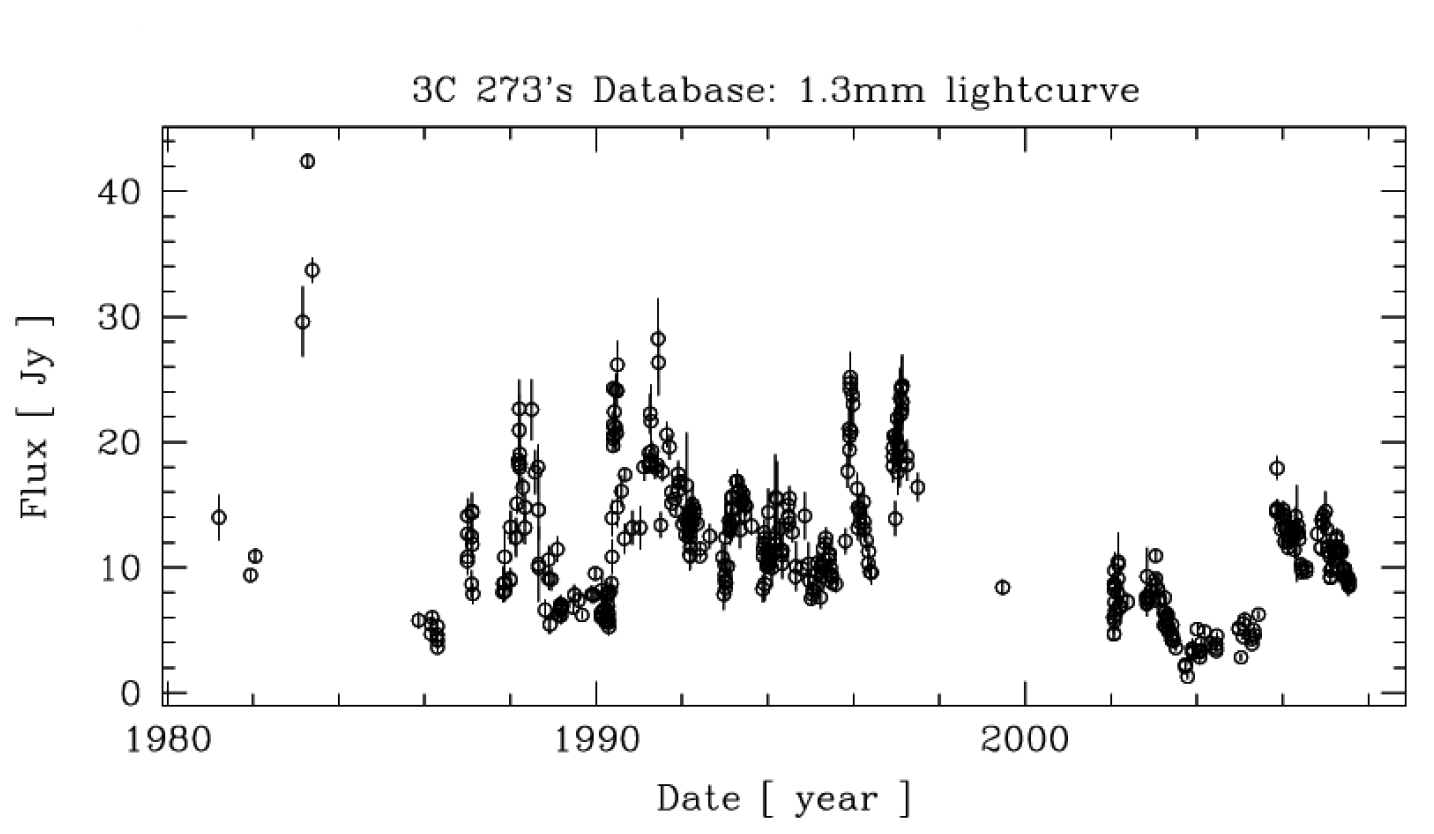}
\includegraphics[scale=0.253]{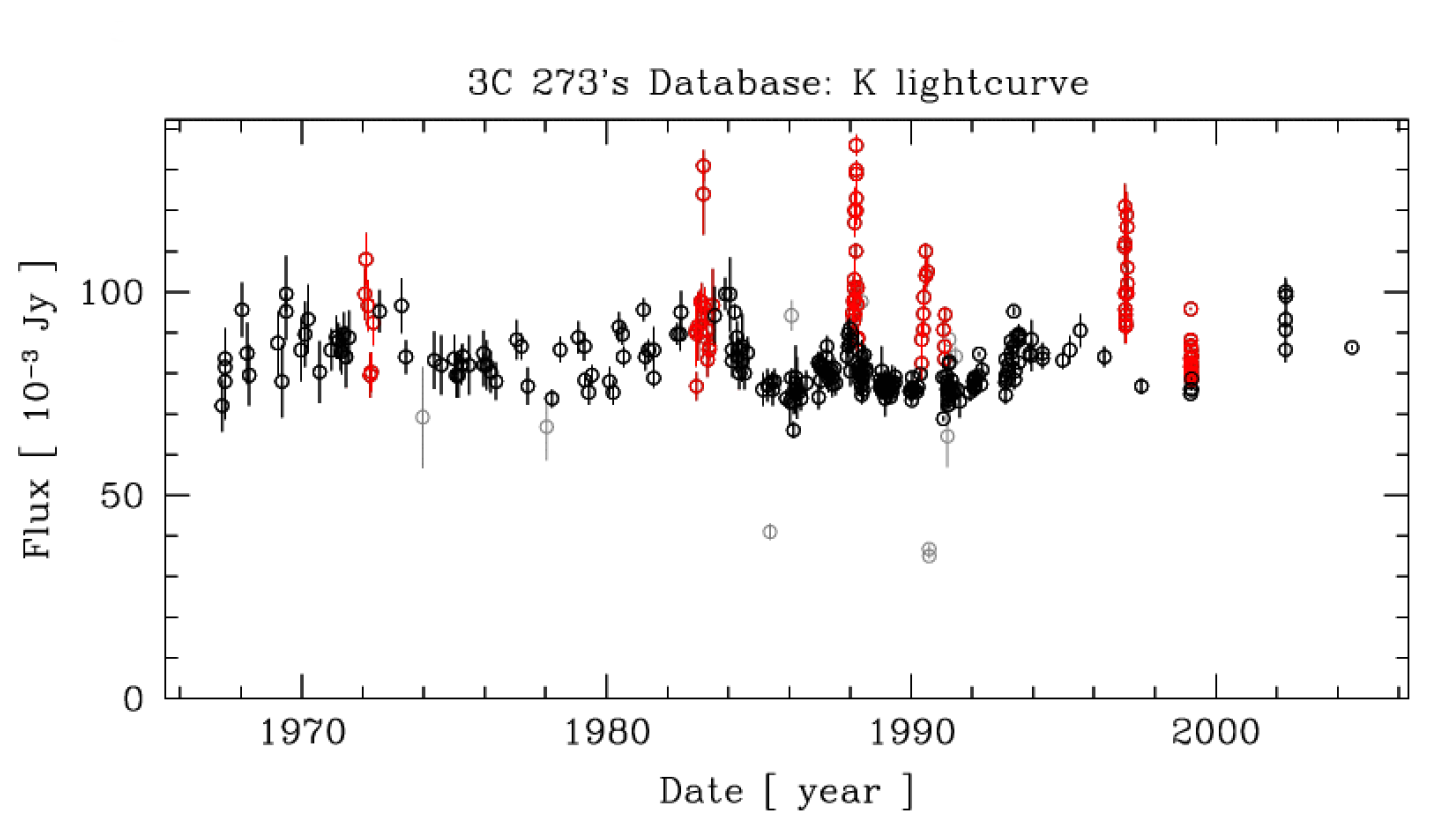}
\includegraphics[scale=0.253]{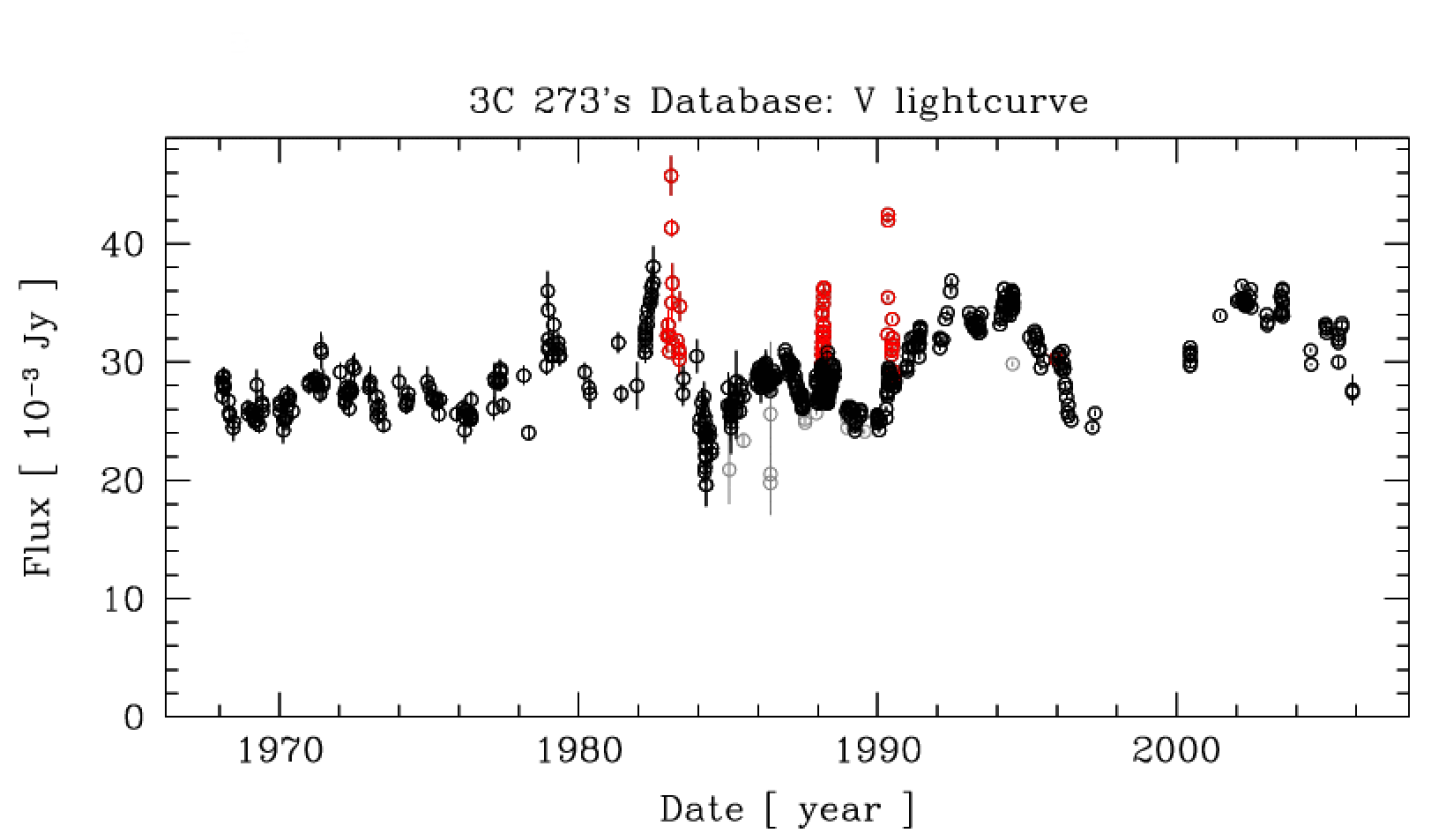}
\includegraphics[scale=0.253]{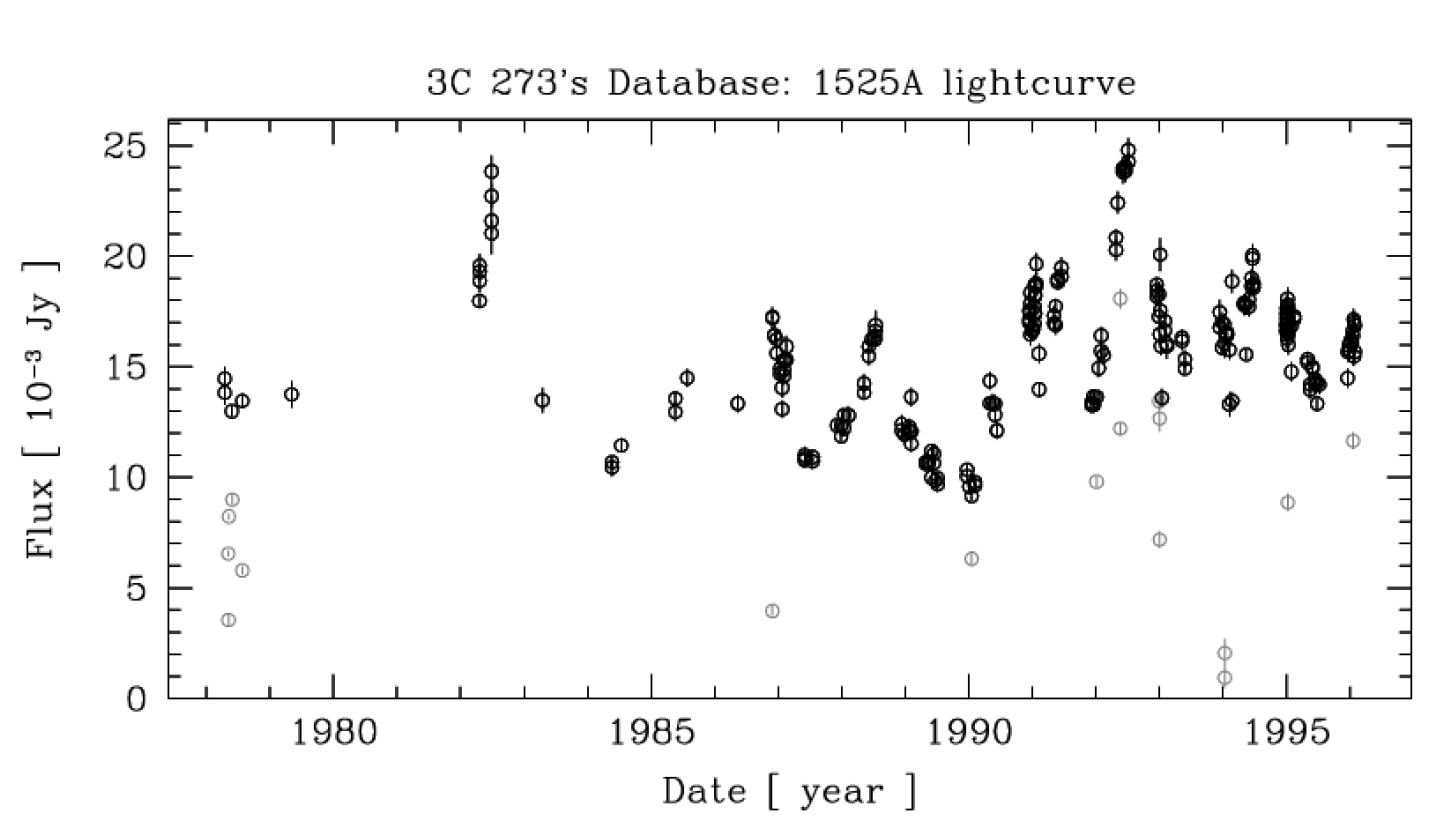}
\includegraphics[scale=0.253]{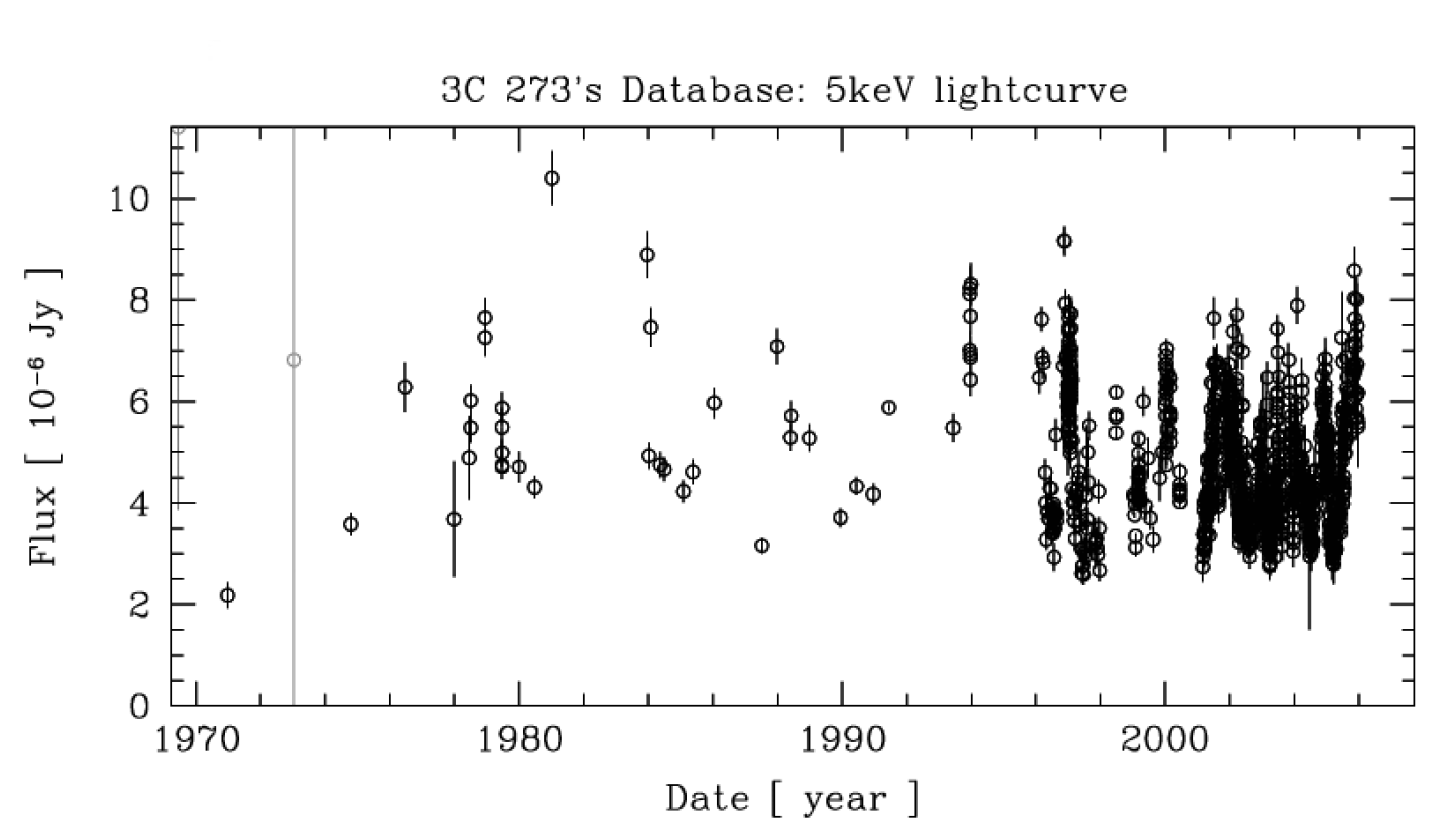}
\caption{3C 273 light-curves. From top to bottom: radio, mm, IR, optical, UV and X-ray observations at 8 GHz, 1.3 mm, 2.2 $\mu$m (K band), 5479 {\AA} (V band), 1525 {\AA}, 5 keV, respectively. Reprint from the original source at the Integral Science Data Centre  (ISDC database).}
\label{fig1}
\end{figure}

For the present study, we have chosen some of the 3C 273 light-curves provided by Integral Science Data Centre (ISDC)\footnote{\label{note1} \url{http://www.isdc.unige.ch/3c273/}} linked to the Astronomical Observatory of the University of Geneva \citep{2008A&A...486..411S,1999A&AS..134...89T}. Specifically, we have selected the time series covering the following wavebands: radio at 8 GHz, millimetre (mm) at 1.3 mm, infrared (IR) at 2.2 $\mu$m, optical at 5479 {\AA}, UV at 1525 {\AA}, and X-ray at 5 keV. The referred data are displayed in Fig. \ref{fig1}, where the behaviour of Flux (in Jansky) versus Time (in days) is given for all the wavebands in consideration. The reason why we chose this candidate is that it is suitable for multifractal analysis due to the well-established flux fluctuation in its light-curves, in all observed spectrum regions. For the light-curves considered, we only have selected  good data points (points with flag $\geq 0$), and disregarded those with negative flag which are considered as useless, uncertain, and dubious. We have performed no special treatment in our selection of the light-curves but we have considered stability (continuity) in the distribution of data points as the first criteria and the number of data points as the second criteria for all the time series considered, except for the 1.3 mm which we selected deliberately to determine whether a gap in data points affects our multifractality analysis or not. 
\medskip

Though the emissions of AGNs in general, and of 3C 273 in particular, are known to be complex due to the presence of many emission components of different nature,  there are different physical mechanisms known to produce radiation at different energy spectra in such astronomical objects. In 3C 273, synchrotron flaring emission from a relativistic jet (non-thermal radiations) dominate the radio to millimeter energy output and known to contribute in the infrared - optical domains \citep{2000A&A...361..850T,1993MNRAS.262..249R,1988Natur.335..330C}, in addition, thermal emission from dust grains by a UV source is also in part responsible for the production of infrared continuum \citep{2006A&A...451L...1T,1993MNRAS.262..249R}. Matter accreted in clumps is a possible mechanism for the production of UV radiation rather than through an accretion disc \citep{2005A&A...444..417C}. The interaction of these clumps generates optically thick shocks producing the UV emission, whereas optically thin shocks closer to the black hole give origin to the X-rays. Comptonisation of thermal plasma, for example, in an accretion disc, would produce X-ray emission similar to that observed in Seyfert galaxies \citep{2004Sci...306..998G}.  It has been indicated that the soft-excess could be due to thermal Comptonisation of cool-disc photons in a warm corona \citep{2004MNRAS.349...57P}. Inverse Compton processes of a thermal plasma in the disc, or in a corona, and of a non-thermal plasma associated to the jet are believed to generate the X-ray to gamma-ray emission \citep{2004Sci...306..998G,2002MNRAS.336..932K}.

\subsection{Method and Procedures}

In this work, we use the Multifractal Detrended Moving Average algorithm, an extended version of DMA algorithm, which consists of a multifractal characterization of non-stationary time series \citep{2010PhRvE..82a1136G}. Our choice of MFDMA is because of the fact that it details the signal on a wide Hurst exponent spectrum multifractality in the time series \citep{2010PhRvE..82a1136G}. We apply the backward ($\theta=0$) MFDMA algorithm as suggested by \citet{2010PhRvE..82a1136G} for better computation accuracy of the multifractal scaling exponent $\tau(q)$.
In multifractality study the most crucial parameters to describe the structural properties of a time series \textit{x(t)} are (i) a $q^{th}$-order fluctuation function $F_q(n)$, (ii) the multifractal scaling exponent $\tau(q)$, and (iii) the multifractal spectrum \textit{f}(\textit{$\alpha$}). These parameters are obtained according to the procedure introduced by \citet{2010PhRvE..82a1136G} as given below:

\begin{enumerate}[1.]
\item First, the time series is reconstructed as a sequence of cumulative sums given by:

\begin{equation}\label{eq1}
y(t)=\sum_{i=1}^t x(t),   \quad  t=1,2,3,...,N,                 
\end{equation}
where $N$ is the length of the signal.\\

\item Second, we calculate the moving average function $\tilde{y}(t)$ of Eq.~\ref{eq1} in a moving window using the relation \citep{2010PhRvE..82a1136G,1986PhRvA..33.1141H}:
\begin{equation}\label{eq2}
\tilde{y}(t)={1\over n}\sum_{k=-\lfloor(n-1)\theta \rfloor}^{\lceil(n-1)(n-\theta)\rceil}y(t-k),
\end{equation}
where $n$ is the window size, $\lceil$ (.) $\rceil$ is the smallest integer not smaller than argument (.), $\lfloor$ (.)  $\rfloor$ is the largest integer not larger than argument (.). The reference point of the moving average window can be changed by the introduction of a parameter $\theta$ with $0\leq \theta\leq 1$. In the present work, $\theta$ is adopted as zero, referring to the backward moving average. The moving average function $\tilde{y}(t)$ is calculated by averaging the $n$-past value in each sliding window of length $n$ and the reference point of the averaging process is the past point of the window. Hence, the moving average function considers ${\lceil(n-1)(n-\theta)\rceil}$ data points in the past and $\lfloor(n-1)\theta \rfloor$ points in the future : $\tilde{y}(t)$  contains half-past and half-future information in each window. The index $k$ in $\tilde{y}(t)$  is accordingly set within the segment $n$. \\

\item The trend (the change in the mean of the signal series over time) is removed from the reconstructed time series $y(t)$ using the function $\tilde{y}(t)$ and the residual sequence $\epsilon(t)$ is obtained from:
\begin{equation}\label{eq3}
\epsilon(i)= y(i)-\tilde{y}(i),
\end{equation}
where $n-\lfloor(n-1)\theta\rfloor\leq{i}\leq{N}-\lfloor(n-1)\theta\rfloor$. The residual time series $\epsilon(t)$ is subdivided into ${N_n}$ disjoint segments with the same size $n$ given by $N_{n} = \lfloor{N\over n}-1\rfloor$. In this sense, the residual sequence $\epsilon(t)$ for each segment is denoted by $\epsilon_v$, where $\epsilon_v(i)=\epsilon(l+1)$ for $1\leq{i}\leq{n}$ and $l=(v-1)n$.\\

\item Then we calculate the root-mean-square function as follows:
\begin{equation}\label{eq4}
{F_v}(n)= \left\{{1\over n}\sum_{i=1}^n {\epsilon_v}^2(i)\right\}^{1/2}.
\end{equation}

\item The $q^{th}$-order overall generating (fluctuation function) $F_q (n)$ is determined using the following relation:
\begin{equation}\label{eq5}
{F_q}(n)= \left\{{1\over {N_n}}\sum_{v=1}^{N_n} {F_v}^2(n)\right\}^{1/q}
\end{equation}
for all $q\neq0$. When $q=0$, according to L'H\^{o}spital's rule, we have:
\begin{equation}\label{eq6}
{F_q}(n)= exp\left\{{1\over 2{N_n}}\sum_{v=1}^{N_n}ln[{F_v}^2(n)]\right\}.
\end{equation}

\item By varying the values of segment size $n$, we determine the power-law relation between the function $F_q (n)$ and the size scale $n$. The scaling relation between the detrended fluctuation function $F_q(n)$ and the size scale $n$ can be calculated as:
\begin{equation}\label{eq7}
F_q(n)=n^{h(q)},
\end{equation}
where the exponent $h(q)$ is known as generalized Hurst exponent. This relation allows to estimate the scaling exponent of $h$($q$) and thus of the Hurst exponent of the series $y$($t$). In general, it tells us how the fluctuations of the profile (the cumulative sum) in a given time window of size $n$, increase with $n$. For $q=2$ it reduces to the ordinary Hurst exponent ($H$) \citep{1988Feder}. $H \neq 0.5$ indicates the presence of long-range correlation in the system. An exponent $0.5<H<1$ corresponds to a positive correlation or persistent in the system and $0<H<0.5$ corresponds to the process which is long-range dependent with negative correlations or anti-persistent. \citep{2004PhRvE..69b6105C}.\\

\item Using the calculated $h(q)$, we determine the multifractal scaling exponent $\tau(q)$ as follows: 
\begin{equation}\label{eq8}
\tau(q)=qh(q)-D_f,
\end{equation}
where $D_f$ is the fractal dimension of the geometric support of the multifractal measure \citep{2002PhyA..316...87K}. For our case, $D_f = 1$, since we are applying the MFDMA for one-dimensional time series analysis.\\

\item At last, we determine the singularity strength function (the Holder exponent) $\alpha(q)$ which is related to $\tau(q)$ via the Legendre transform \citep{1986PhRvA..33.1141H}, as shown below, and the multifractality spectrum $f(\alpha)$ as follows:
\begin{equation}\label{eq9}
\alpha(q)=d\tau(q)/dq,       
\end{equation}
and
\begin{equation}\label{eq10}
f(\alpha)=q\alpha-\tau(q). 
\end{equation}

\end{enumerate}

The singularity strength, $\alpha$, is extracted from the slope of the scaling exponent curve, i.e., its value is dependent on the nature of the scaling exponent curve (on the relationship between $\tau$($q$) and $q$ ) which in turn depends on the local fluctuations or the $q$ dependency of the local Hurst exponent $h$($q$). The $\Delta\alpha$ (the range of singularity strength $\alpha$) is the measure of the range of multifractal singularity strength. If there is strong non-linearity in the scaling exponent curve, i.e., different slope for the negative and positive $q$ values, then we can have wider $\Delta\alpha$, which reflects strong multifractality behaviour in the time series considered. In general, the wide range corresponds to strong multifractality, and contrarily the narrow range corresponds to weak multifractality or monofractal. The multifractal spectrum provides detailed information about the relative importance of several types of fractal exponents present in the signal \citep{2006Tectp.423..115T}, and also the detection of multifractality behaviour indicates the presence of intermittency in the data under consideration \citep{2007MNRAS.375.1521M}. A physical system is said to be intermittent if it has a wide multifractal spectrum or concentrates into a small-scale features with large magnitude of fluctuations enclosed by extended areas of less strong fluctuations \citep{monin2007statistical,1995tlan.book.....F,moffatt_1994}.

In our multifractality analysis, we have characterized the multifractal spectrum based on the behaviour of (i) the width of $\alpha$ (given by $\Delta\alpha=\alpha_{max}-\alpha_{min}$) and (ii) the symmetry in the shape of $\alpha$ defined as $A = (\alpha_{max}-\alpha_{0})/(\alpha_{0}-\alpha_{min})$ , where  $\alpha_{0}$ is the value of  $\alpha$ when $f(\alpha)$ assumes its maximum value. Concerning to the width of $\alpha$, \citet{2002SHIMIZU} and \citet{2003GeoRL..30.2146A} proposed that the width of a multifractal spectrum is the measure of the degree of multifractality. Broader the spectrum (larger the value of $\Delta\alpha$), richer the multifractality \citep{2004PCE....29..295T}. Smaller values of $\Delta\alpha$ (i.e., $\Delta\alpha$ gets close to zero) indicates the monofractal limit whereas larger values indicates the strength of the multifractal behaviour in the signal \citep{2010PhRvE..82a1136G}. 
For the symmetry in the shape of $\alpha$, the asymmetry presents three shapes: asymmetry to the right-skewed ($A >1$), left-skewed ($0< A <1$), or symmetric ($A = 1$). It has been reported by \citet{2012Ihlen} that the symmetric spectrum is originated from the levelling of the $q^{th}$-order generalized Hurst exponent for both positive and negative $q$ values. The levelling of $q^{th}$-order Hurst exponent reflects the $q^{th}$-order fluctuation is insensitive to the magnitude of local fluctuation. When the multifractal structure is sensitive to the small-scale fluctuation with large magnitudes, the spectrum will be found with right truncation; whereas, the multifractal spectrum will be found with left-side truncation when the time series has a multifractal structure that is sensitive to the local fluctuations with small magnitudes. Thus, it may be noted that the width and shape of the multifractal spectrum are able to classify a small and large magnitude (intermittency) fluctuations in the considered time series.
It has been understood that there are two possible sources of multifractality in a time series data \citep{2002PhyA..316...87K}, namely (i) multifractality due to long-range time correlations of the small and large fluctuations and (ii) multifractality due to a fat-tailed probability distribution function of the values in the series. The first kind of multifractality can be removed by random shuffling of the given series. Shuffling a time series destroys the long-range temporal correlation for small and large fluctuations. The corresponding shuffled series will exhibit monofractal scaling, i.e., leaves the original time series the same but without memory, $\alpha = 0.5$. Obviously, the probability distribution will not be changed by random shuffling and hence the multifractality of the second kind will remain intact. For the latter case, the non-Gaussian effects can be weakened by creating phase-randomized surrogates. In this context, the procedure preserves the amplitudes of the Fourier transform and the linear properties of the original series but randomizes the Fourier phases while eliminating nonlinear effects \citep{2007IJMPC..18.1071N}. On the other hand, the surrogate (phase randomization) analysis is an empirical technique of testing non-linearity for a time series. First, we interpolate our data to get evenly sampled data and generate surrogate series using Fourier transform method (phase randomization). We have applied the same procedure for the shuffled and surrogate time series as the original one.  These procedures are used to study the degree of complexity of time series to distinguish different sources of multifractality in the time series. The basic idea of the surrogate data method is to first specify some kind of linear stochastic process that mimics linear properties of the original data. Thus, if the shuffled signal only presents long-range correlations, we should find that $h_{shuff}(q) = 0.5$. However, if the source of multifractality is due to heavy-tailed distributions obtained by the surrogate method, the values of $h_{surr}(q)$ will be independent of $q$. If a given series contains both kinds of multifractality, the corresponding shuffled series will exhibit weaker multifractality than the original one \citep{2006JSMTE..02..003S}. If $h(q)$ is not enough to identify the source of multifractality, an alternative method to assess the behaviour of multifractality is to compare the non-linearity in the multifractal scaling exponent $\tau(q)$ for the original, shuffled, and surrogate (phase-randomized) data. Differences between these two scaling exponents and the original exponent reveal the presence of long-range correlations and/or heavy-tailed distributions. These comparisons can be shown by the scaling exponent $\tau(q)$ versus $q$ plots, which present the following relations \citep{2010PhRvE..82a1136G}:
 \begin{equation}\label{eq11}
  \tau(q)-\tau_{shuff} (q)=q[h(q)-h_{shuff} (q)]=qh_{corr} (q)
   \end{equation}
and
\begin{equation}\label{eq12}
\tau(q)-\tau_{surr}(q)=q[h(q)-h_{surr}(q) ]=qh_{tail}(q).
\end{equation}
The linear behaviour of $\tau(q)$ (and the narrow width of the singularity spectrum, $\alpha(q)$) indicates the presence of a monofractal, while non-linear behaviour (and a wide singularity spectrum, $\alpha(q)$) indicates multifractality.

\section{Results and Discussion}\label{res}

In this work, we perform the backward MFDMA for one-dimensional time series and analyse the non-linear and multifractal properties of the quasar 3C 273 time series covering all bands at different observation points. The scenario emerging for each time series is discussed separately as follows. In Fig. \ref{fig2}, we present the multi-scaling behaviour of the fluctuation functions $F_{q}(n)$ corresponding to the time series by the log-log plots of the fluctuation function $F_{q}(n)$ against the time scale (segment) $n$ for the original, shuffled and surrogate time series. The $q^{th}$-order weights the influence of segments with large and small fluctuations. For negative $q$ values, $F_{q}(n)$ is influenced by segments with small fluctuation whereas for positive $q$'s, $F_{q}(n)$ is influenced by segments with large fluctuation. The midpoint $q=0$ is neutral to the influence of segments with small and large fluctuations. The slope of the regression line of $q^{th}$-order fluctuation function and time segment gives the well-known generalized Hurst exponent $h(q)$. The slopes $h(q)$ of the straight lines obtained by a least-square fitting method are given in Table \ref{tab1}. For monofractal data sets, the fluctuation function has a constant slope at all $q$ values \citep{2002PhyA..316...87K}.

\subsection{Radio observations}

In Fig. \ref{fig2} (top panels), we observed a power-law relationship between the fluctuation function $F_{q}(n)$ and the segment $n$. We remark that the $q$ dependency of $h(q)$ reveals the presence of multifractality in the time series. The slope of the best fit line decreases by a small amount from the negative to the positive moments indicating the \textit{q} dependency of the slope $h(q)$ which in turn implies the presence of weak multifractality signature in the time series. The difference in the slope $h(q)$ between the original, shuffled, and surrogate data shows that the three data have different multifractality strengths. As we have discussed in Sect. \ref{data}, since  $h_{shuff}(q) \neq 0.5$ and $h_{surr}(q)$ is $q$-dependent, it is not possible, at this stage, to determine whether the multifractal signature (though it is weak) is due to the temporal correlation or the fat-tailed probability distribution in the time series considered; and therefore, we referred to the relations given by Eqs. \ref{eq11} and \ref{eq12}. As shown in Fig. \ref{fig3} (top panels) for all values of $q$, the three time series for the radio waveband exhibit almost a straight form indicating that the multifractal behaviour in this time series is weak (close to monofractal) which is in support of our $h_{shuff}(q)$ based analysis. 

For the time series under consideration, we have calculated the width $\Delta\alpha$ value and obtained $\Delta\alpha = 0.0434 \slash 0.0085 \slash 0.0501$ for the original, shuffled, and surrogate data, respectively as shown in Table \ref{tab1}. These results of $\Delta\alpha$, with respect to zero, further strengthened our analysis based on the slopes $h(q)$ and scaling exponent $\tau(q)$ that the multifractality strength in this time series is weak (close to monofractal) and the temporal correlation is the possible source of the detected weak multifractality. 

\subsection{Millimetre observations}

Following same analysis techniques and procedures used for the radio observation, we have detected multifractality behaviour in this time series. The difference between the $q^{th}$-order fluctuation function for positive and negative $q$ values or/and at the smallest segment sizes compared to the large segment sizes reveal the multiscaling properties of the time series considered \citet{2012Ihlen}.
The curves of the original, shuffled and surrogate time series decrease with $q$ which show that the slopes $h(q)$ of the regression lines are $q$-dependent indicating the presence of a multifractal signature in the time series under consideration \citep{2006JSMTE..02..003S}. 

The $q$ dependency of $h(q)$ leads to a non-linear $\tau(q)$ dependence on $q$ \citep{2006JSMTE..02..003S}. The non-linearity indicates that there is a non-linear interaction between the different scale events and multifractal nature of the system \citep{2001PhyA..295..441K,moffatt_1994}. The $q$ dependency of $\tau(q)$ shown in Fig. \ref{fig3} (2nd panels) confirms multifractality in the time series. Also notice that in the parts of the curves corresponding to small variance regimes $q < 0$, in Fig. \ref{fig2} (2nd panels), the shuffled series seem deviated more (i.e., close to linearity) from the original data with respect to the surrogate data, revealing that the shuffling process affects the original time series. The difference in the non-linearity between the shuffled and surrogate data with respect to the original time series tells us about the possible source of the observed multifractal behaviour. In this case, the temporal correlation is the main source of the observed multifractal signature than the fat-tail distribution. The presence of crossover in  $\tau(q)$ versus $q$ plot (Fig. \ref{fig3}), i.e., different slopes for $q < 0$ and $q > 0$, shows that the multifractality in the time series is strong. 

The calculated width ($\Delta\alpha$) values for the original time series and the corresponding shuffled, and surrogate data are given by $\Delta\alpha = 0.5234 \slash 0.1227 \slash 0.765$, respectively (Table \ref{tab1}). Therefore, multifractal parameters $h(q)$, $\tau(q)$, and the multifractal spectrum for the shuffled data clearly demonstrate that the observed multifractality originates more from the temporal correlation in the time series rather than the fat-tailed probability density distribution. To control the effect due to some gaps in the data points of the time series considered, we interpolate the data and produced evenly distributed points. Applying same procedures for the data points evenly distributed (interpolated data) we obtained similar results. The wider $\Delta\alpha$ with regard to $\Delta\alpha=0$ richer and stronger the multifractality is. The left-side truncated spectrum reflects the existence of a multifractal structure sensitive to the local fluctuations with small magnitudes (intermittency).

\subsection{IR observations}

Here also, there is a $q$ dependence of $h(q)$ though it changes slowly as seen in the radio observation (Fig. \ref{fig2}, 3rd panels). The slope of the surrogate data is also changing with $q$ showing the presence of temporal correlation in the time series. To support our $h(q)$ based analysis, we can see in the scaling exponent plot $\tau(q)$ in Fig. \ref{fig3} (3rd panels), that all the data have weak non-linear behaviour both for positive and negative $q$ values, indicating the existence of weak multifractality in the time series.

As we have discussed in the methodology part, the relevance of calculating $\Delta\alpha$ for the shuffled and surrogate data in addition to the original one is to identify the possible source(s) for the multifractal signature (if any) in a given time series. In this time series, the slowly changing local Hurst exponent (the local slope), the scaling exponent curve which shows weak non-linearity in $q$, and, of course, the calculated width value for the original time series, which is narrow though not extremely narrow, indicate the presence of weak multifractal signature in this time series. Though the multifractality signature observed in this time series is weak, it is scientific to determine the possible source(s) of this weak multifractal behaviour. For that end, we have applied the same procedure and calculated the width, $\Delta\alpha$, for the shuffled and surrogate data in addition to the original one as we have done for all wave bands to detect the origin(s) of the observed  multifractality behaviour in each time series. The width values for the original, shuffled, and surrogate data are calculated as $\Delta\alpha = 0.0599 \slash 0.0065 \slash 0.0634$, respectively (Table \ref{tab1}), which further shows that the multifractality signature in this time series is weak and it is mainly due to the temporal correlation in the original time series. The shape of the multifractal spectrum reflects the temporal variation of the local Hurst exponent, and is helpful in classifying a small and large scale fluctuations (intermittency) in the time series. Therefore, in this case, the right-side truncated spectrum indicates the existence of small-scale fluctuation with large magnitudes \citep{2014Ap&SS.350...47T} unlike the nature of fluctuations (intermittencies) observed in the mm, UV, and X-ray bands where the spectrum is left-side truncated.

\subsection{Optical observations}
\begin{figure*}
\centering
\includegraphics[scale=0.14]{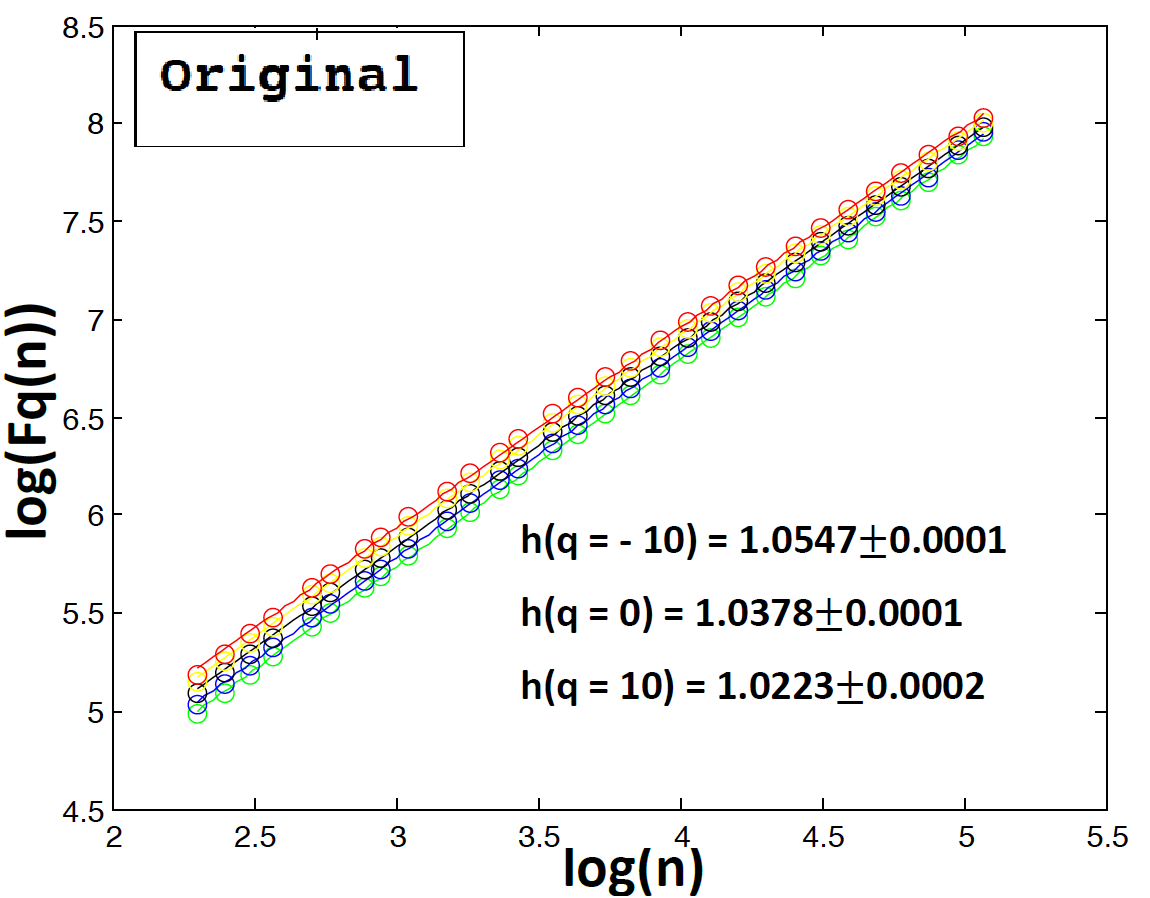} 
\includegraphics[scale=0.14]{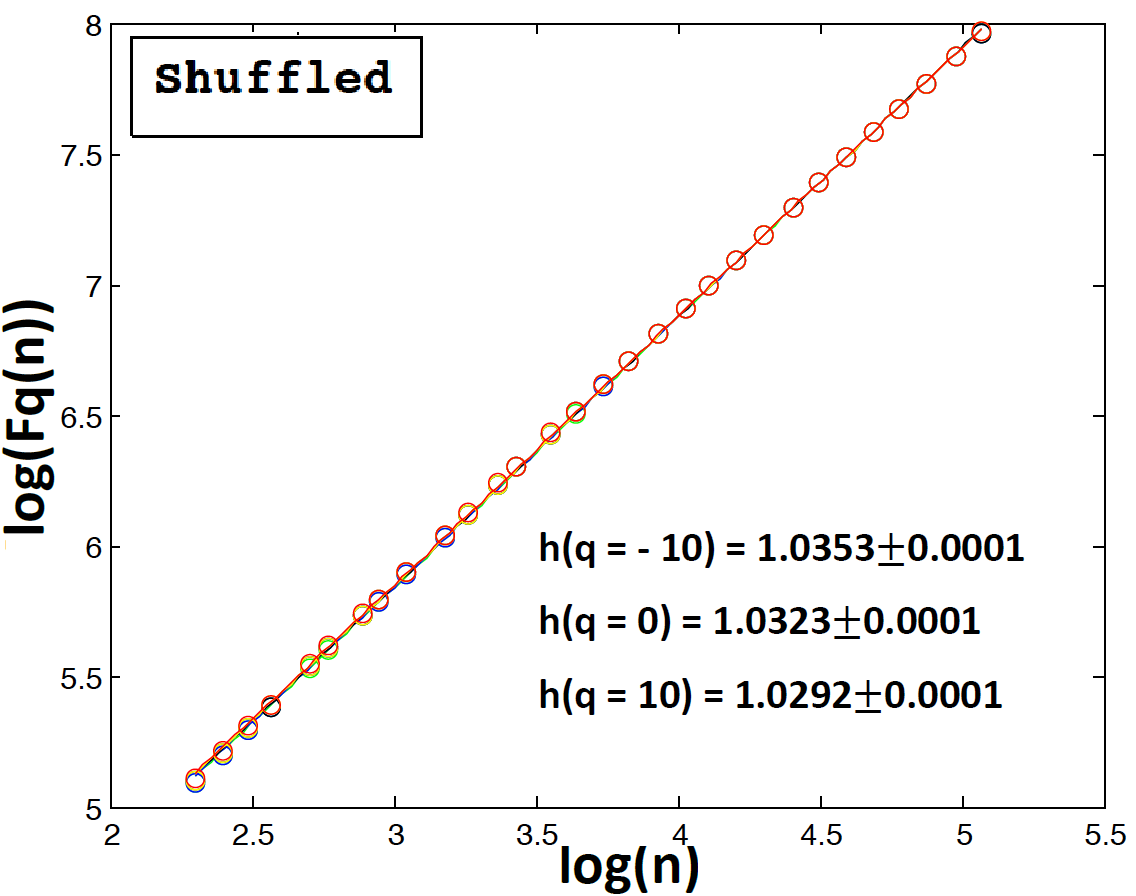}
\includegraphics[scale=0.14]{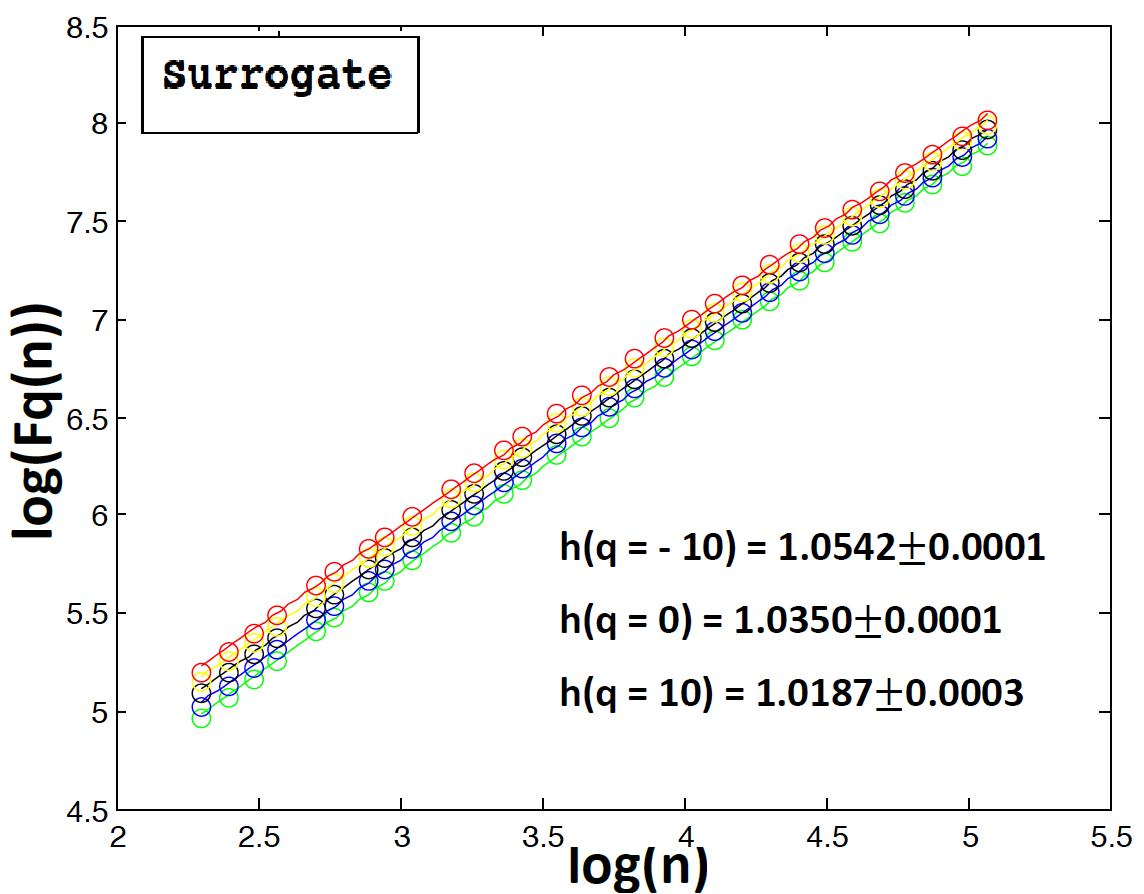}
\includegraphics[scale=0.14]{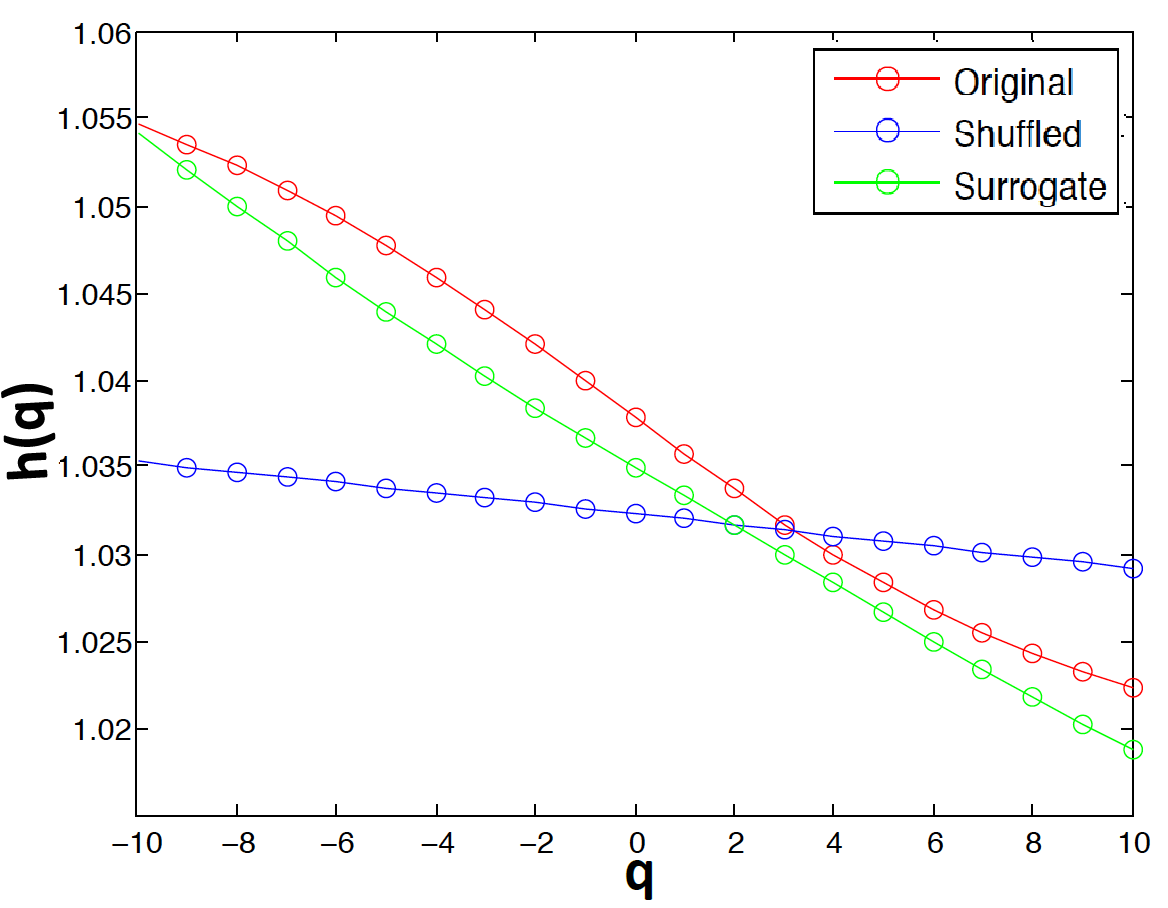}
\includegraphics[scale=0.14]{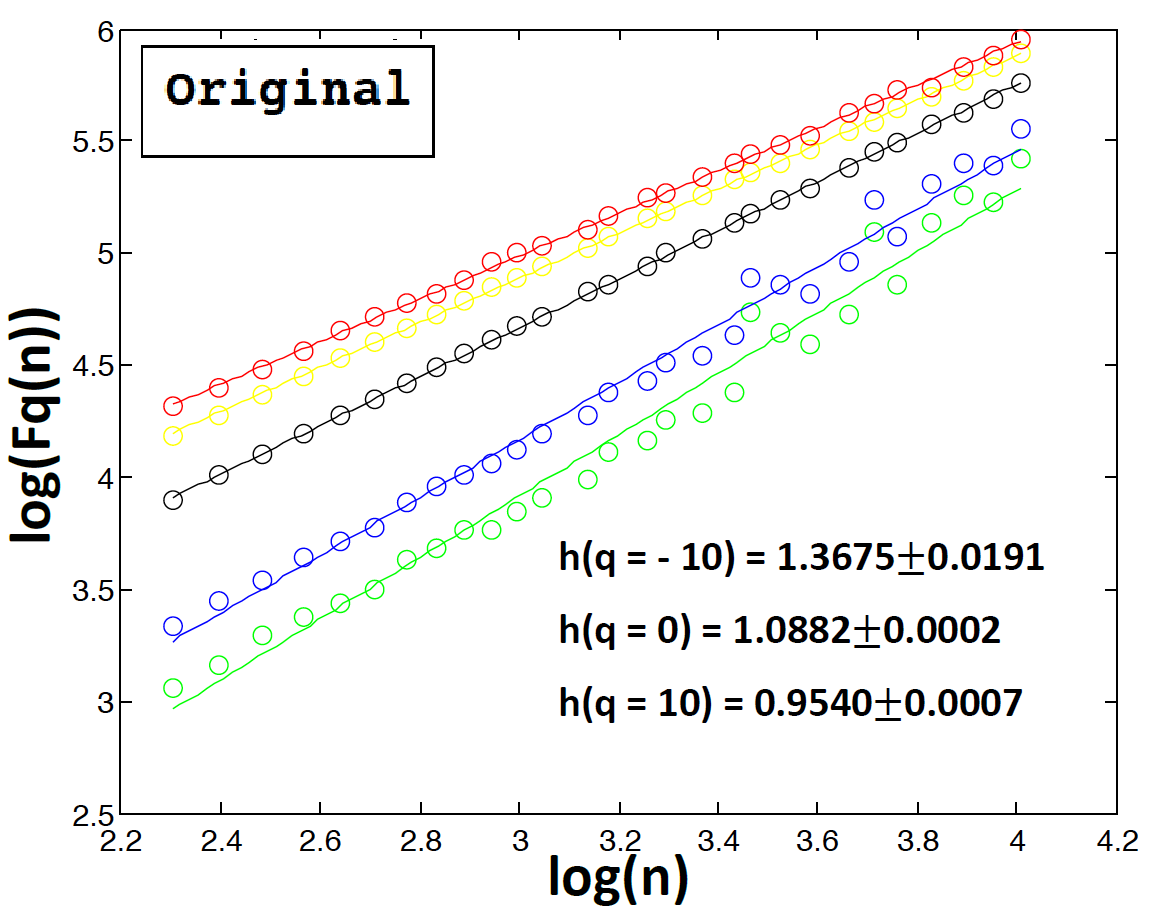} 
\includegraphics[scale=0.14]{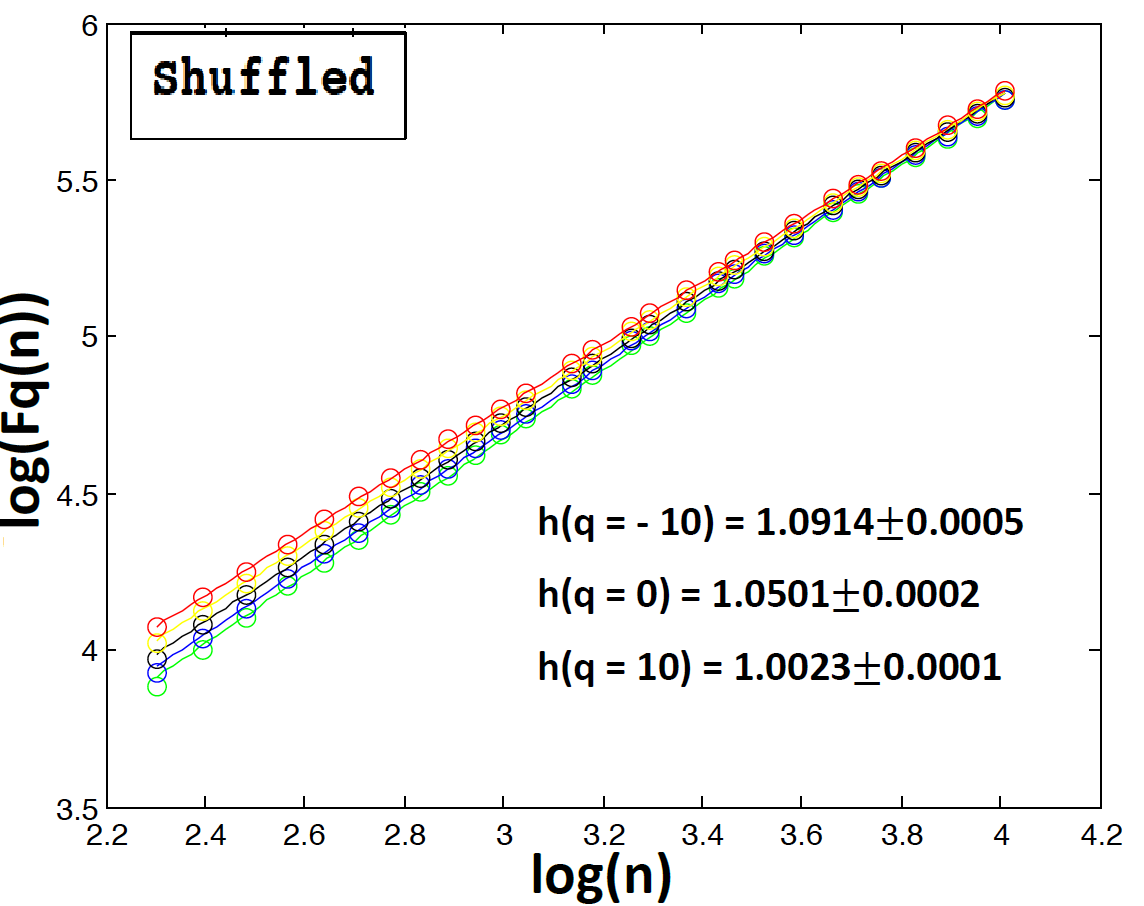}
\includegraphics[scale=0.14]{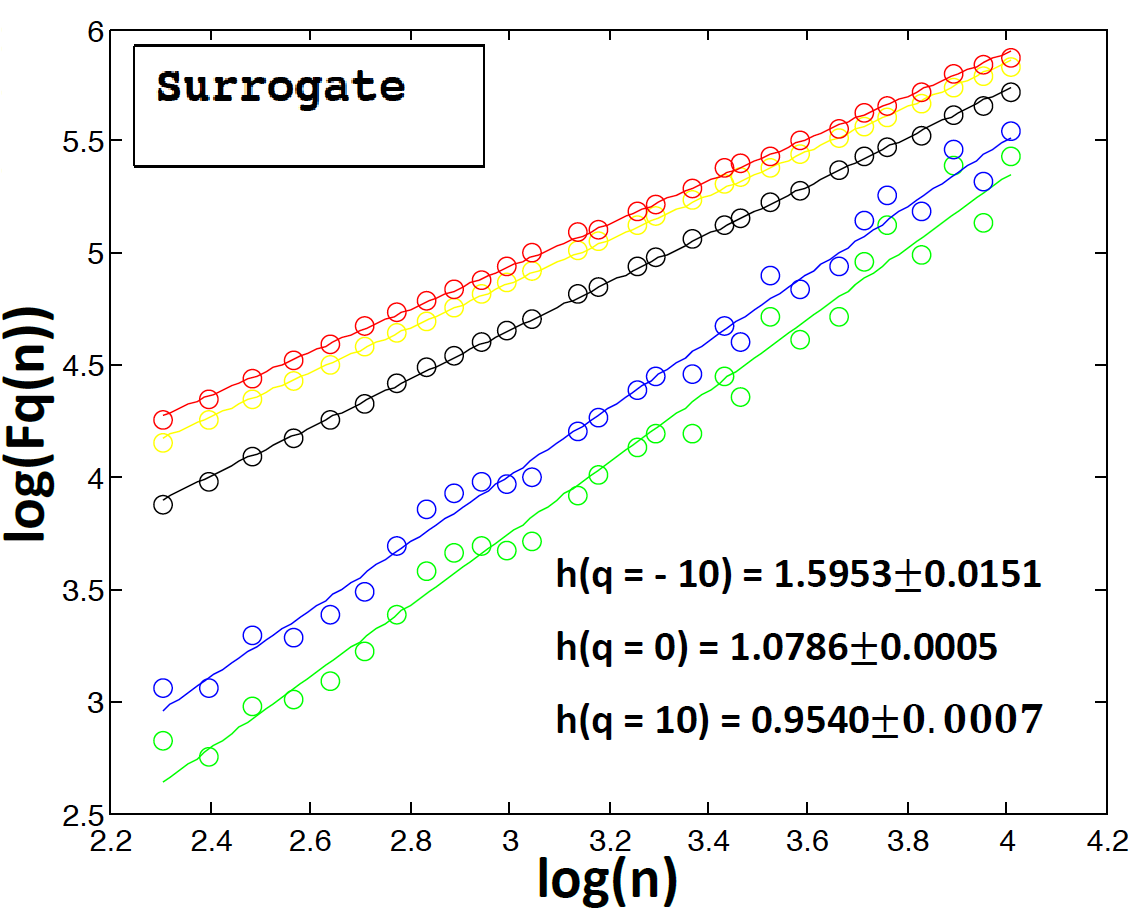}
\includegraphics[scale=0.14]{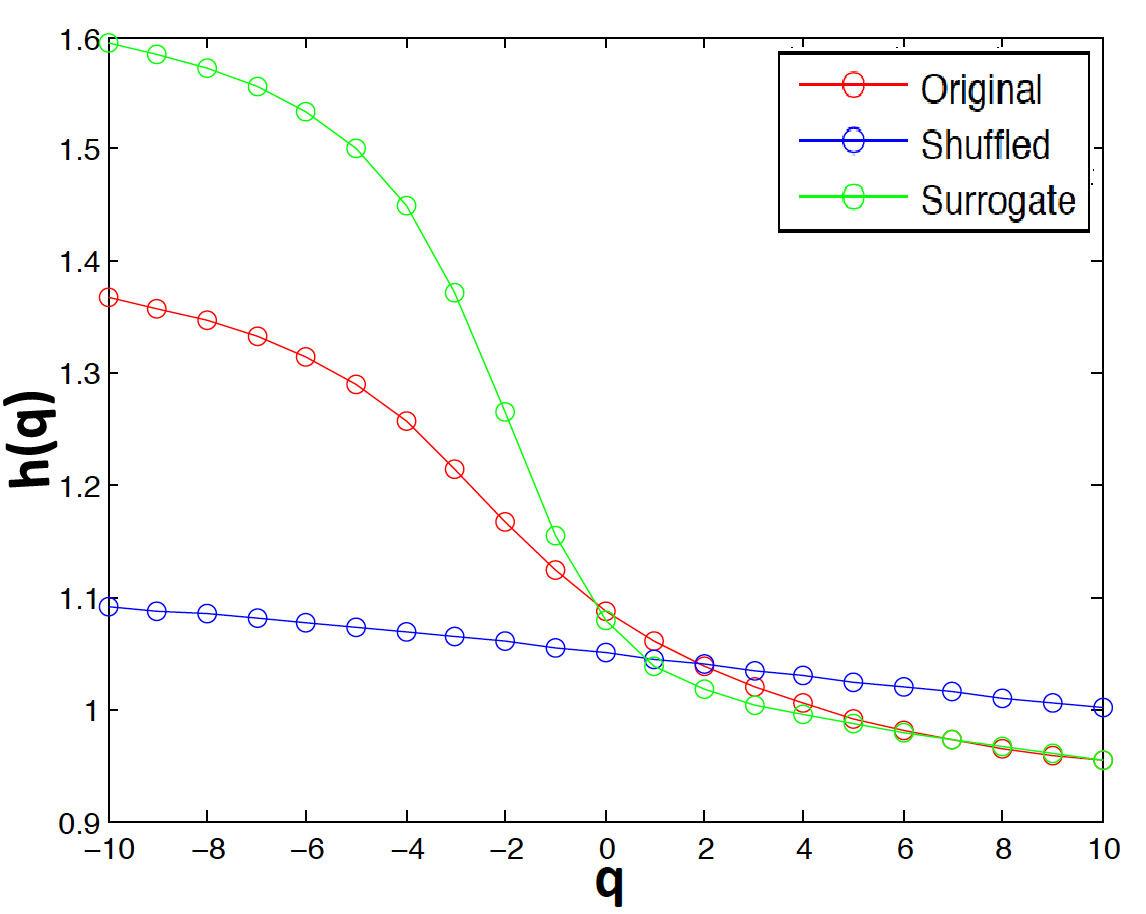}
\includegraphics[scale=0.14]{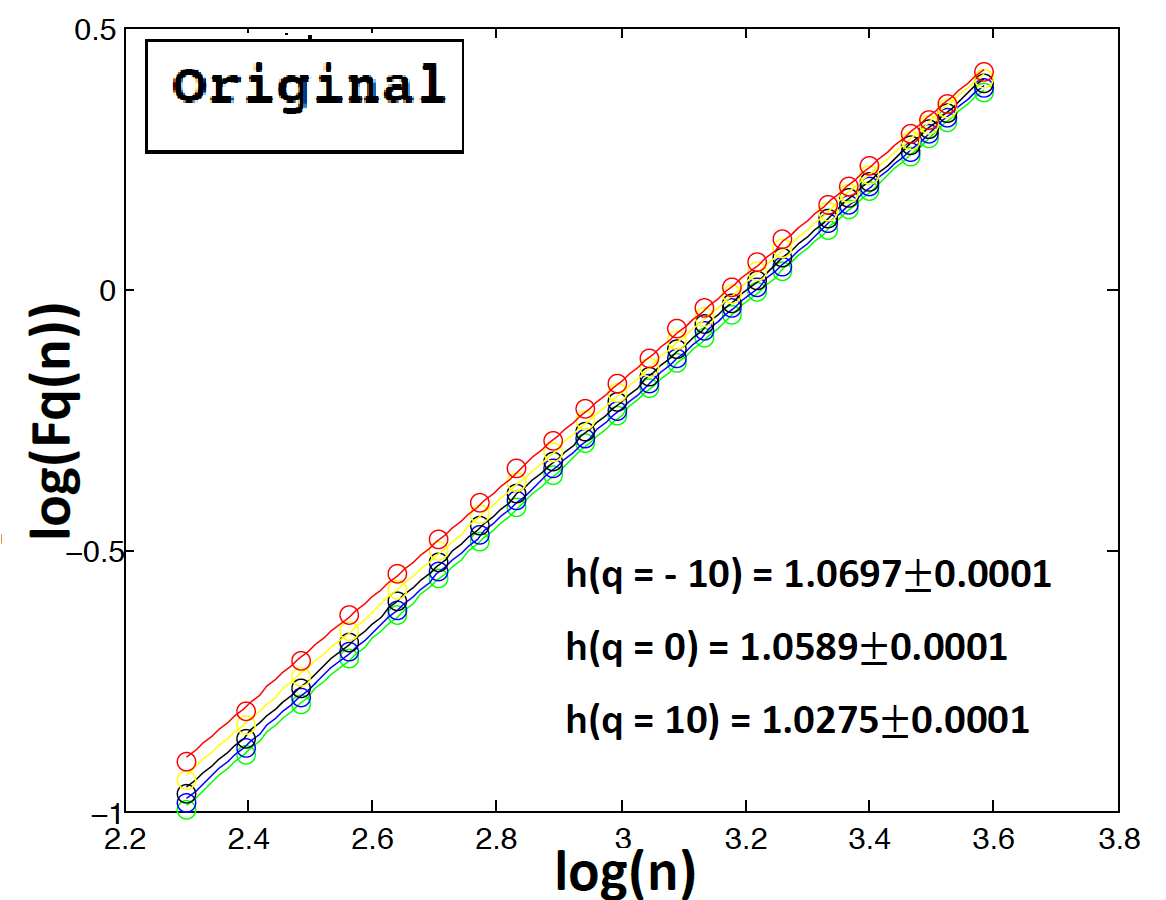} 
\includegraphics[scale=0.14]{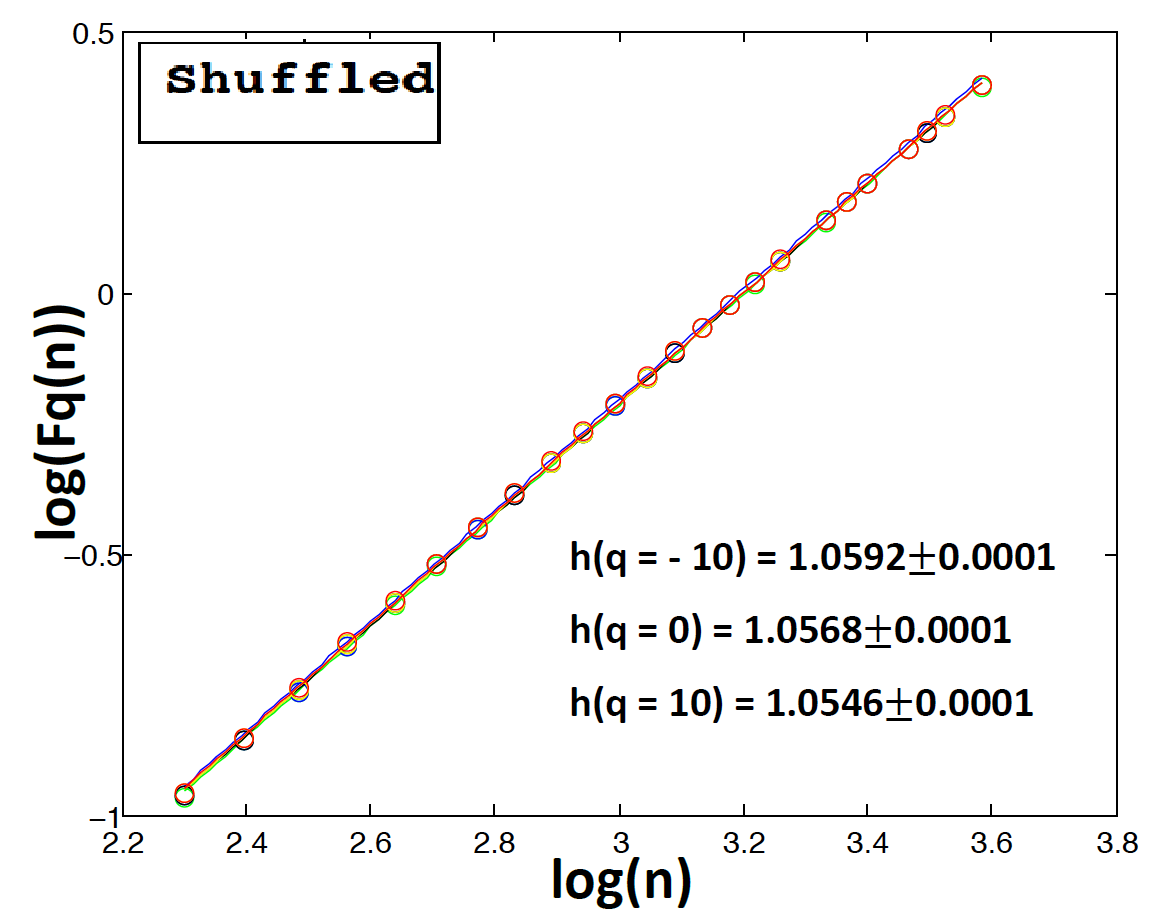}
\includegraphics[scale=0.14]{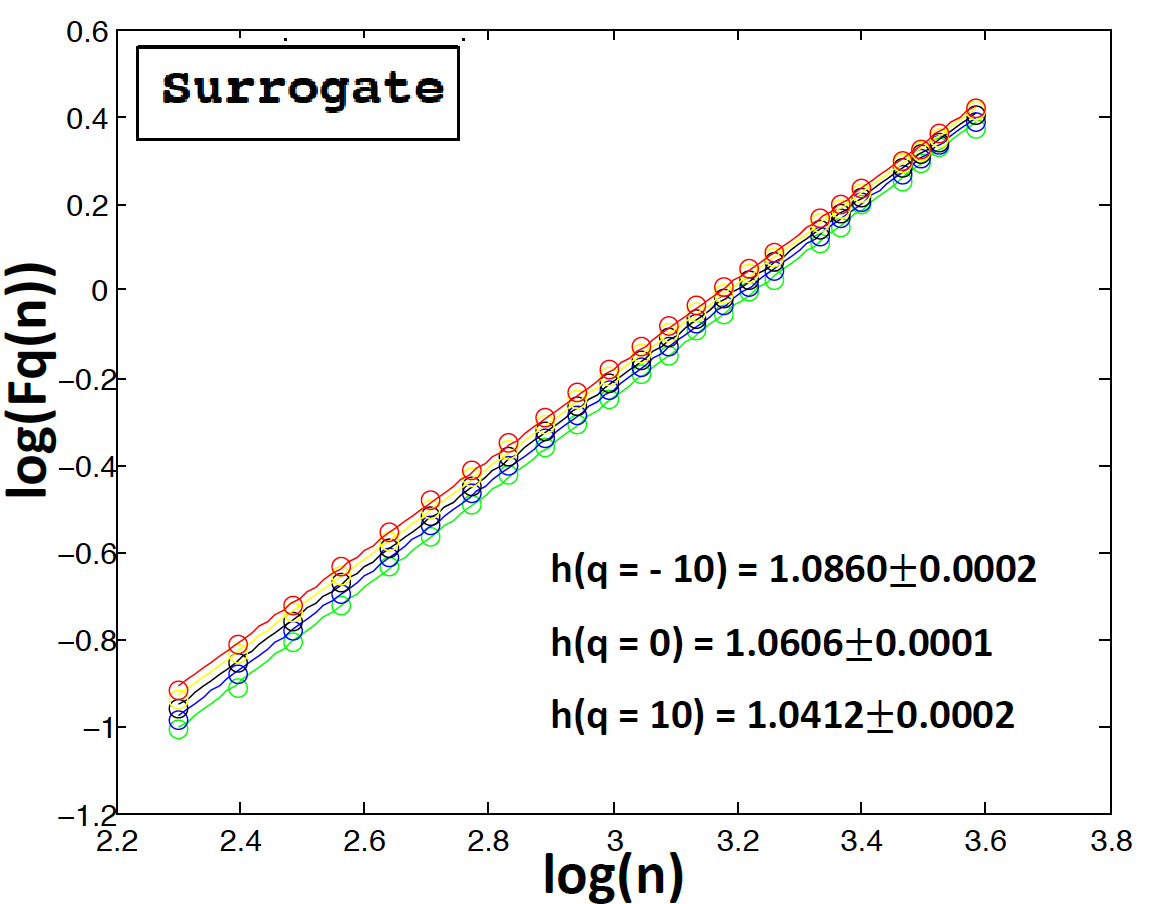}
\includegraphics[scale=0.14]{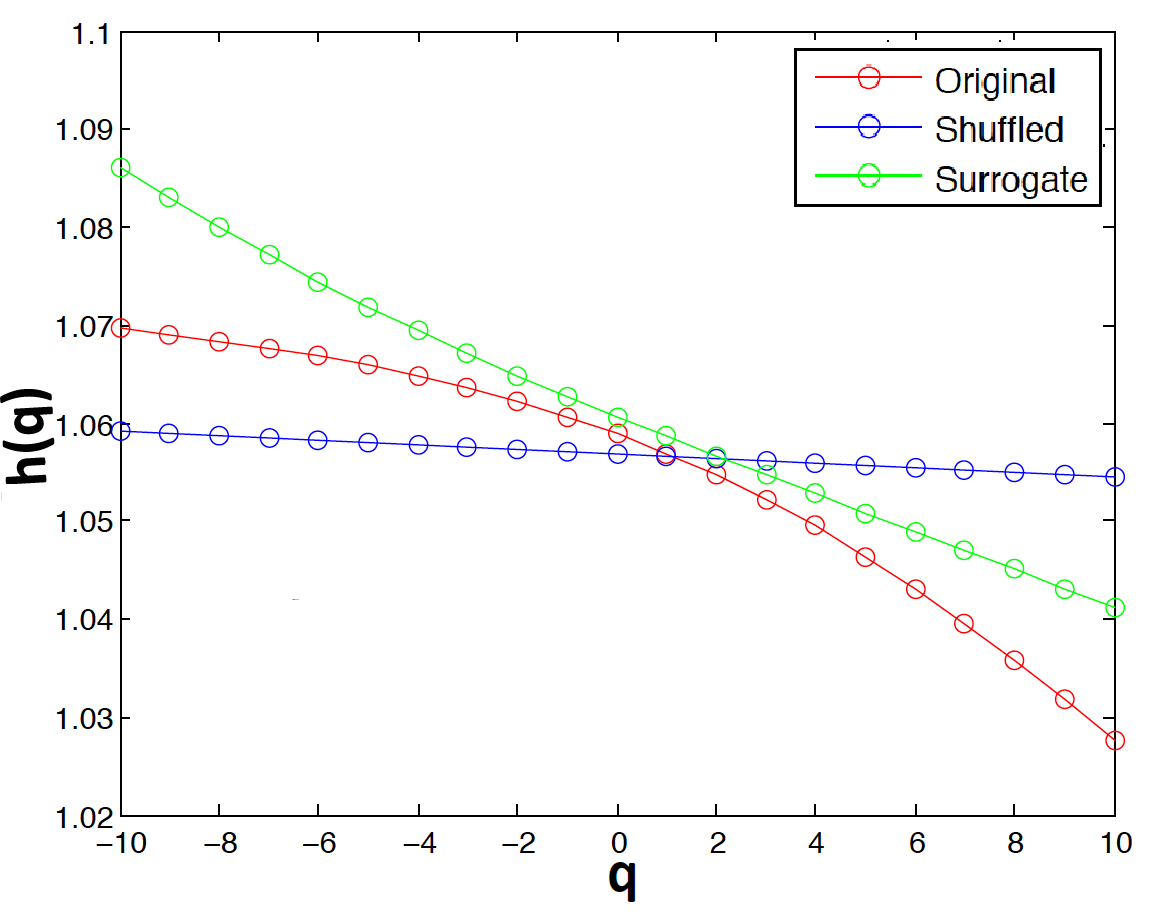}
\includegraphics[scale=0.14]{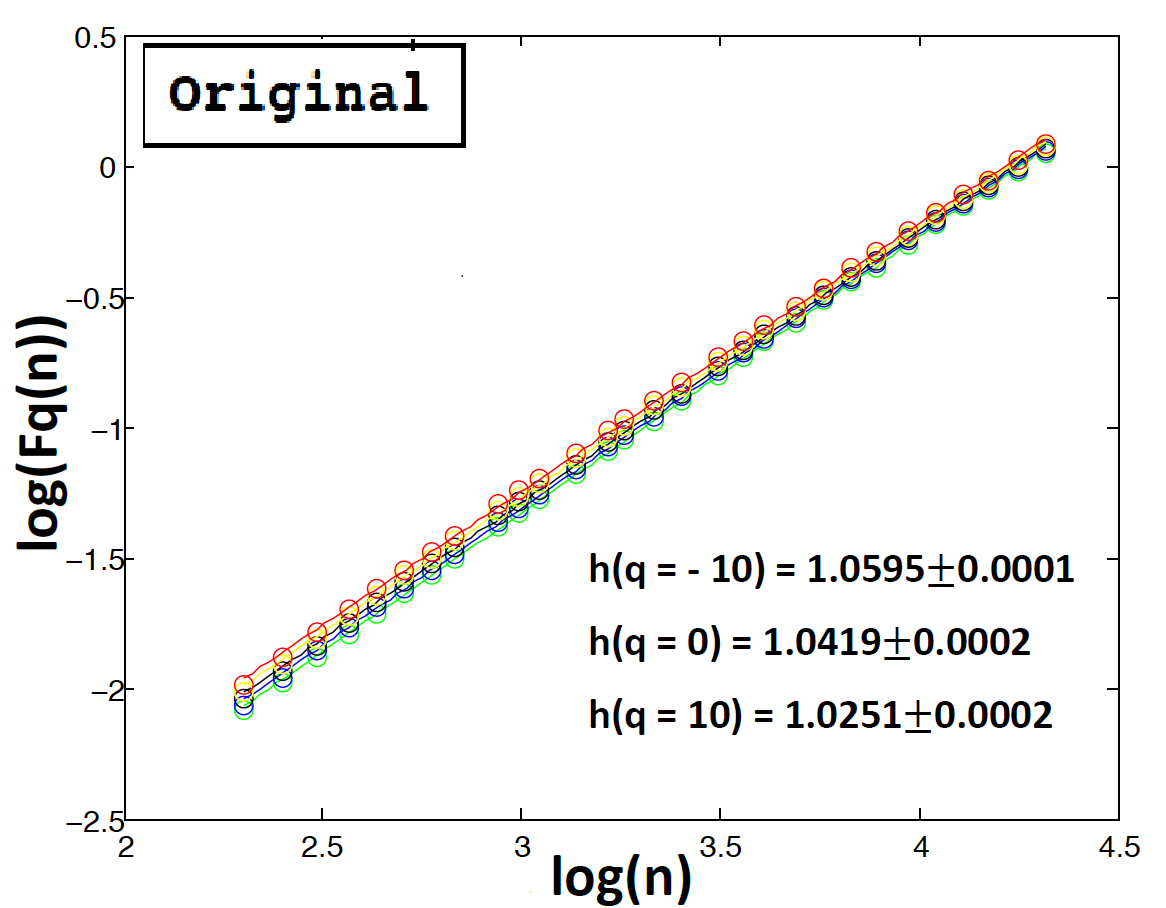} 
\includegraphics[scale=0.14]{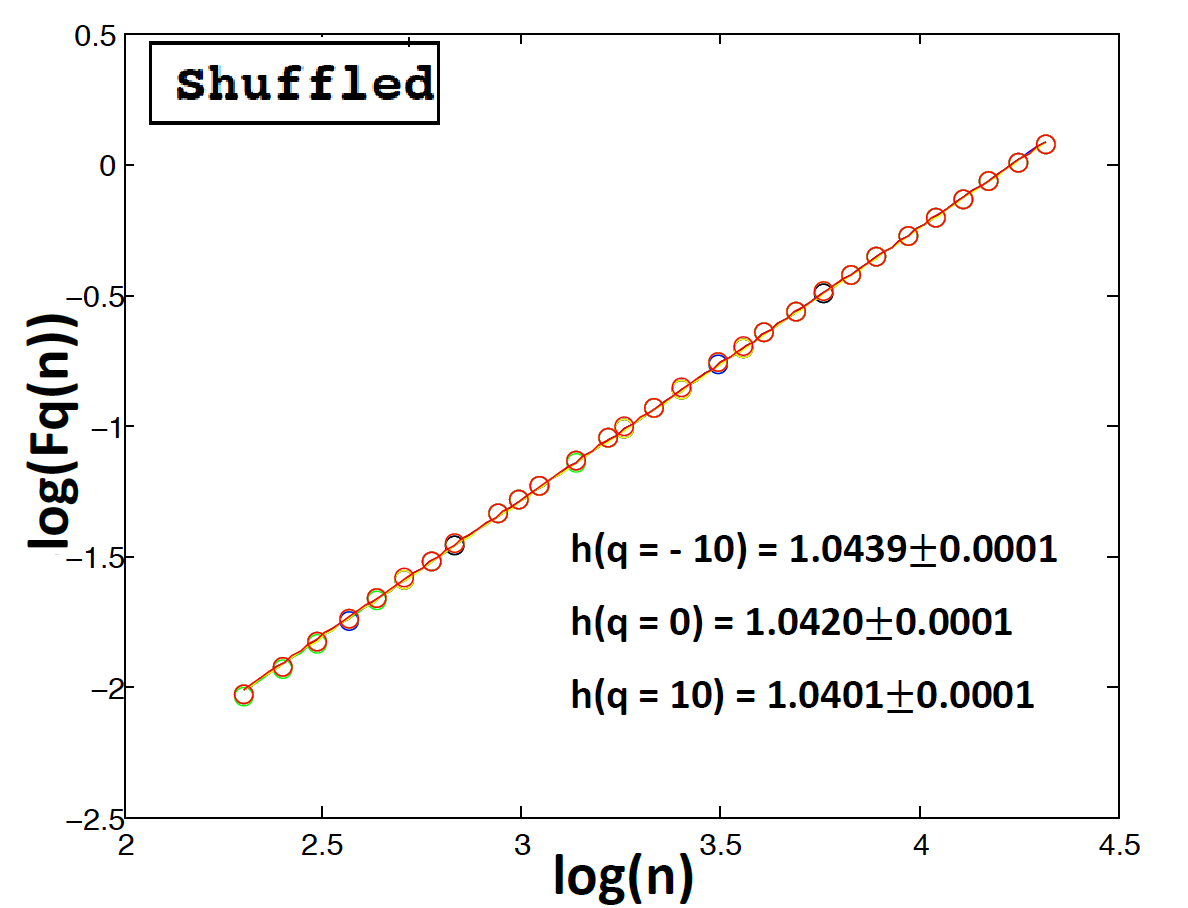}
\includegraphics[scale=0.14]{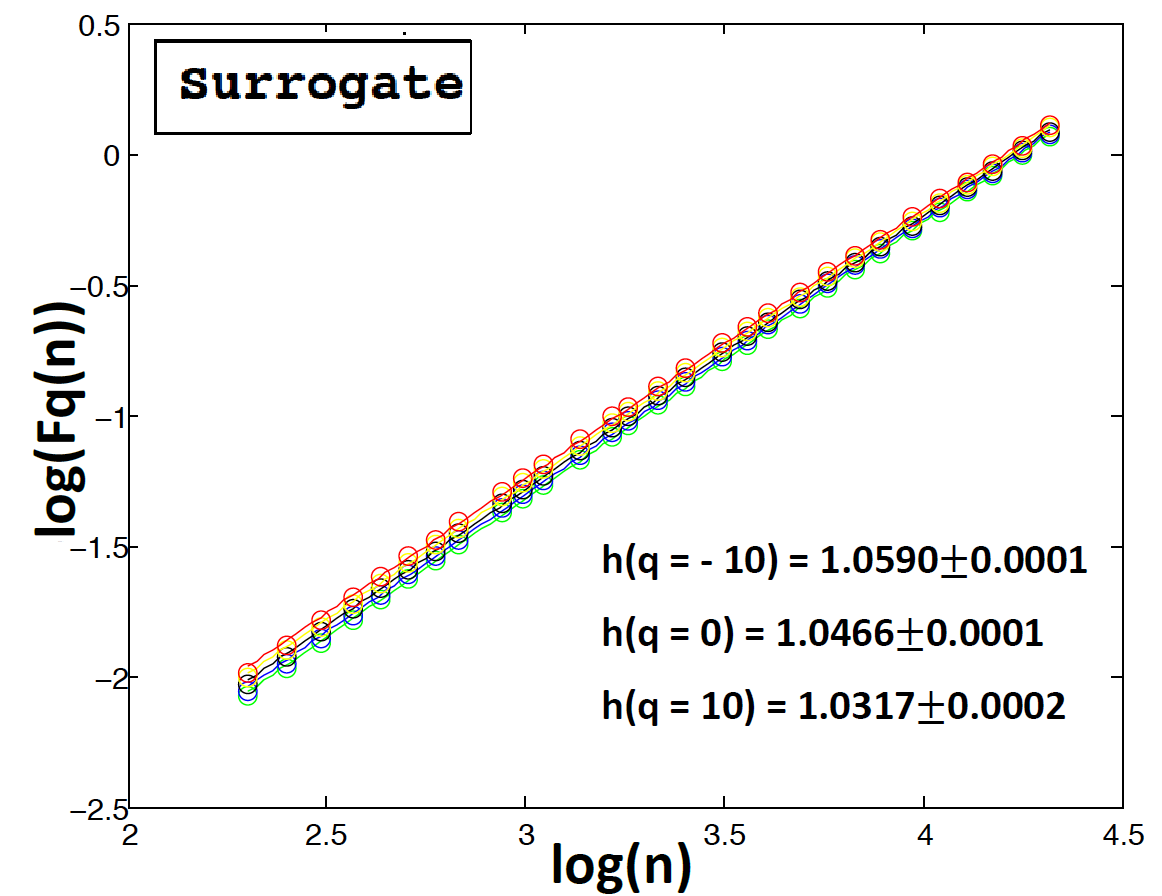}
\includegraphics[scale=0.14]{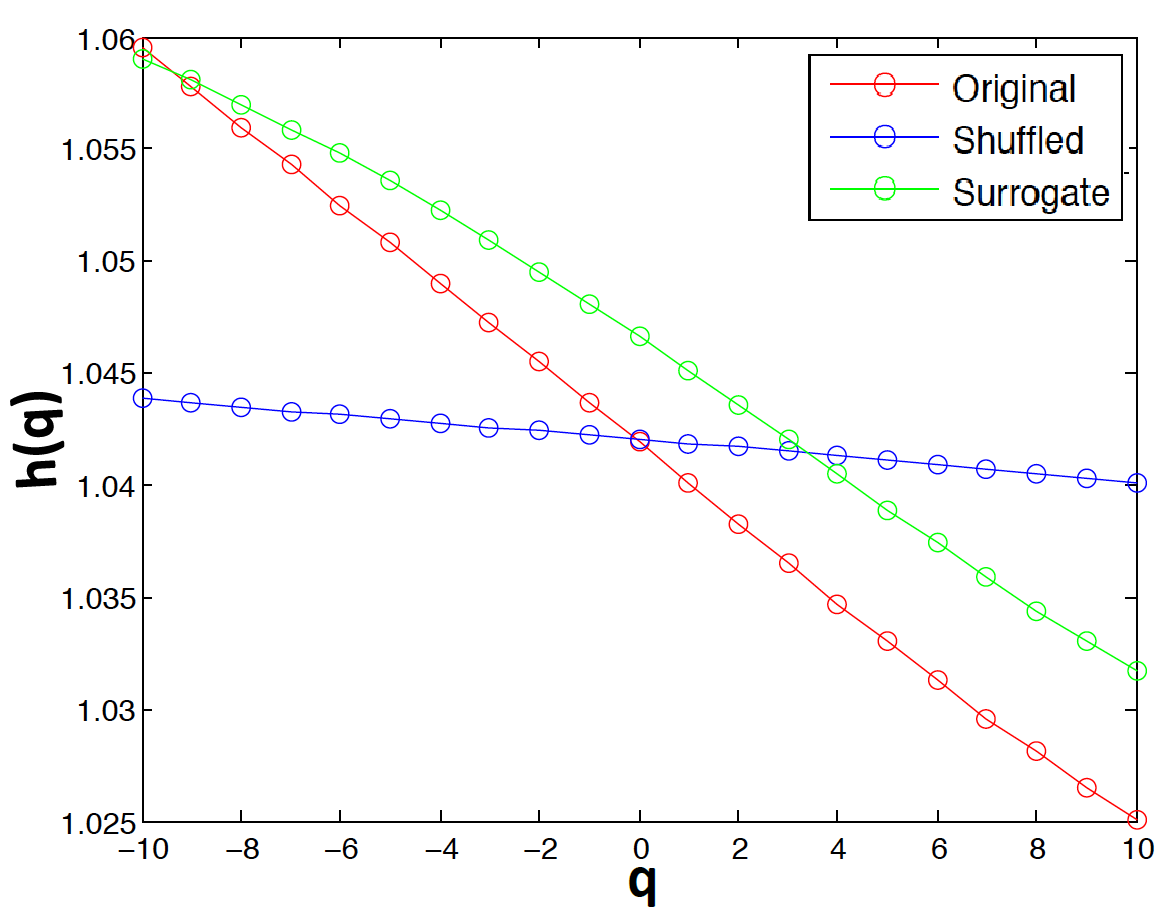}
\includegraphics[scale=0.14]{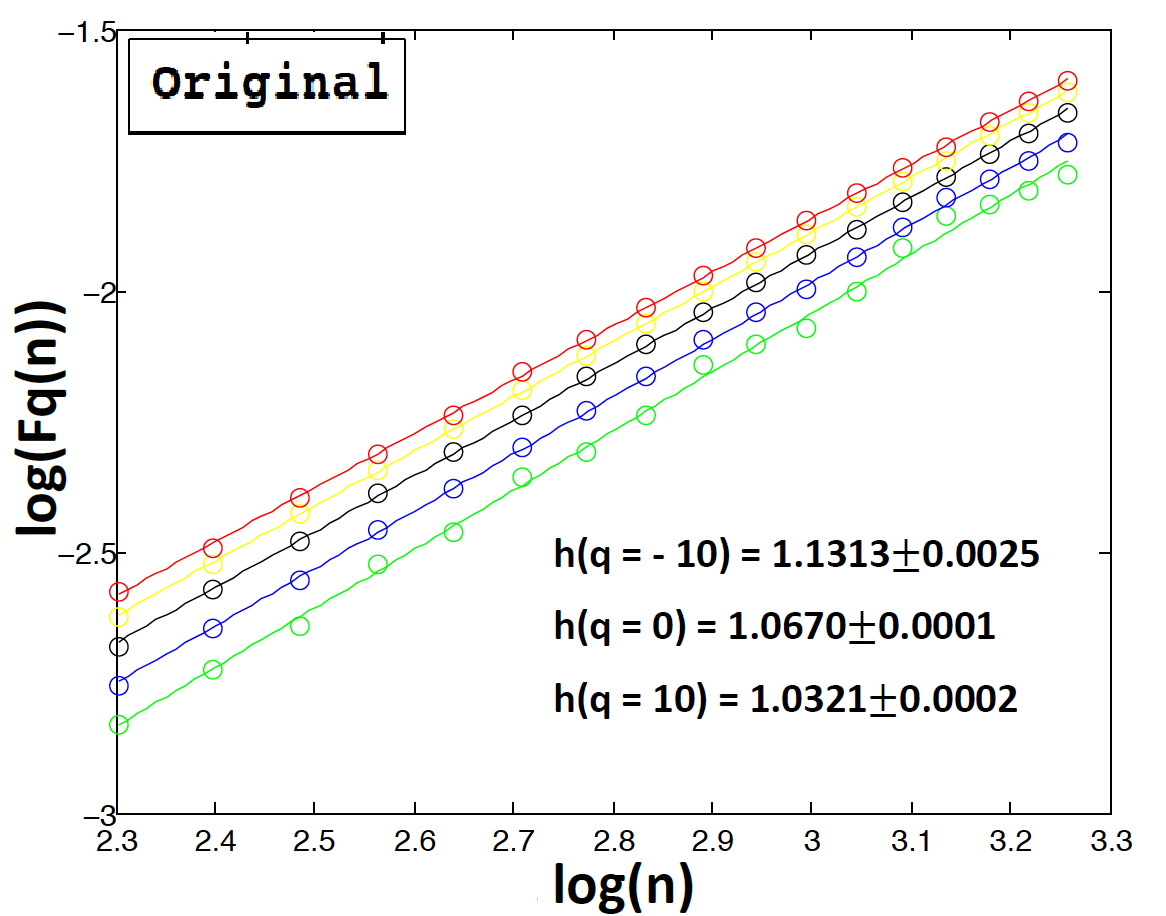} 
\includegraphics[scale=0.14]{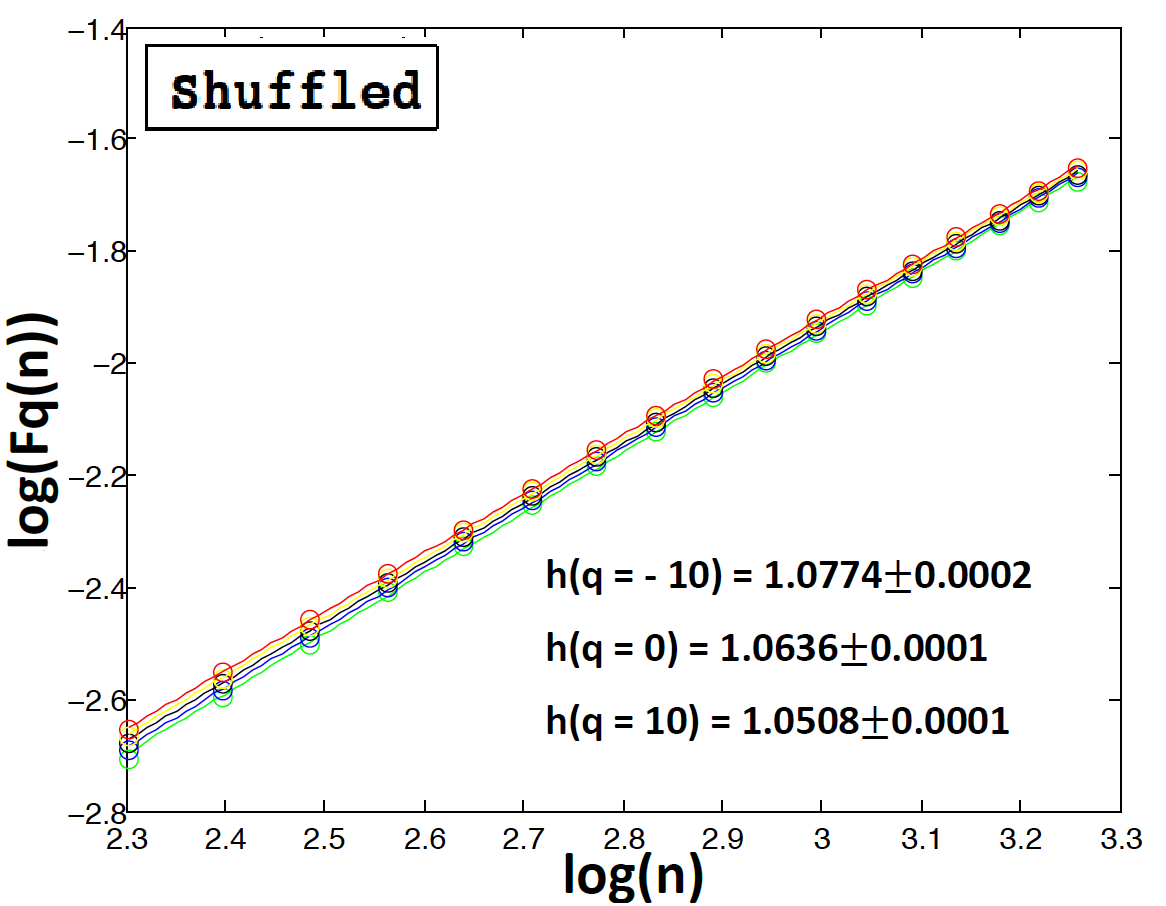}
\includegraphics[scale=0.14]{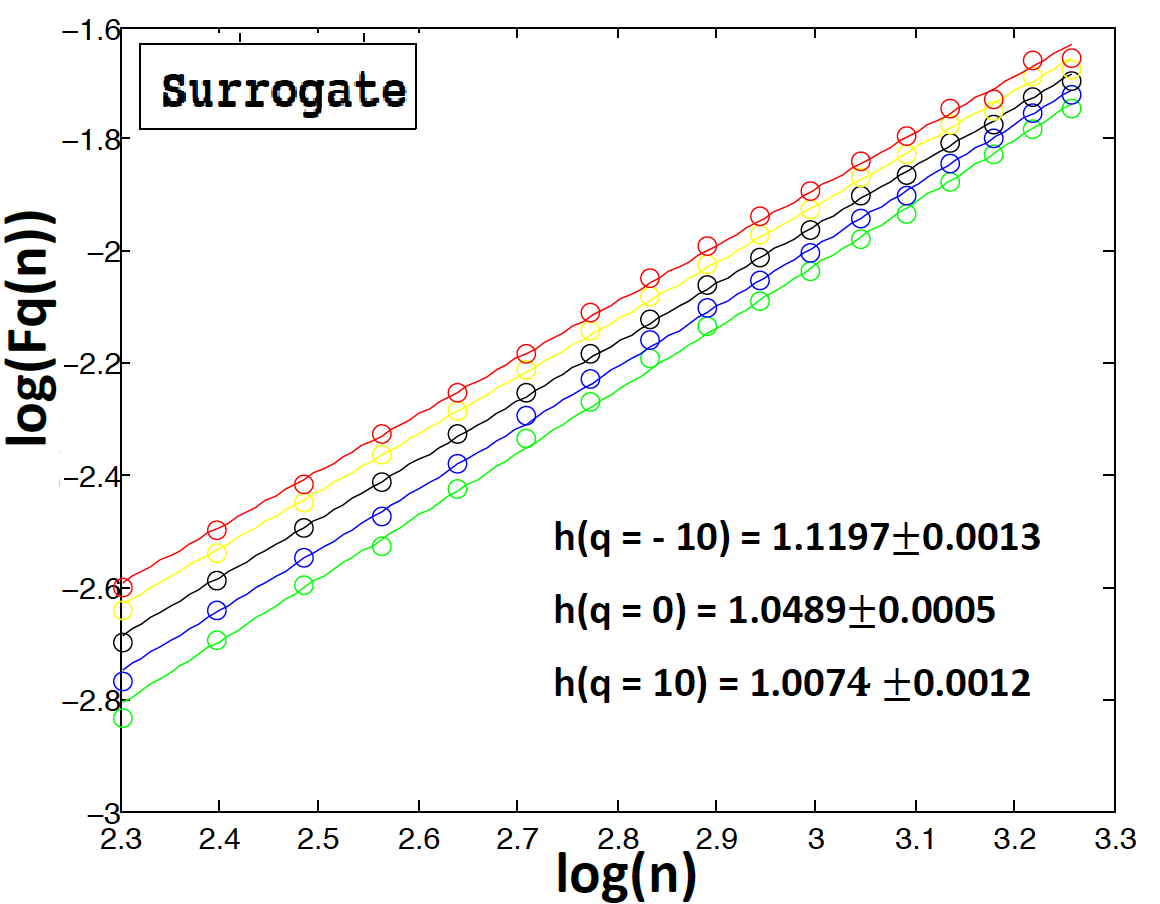}
\includegraphics[scale=0.14]{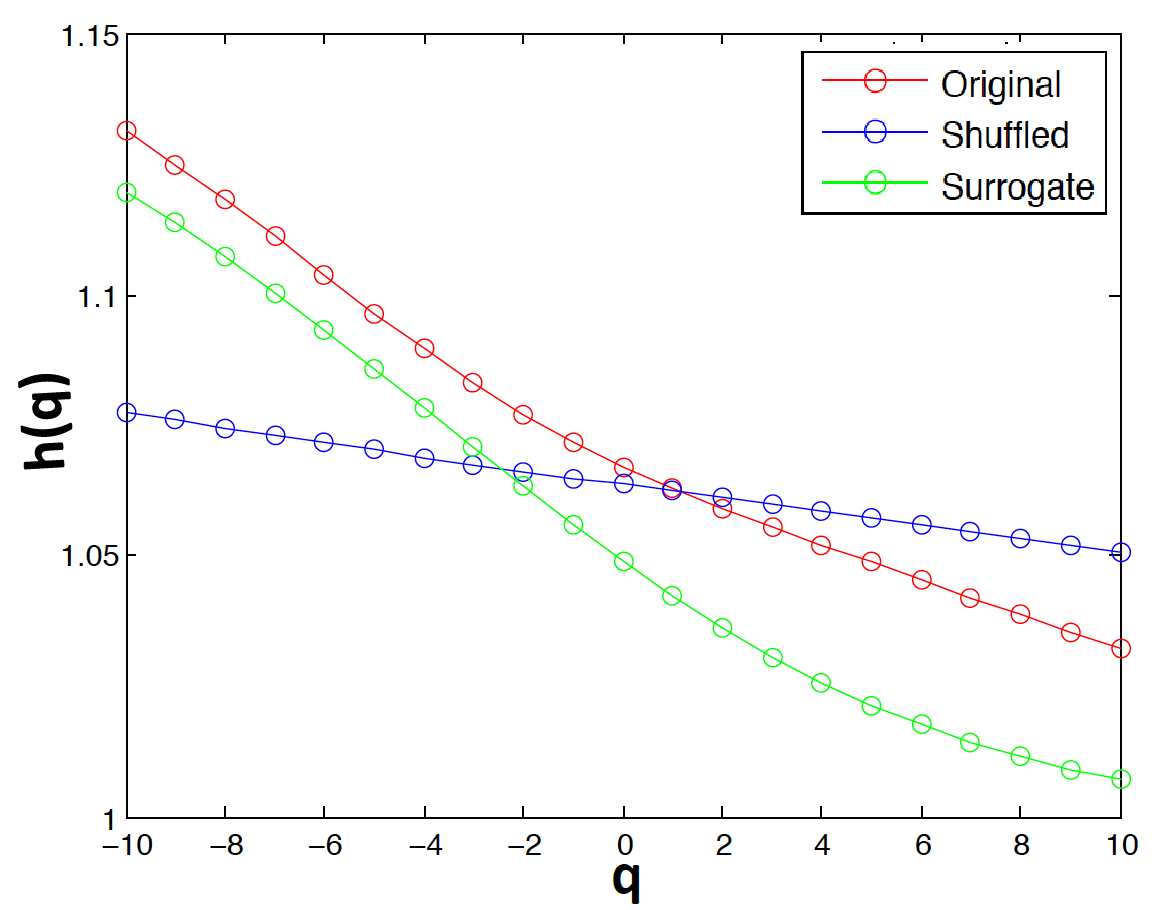}
\includegraphics[scale=0.14]{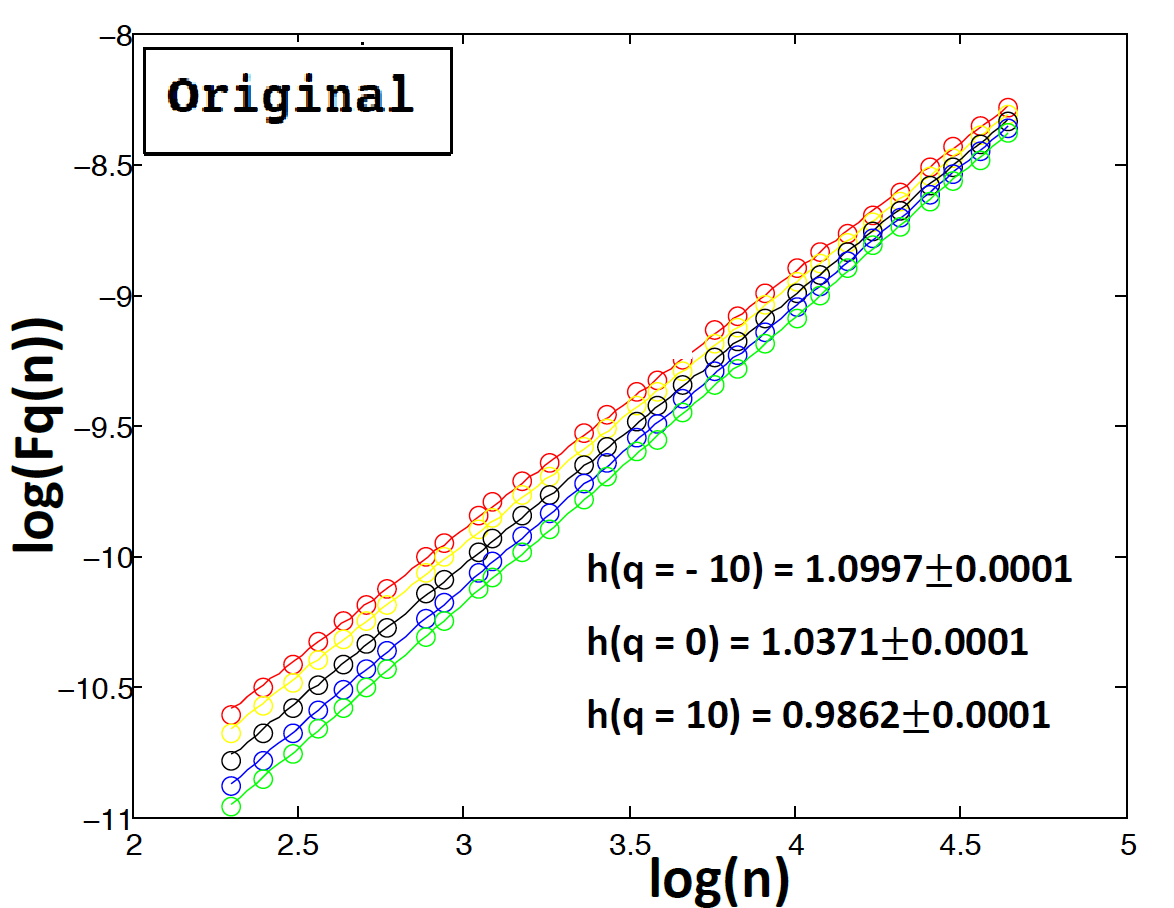} 
\includegraphics[scale=0.14]{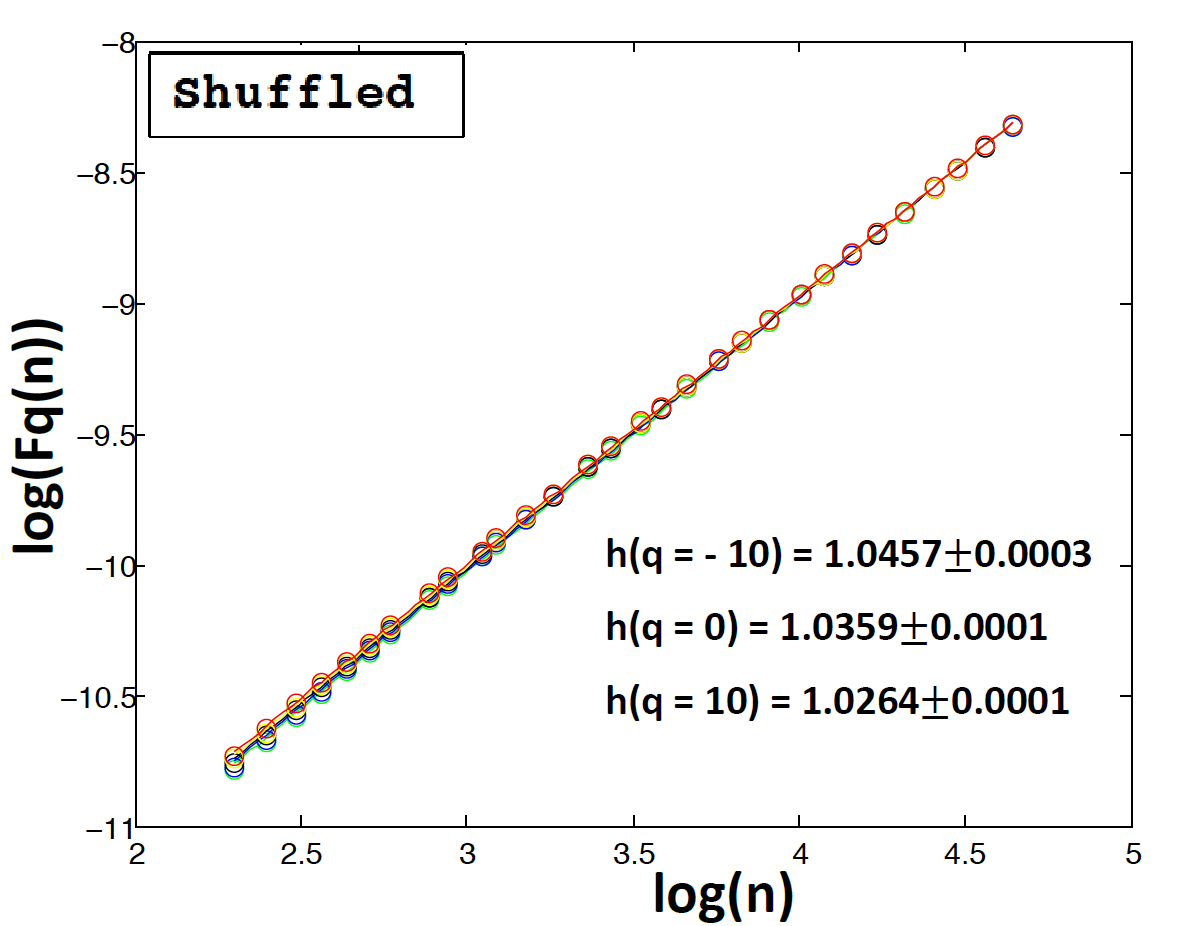}
\includegraphics[scale=0.14]{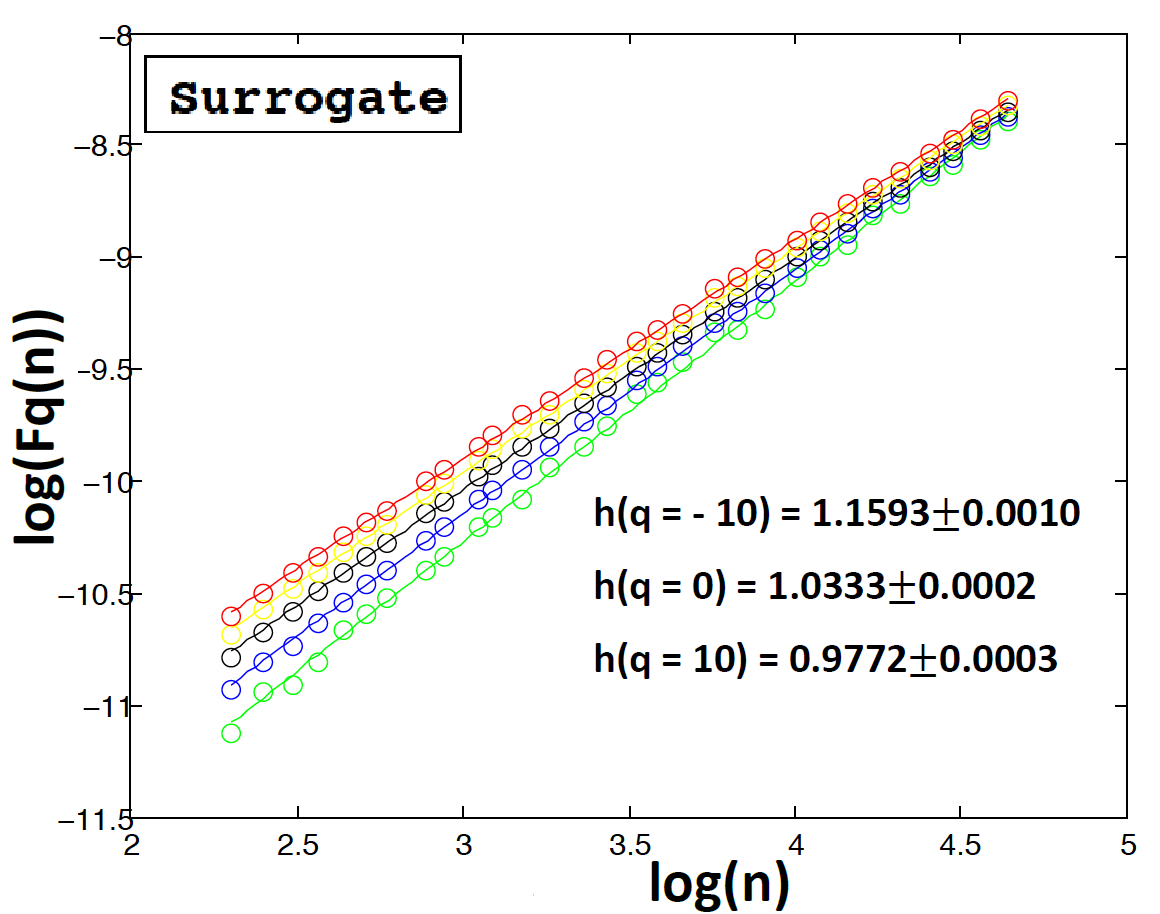}
\includegraphics[scale=0.14]{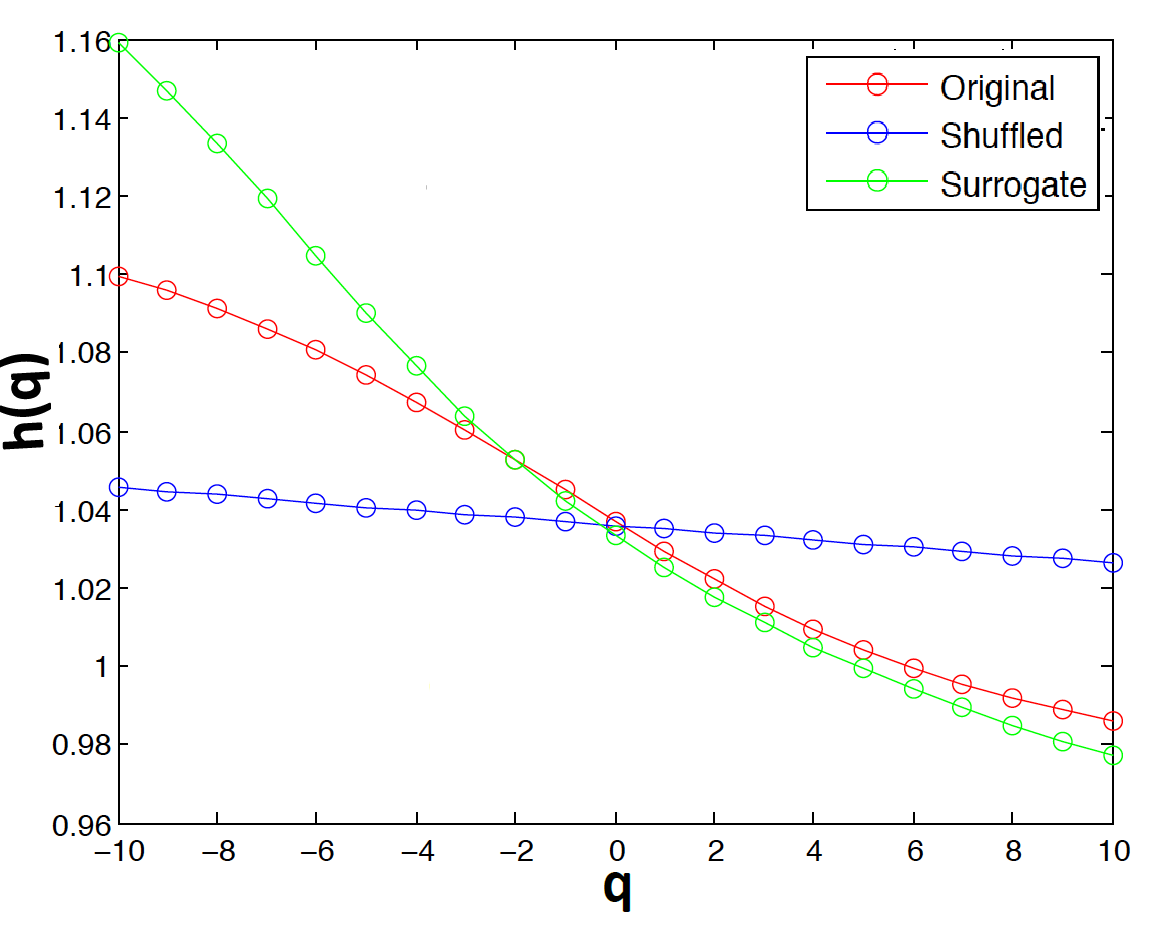}
\caption{The logarithm of the fluctuation function, log($F_{q}(n)$), versus the log of the time scale (segment), log($n$),  for the original, shuffled and surrogate time series considering, from top to bottom, radio, mm, IR, optical, UV and X-ray observations. For all cases: $q = -10$ (green), $q = -5$ (blue), $q = 0$ (black), $q = 5$ (yellow) and $q = 10$ (red). The solid lines are the best fit lines. The slopes $h(q)$ versus $q$ are displayed in the far right side column.}
\label{fig2}
\end{figure*}

\begin{figure*}
\centering
\includegraphics[scale=0.15]{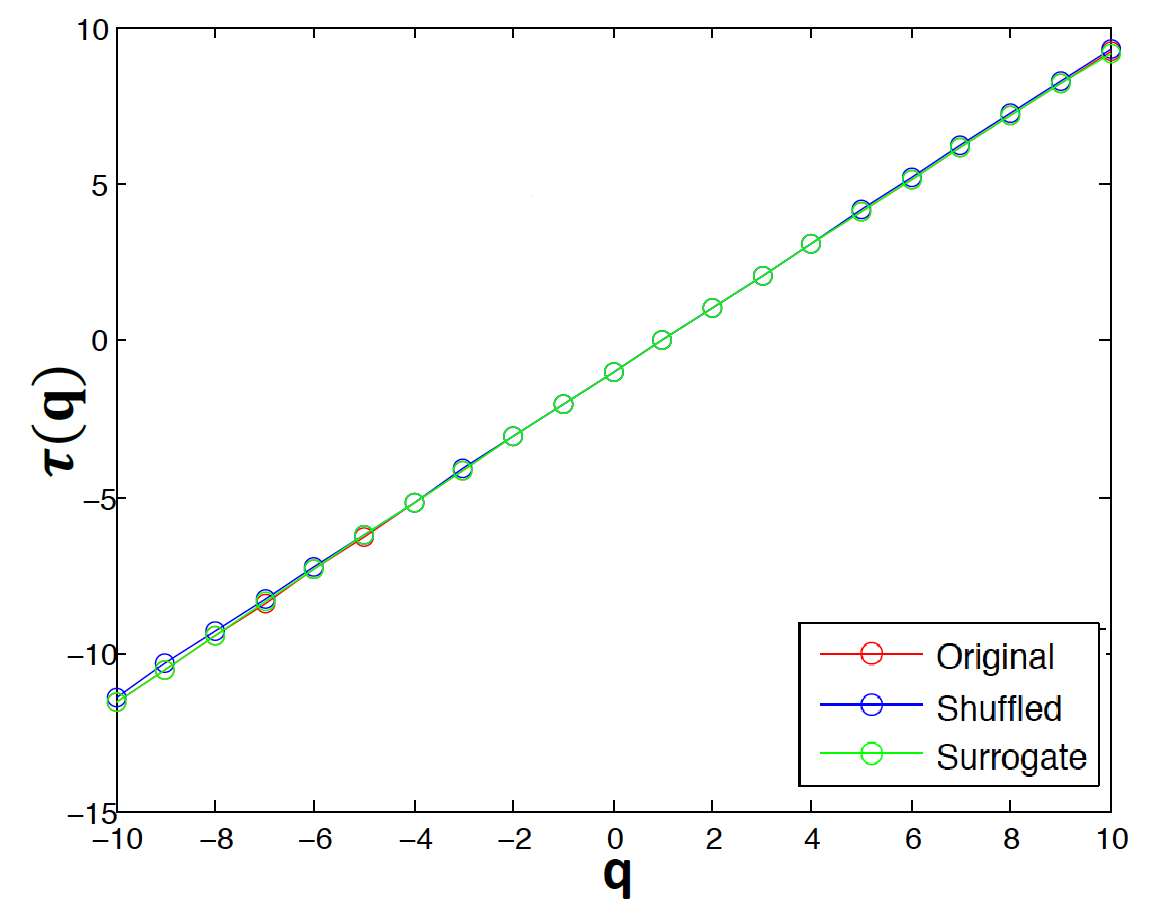} 
\includegraphics[scale=0.15]{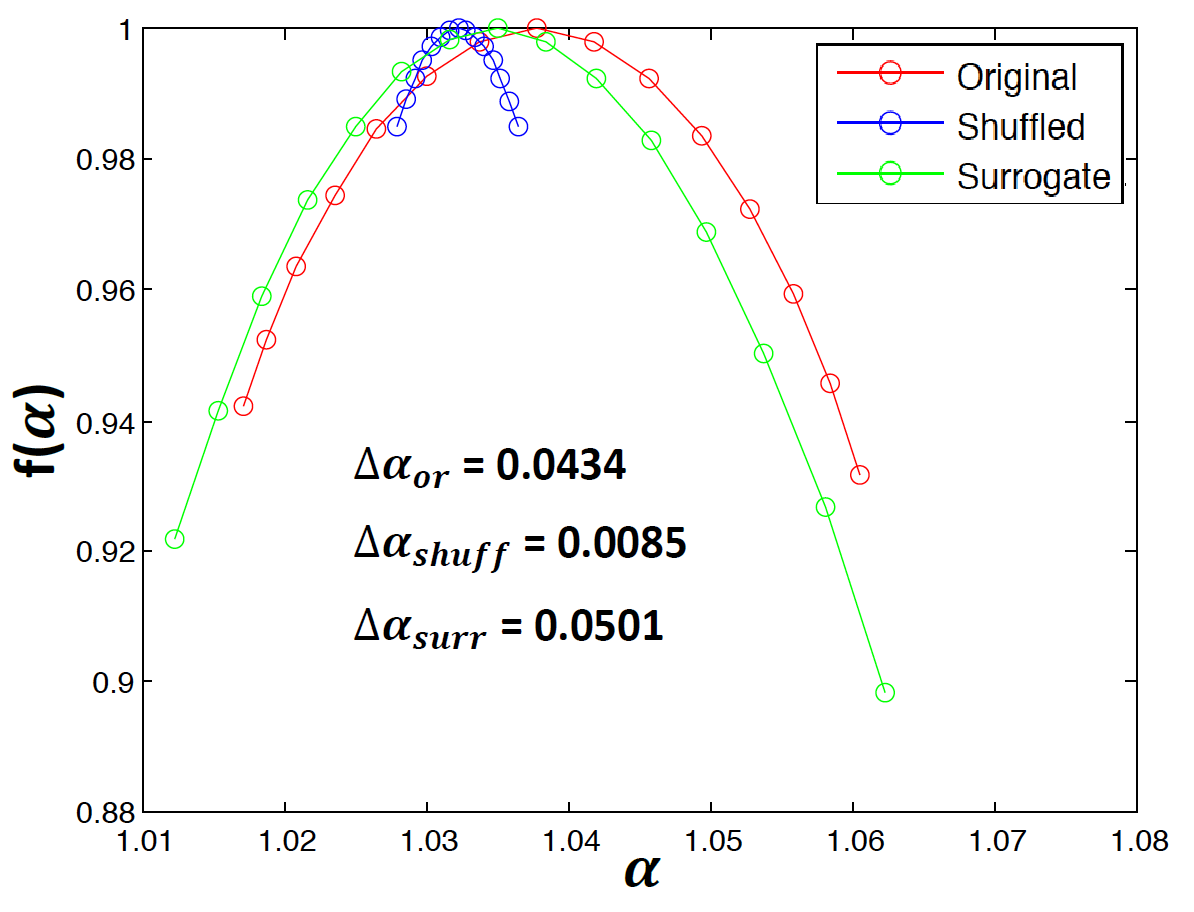}
\includegraphics[scale=0.15]{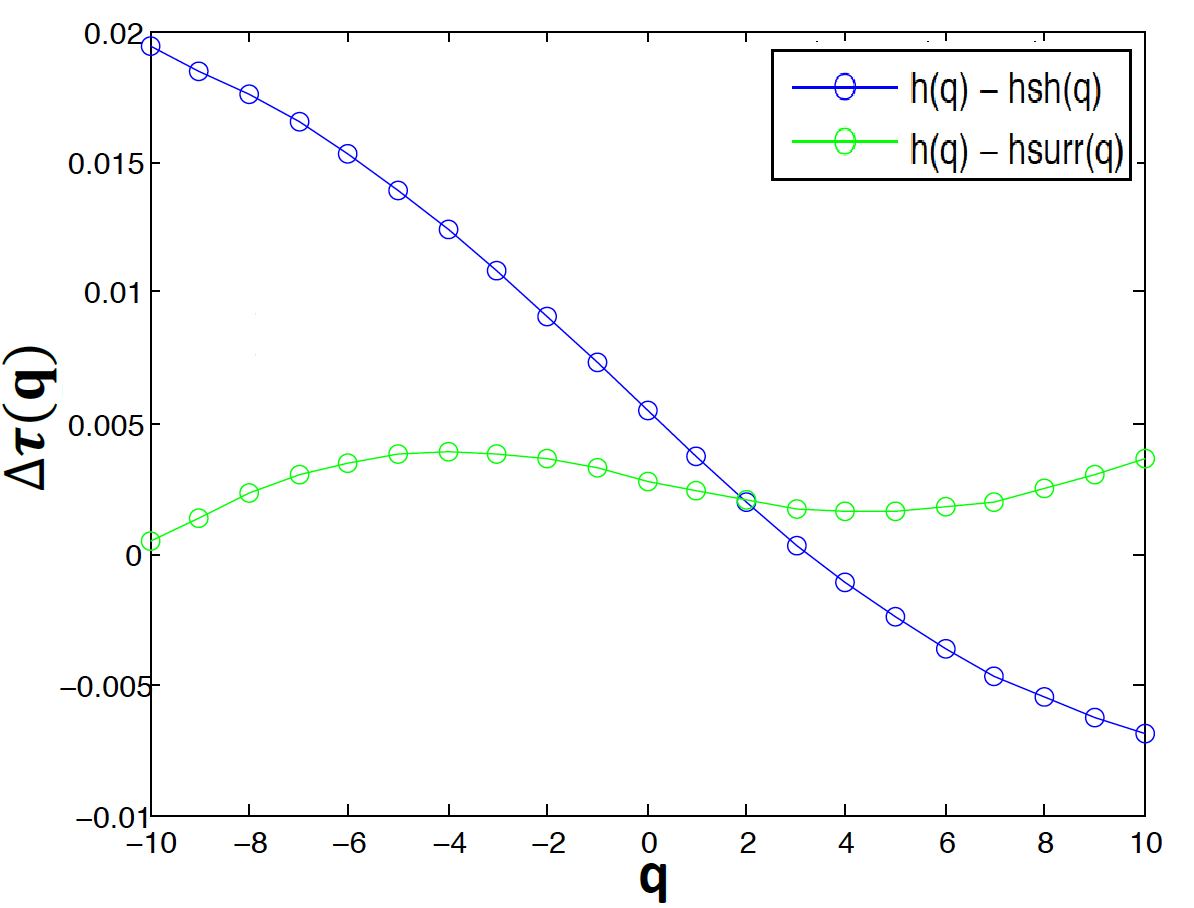}
\includegraphics[scale=0.15]{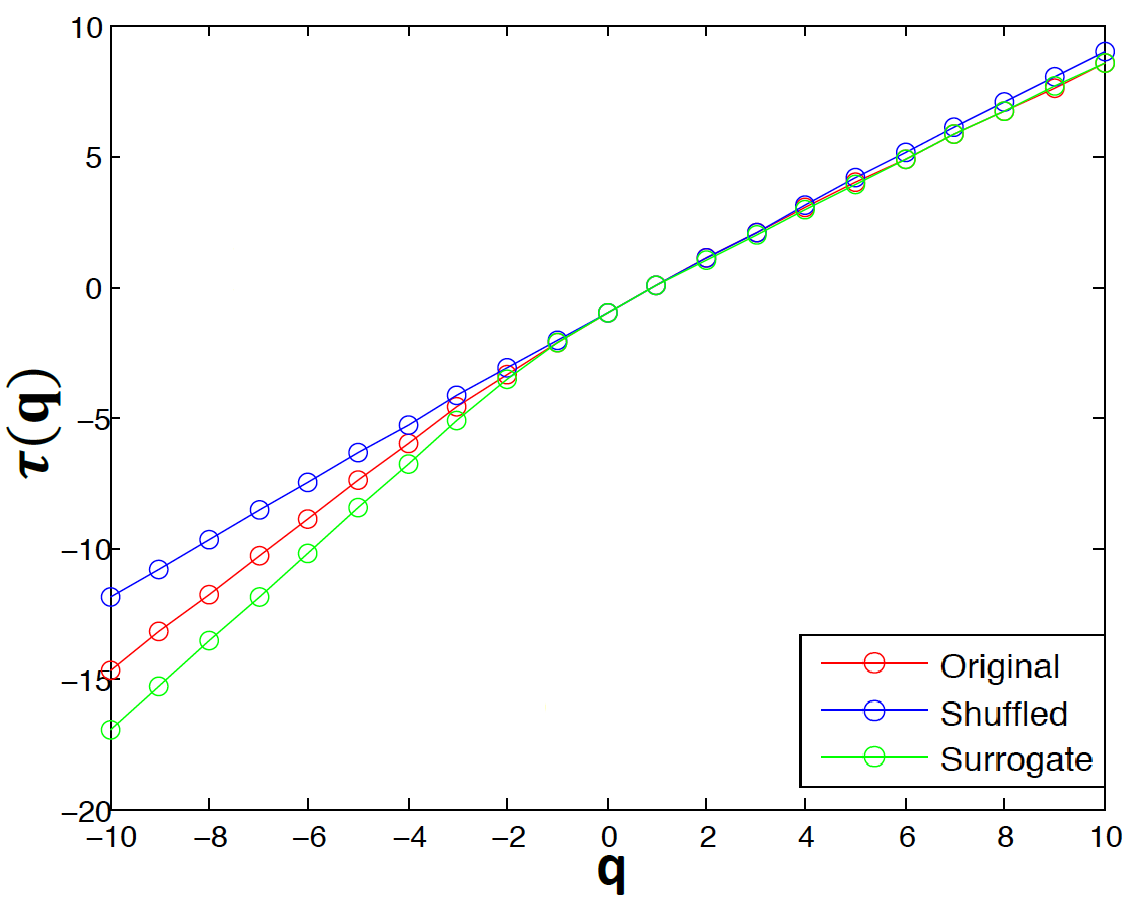} 
\includegraphics[scale=0.15]{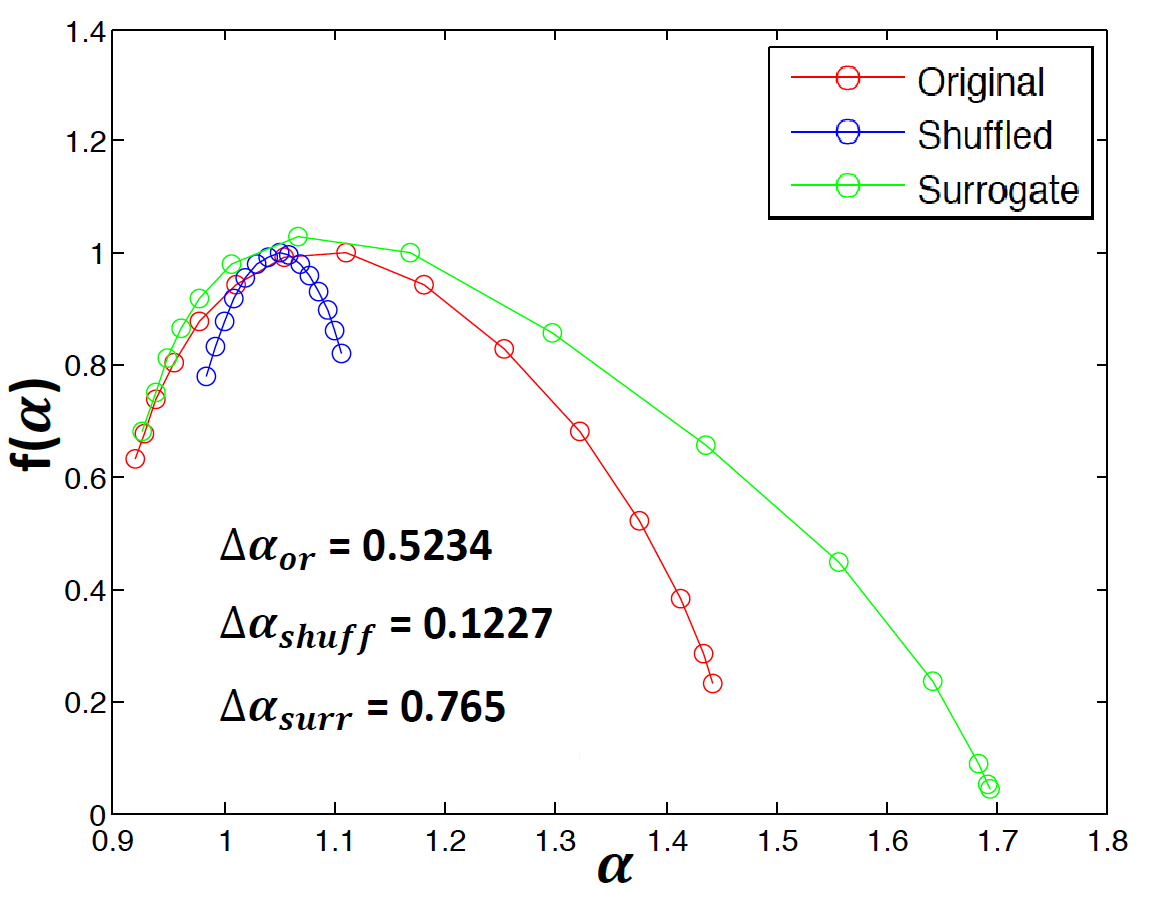}
\includegraphics[scale=0.15]{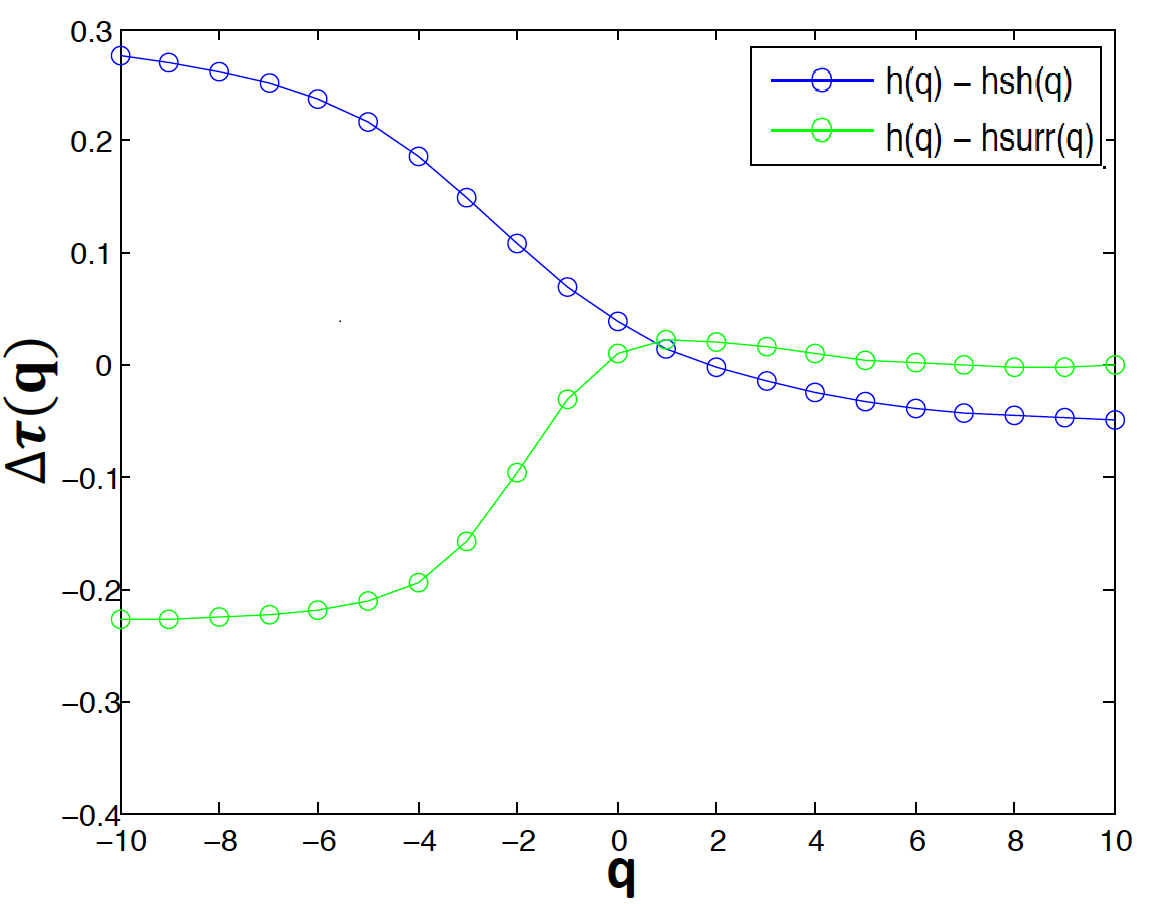}
\includegraphics[scale=0.15]{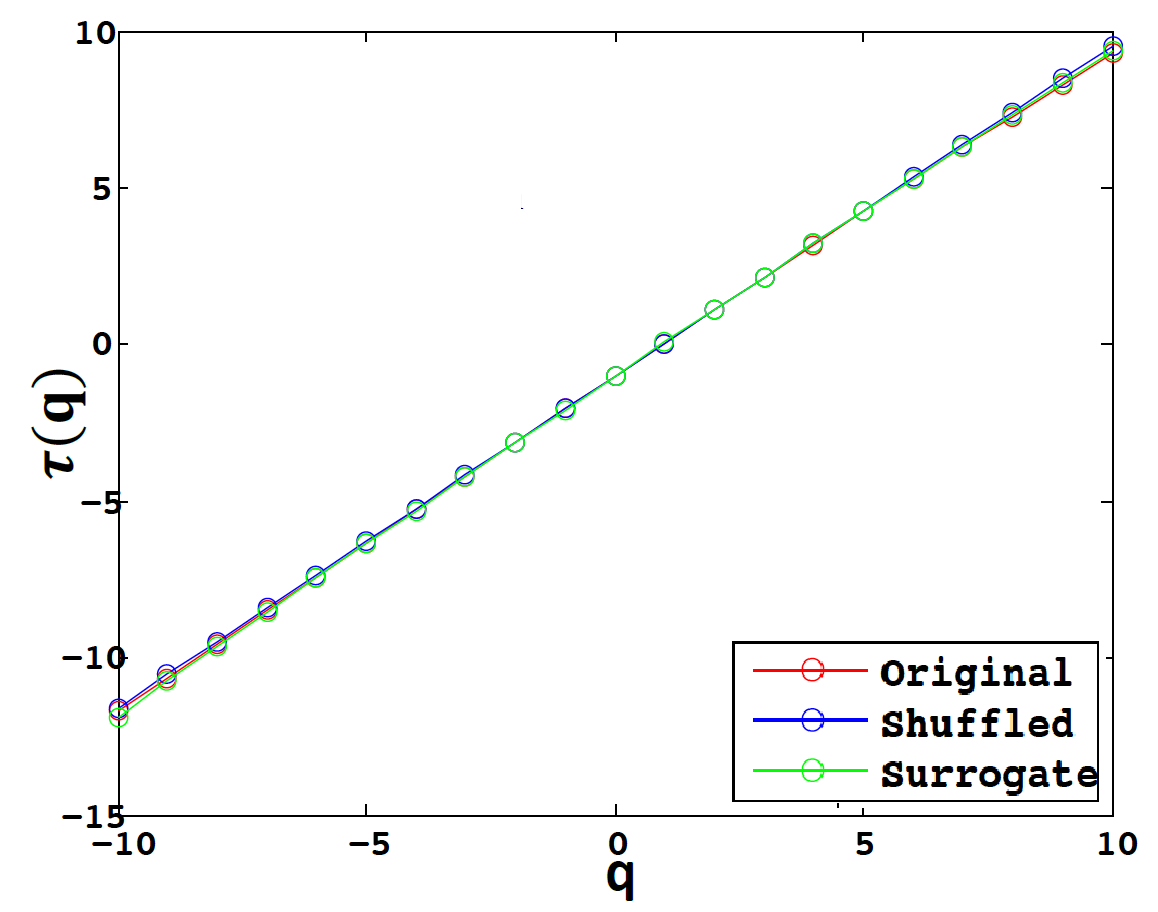} 
\includegraphics[scale=0.15]{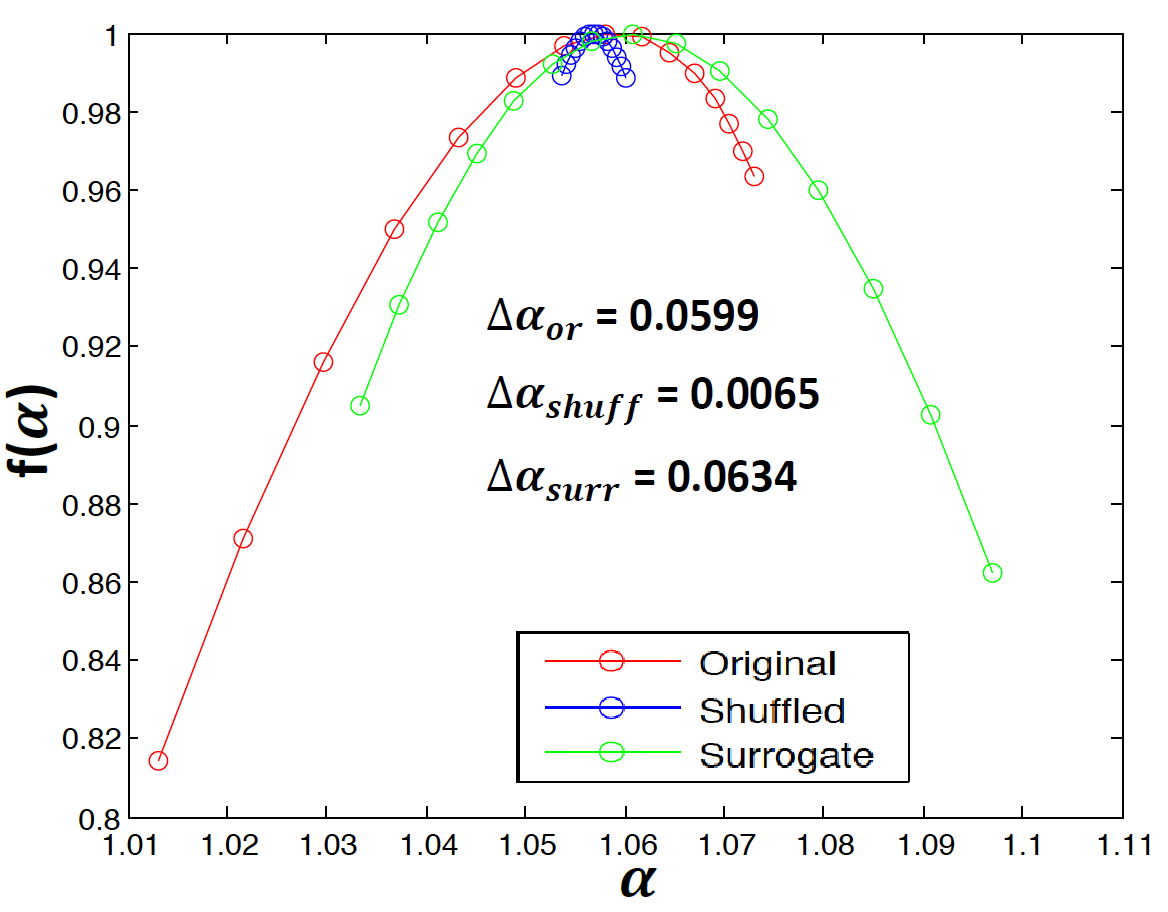}
\includegraphics[scale=0.15]{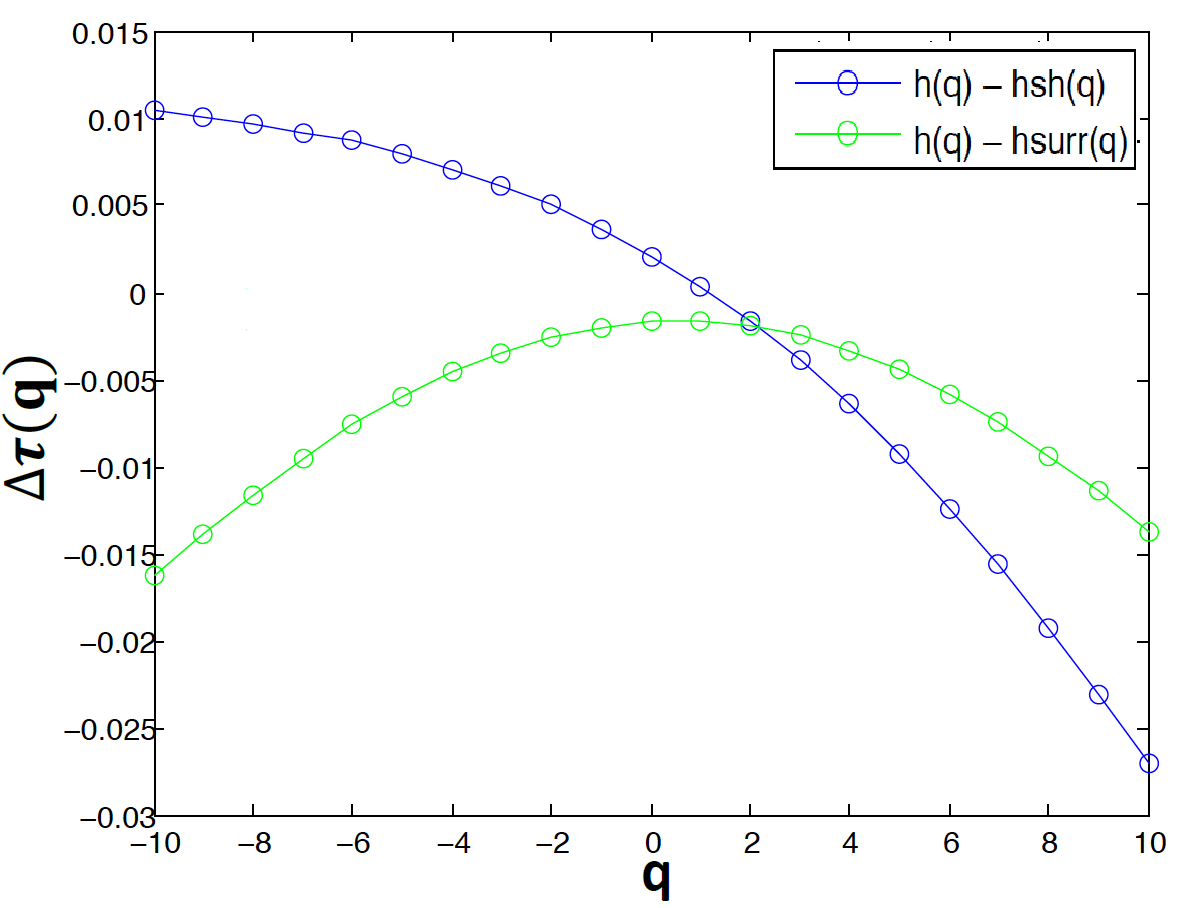}
\includegraphics[scale=0.15]{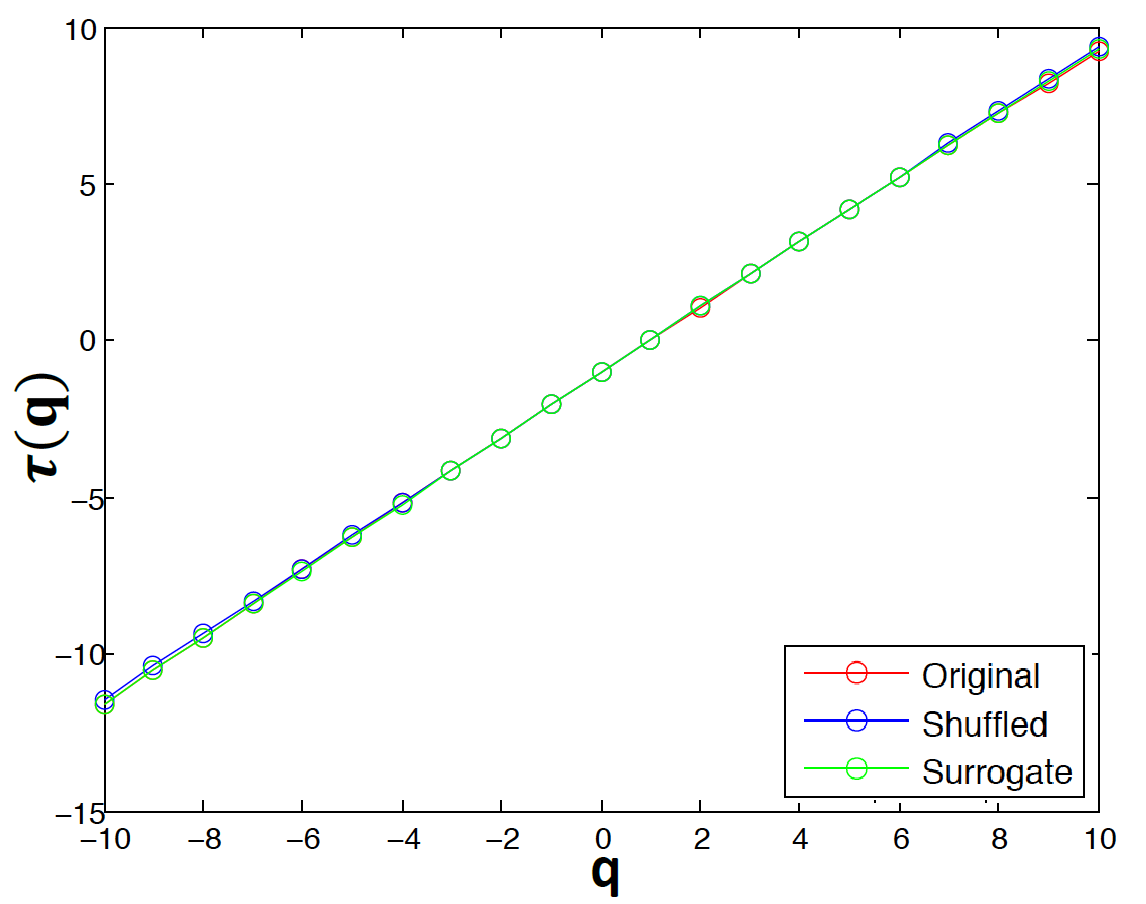} 
\includegraphics[scale=0.15]{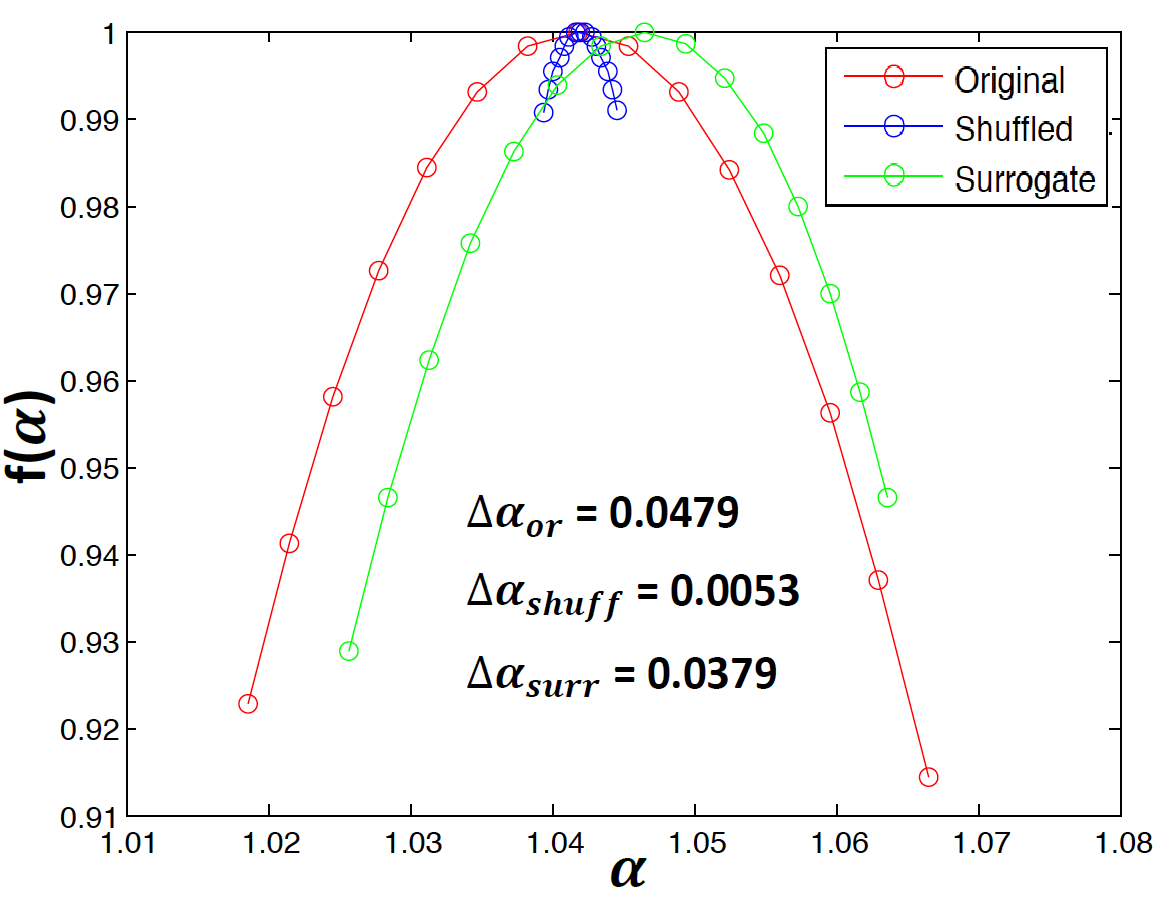}
\includegraphics[scale=0.15]{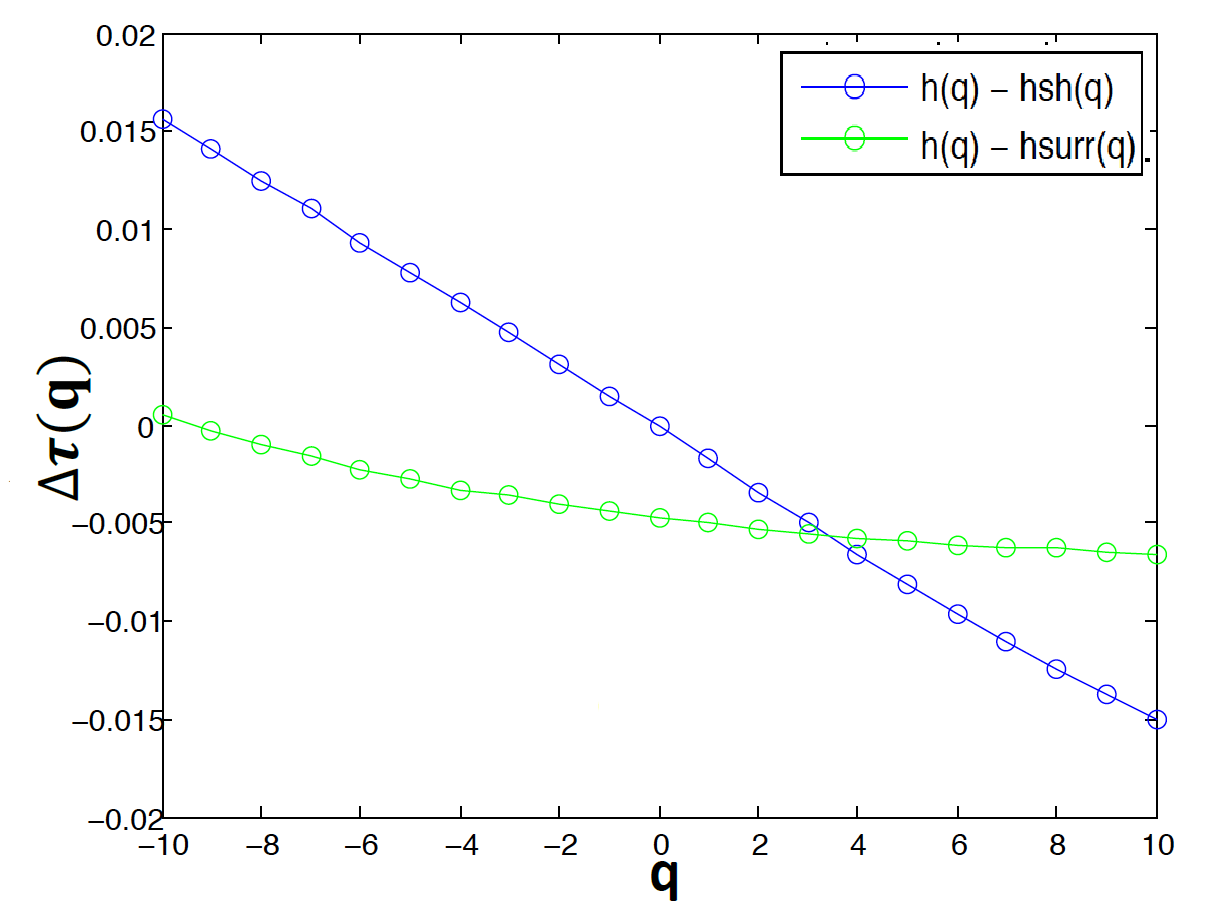}
\includegraphics[scale=0.15]{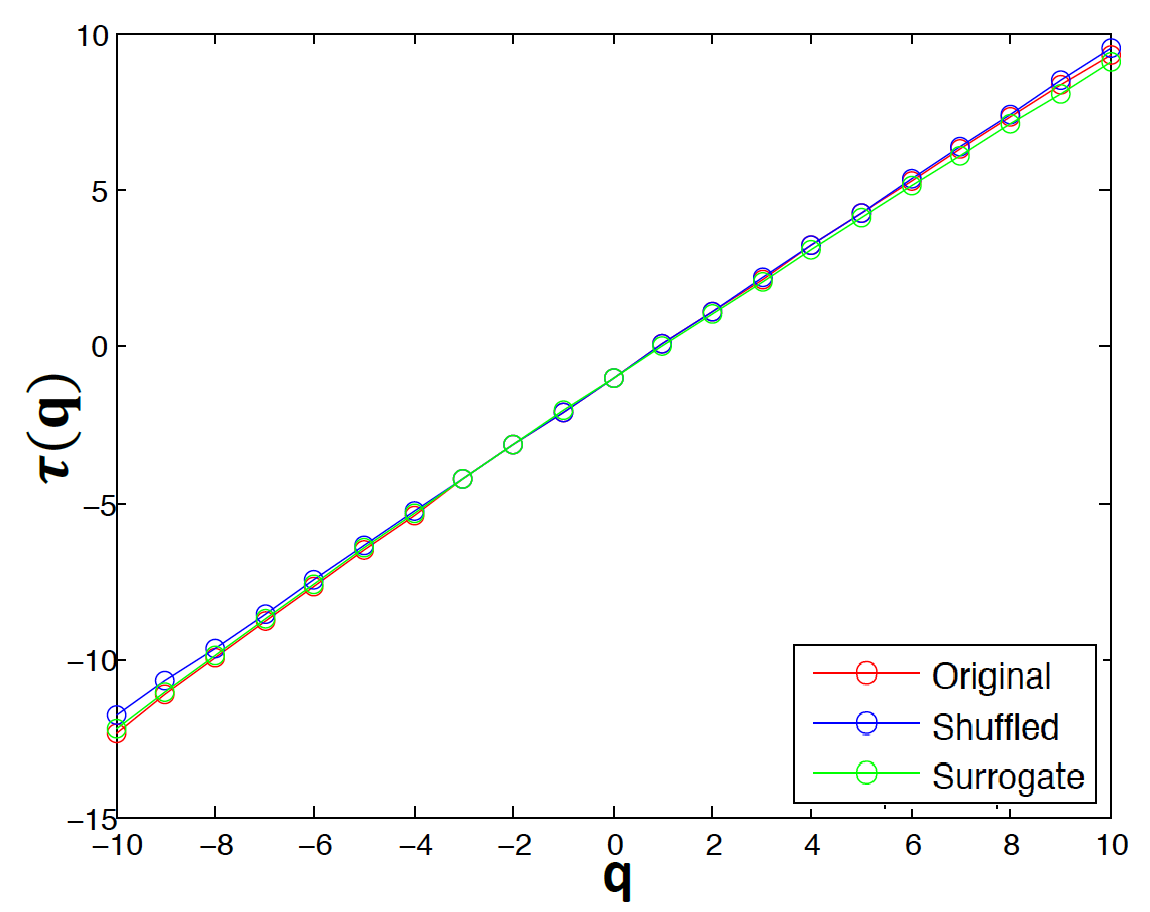} 
\includegraphics[scale=0.15]{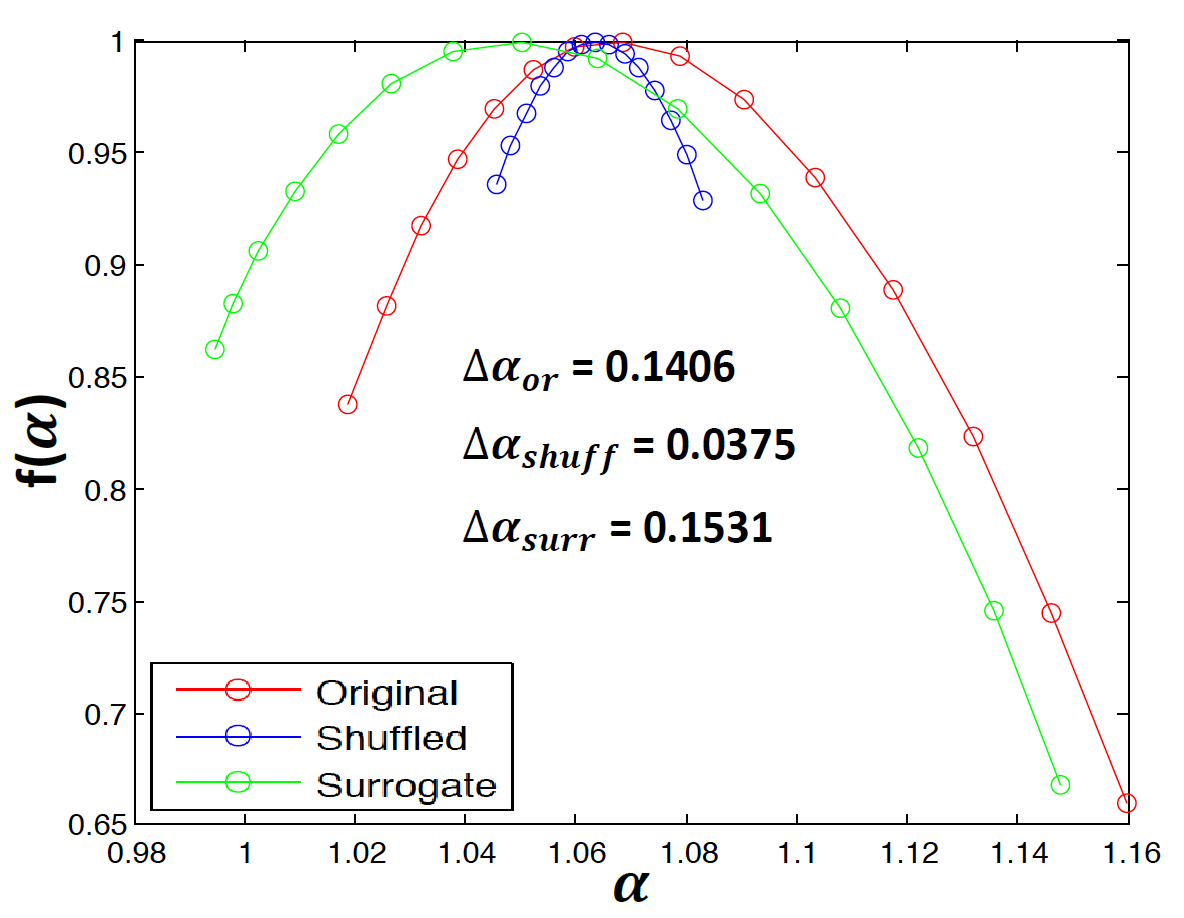}
\includegraphics[scale=0.15]{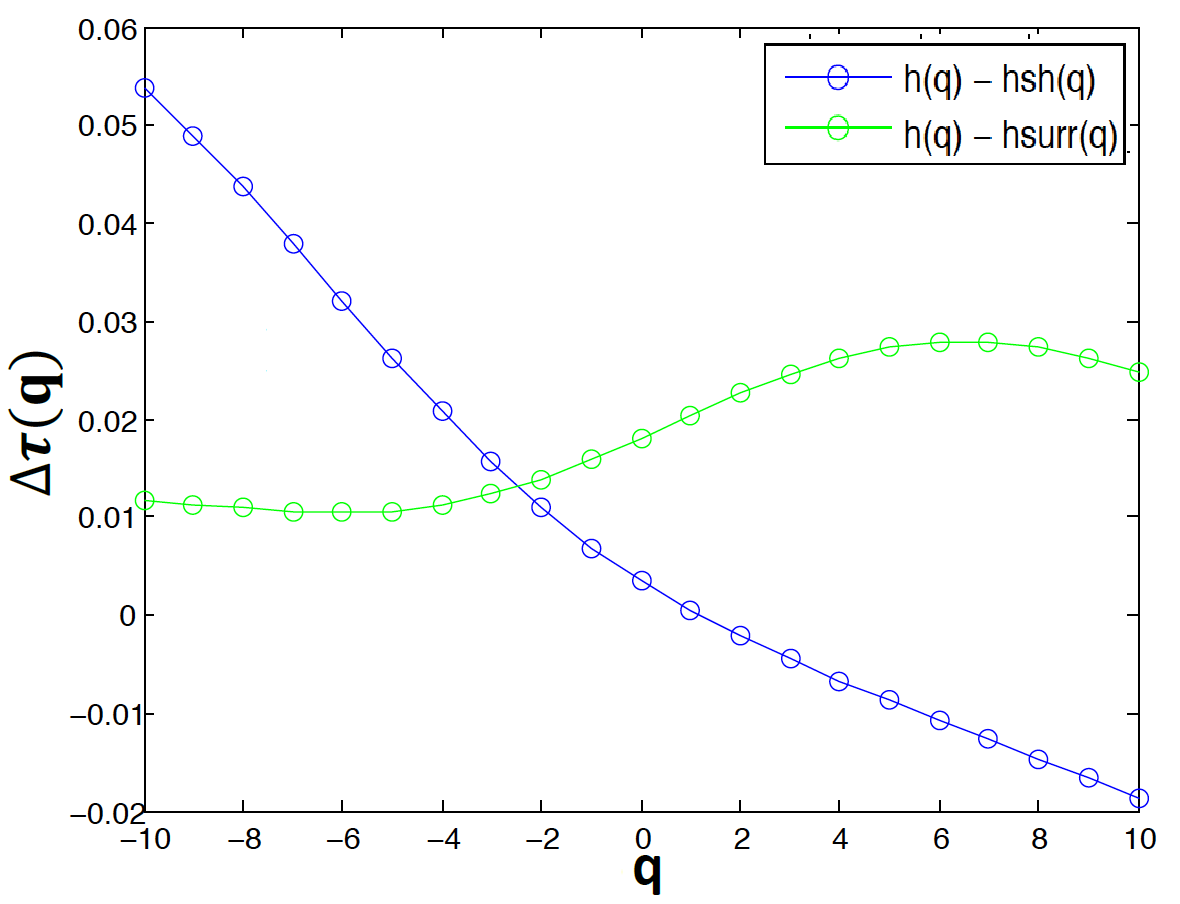}
\includegraphics[scale=0.15]{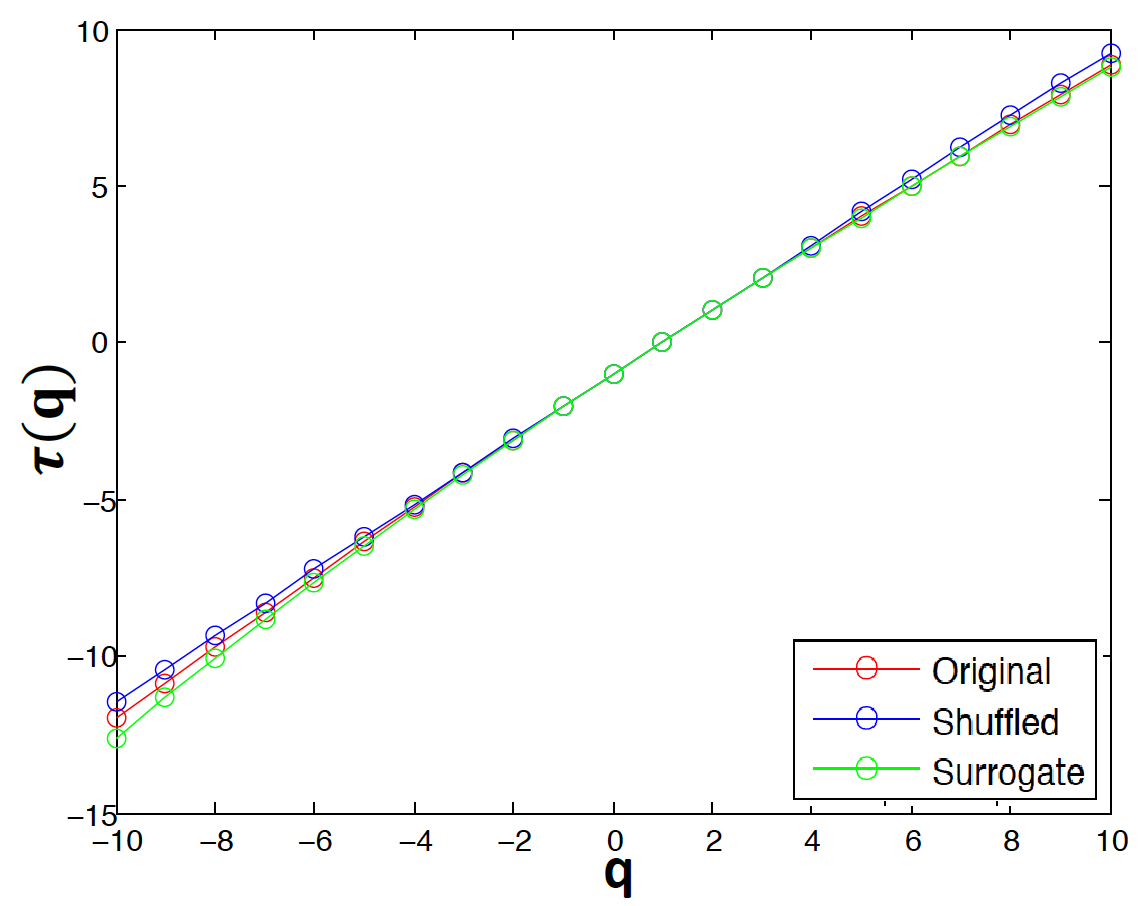} 
\includegraphics[scale=0.15]{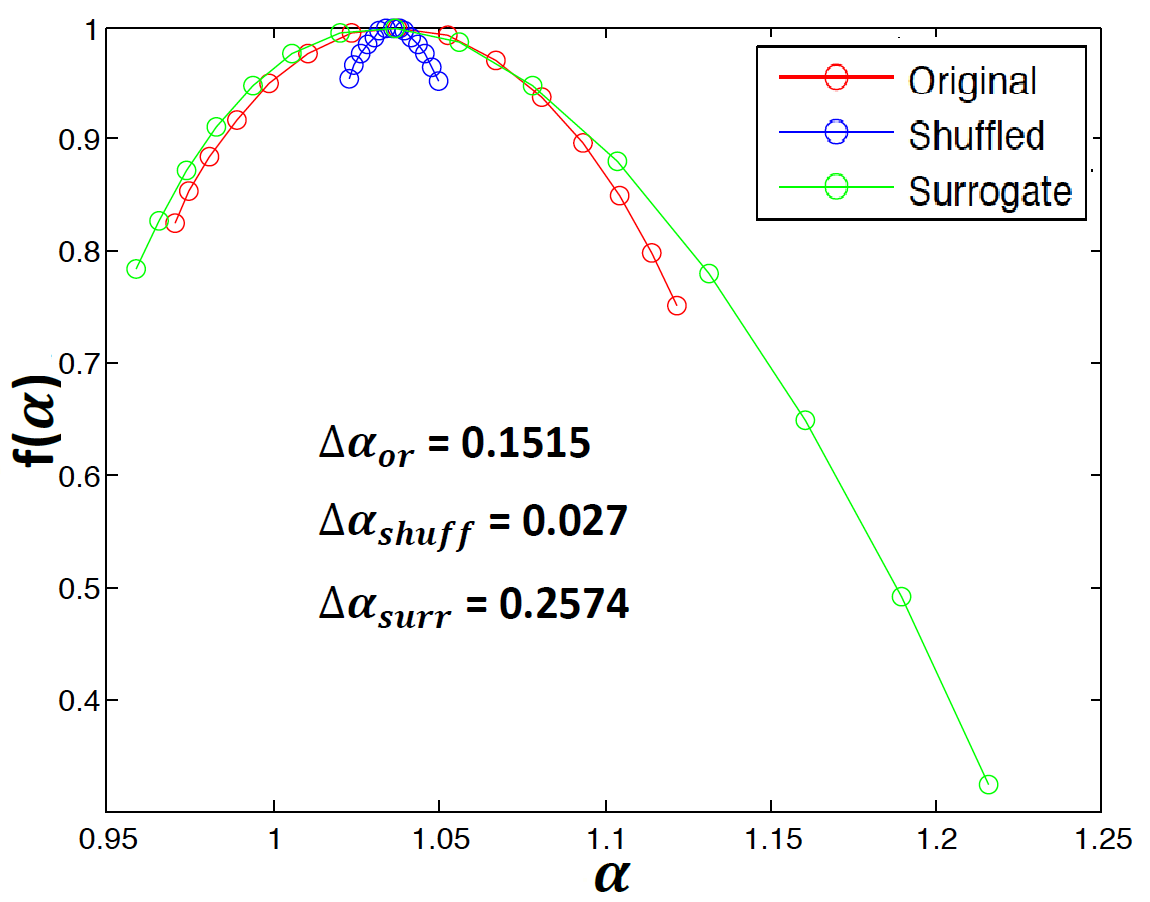}
\includegraphics[scale=0.15]{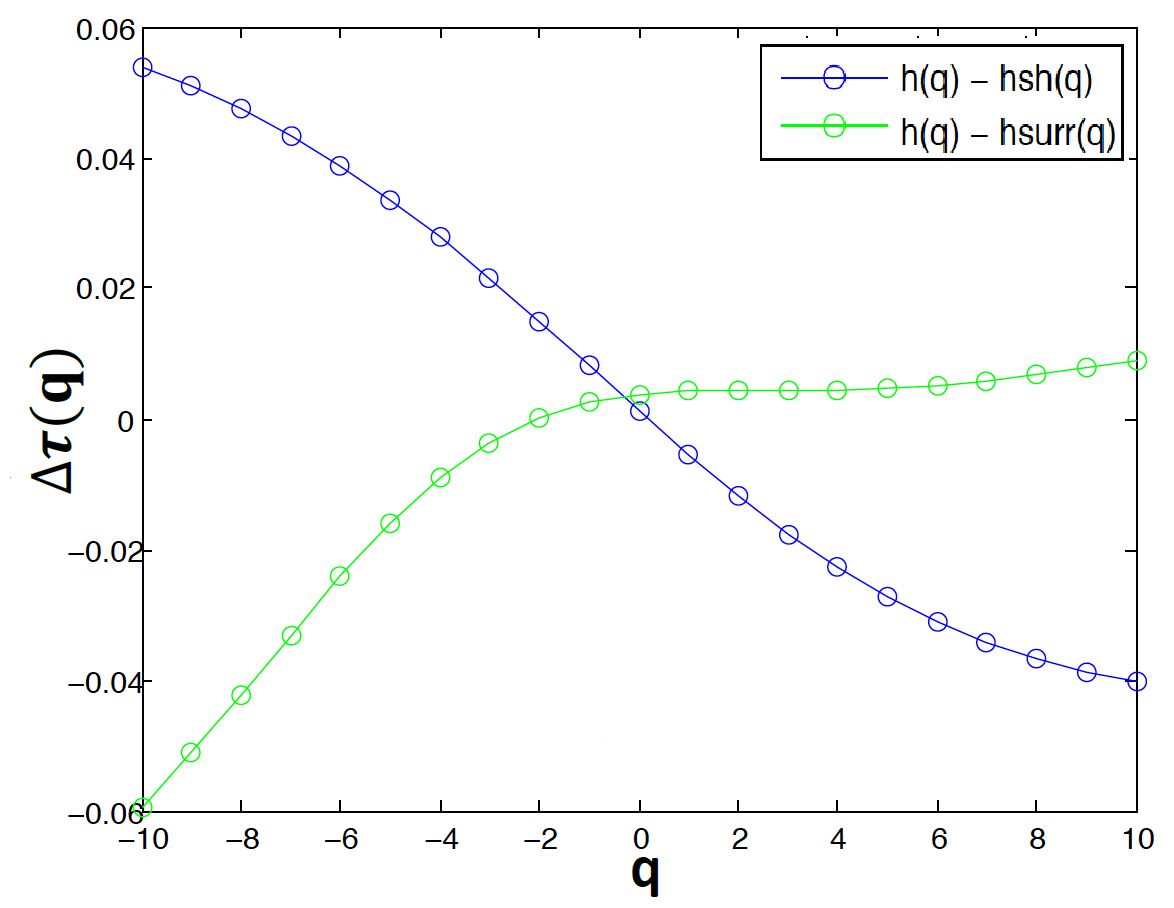}
\caption{The scaling exponent $\tau(q)$ (left), the multifractal spectrum $f(\alpha)$ (middle) and the deviation in the scaling exponent $\Delta\tau(q)$ (right) considering, from top to bottom, radio, mm, IR, optical, UV and X-ray observations.}
\label{fig3}
\end{figure*}

Also in this time series, the slope $h(q)$ is slightly changing with $q$ revealing weak multifractal nature in the time series (Fig. \ref{fig2}, 4th panels). Furthermore, the $q$ dependency of $h_{corr}(q)$ tells us that the temporal correlation is the possible source for the observed weak multifractality.

In Fig. \ref{fig3} (4th panels), the scaling exponent $\tau(q)$ shows no strong $q$ dependency which confirms weak multifractality behaviour in the time series considered. Exceptionally in this time series, the contribution due to the fat-tail is visible despite the fact that the temporal correlation is still the dominant contributor to the observed weak multifractal signature.

We have calculated the degree of multifractality as $\Delta\alpha = 0.0479 \slash 0.0053 \slash 0.0379$  for the original, shuffled, and surrogate, respectively (Table \ref{tab1}). Based on the calculated width value for the original series, we have observed weak multifractal (close to monofractal) behaviour in this time series. This supports our analysis based on the $q$ dependency of $h(q)$ and the scaling exponent $\tau(q)$. 

\subsection{UV observations}

Comparing the generalized Hurst exponent $h(q)$ for the original, shuffled, and surrogate data we observe multifractal behaviour in this time series. In Fig. \ref{fig2}  (5th panels), we present the fluctuation function against the scale, and it shows a clear power-law relationship between them. The slopes for the original, shuffled and surrogate data are $q$-dependent which indicate the presence of multifractality in the time series under consideration and as the degree of multifractality for the three data is different. Since the slope of the shuffled data is not equal to 0.5 and of the surrogate data is $q$-dependent, the information we have is not enough to determine the origin of the observed multifractal behaviour. In this case, we should make use of relations \ref{eq11} and \ref{eq12}.

As shown in Fig. \ref{fig3} (5th panels), the original data is away from linearity which tells us the presence of multifractal behaviour in the time series. The crossover in $\tau(q)$ in Fig. \ref{fig3} (5th panels), further confirms the presence of a multifractal signature in the time series. The shuffled series is almost close to linearity whereas that of the surrogate series is away from linearity with respect to the original data. This difference in the degree of non-linearity between the shuffled and surrogate data with respect to the original one shows that the shuffling process has removed the temporal correlation in the time series; the temporal correlation contributes more to the detected multifractality behaviour. 

Both, $h(q)$ and $\tau(q)$, speak in favour of multifractal behaviour in the time series considered. The calculated spectrum width ($\Delta\alpha$) values for the original time series, the corresponding shuffled and surrogate data are given by $\Delta\alpha = 0.1406 \slash 0.0375 \slash 0.1531$, respectively (Table \ref{tab1}). This is further strengthened the presence of a multifractal signature in the time series and the fact that the temporal correlation is the dominant source of multifractality in the time series. The symmetry is left-truncated reflecting the existence of a multifractal structure sensitive to the local fluctuations with small magnitudes (intermittency).

\subsection{X-ray observations}

Similar to all our previous analysis, in addition to the calculations of $q^{th}$-order fluctuation functions and the generalized Hurst exponent, exploring the scaling exponent and the multifractal spectrum indicates the presence of a multifractal behaviour in this time series.

In Fig. \ref{fig2} (bottom panels), we observe excellent growing similarity (power-law relation) between the $q^{th}$-order fluctuation function and the segment size which leads to a decreasing in $h(q)$ indicating a multifractality signature in the time series. The original time series has the biggest slopes throughout its way while the shuffled time series has the smallest one. Therefore, the three time series reveal a multifractal behaviour with different strengths. As shown in Fig. \ref{fig2} (bottom panels), the $q$ dependency of $ h_{surr}(q)$ confirms the presence of temporal correlation in the time series or shows that the multifractality is reduced more by the shuffling procedure than by the phase-randomization procedure. These show that the temporal correlation contributes mainly to the multifractality behaviour of the time series with respect to the fat-tail distribution. 

The non-linearity in $\tau(q)$ (Fig. \ref{fig3}, bottom panels) clearly confirms the existence of multifractality in the time series. We can see that the original time series has non-linearity shape with respect to the shuffled time series, and the curve for the surrogate data is away from linearity in comparison with the shuffled one. This proves that the shuffling technique reduces the multifractality strength of the original time series. This is in agreement with our previous analysis based on the generalized Hurst exponent $h(q)$ that the temporal correlation is the origin of the multifractal behaviour in the time series. As we have already seen, the time series under analysis has multifractal nature. We measure the degree of multifractality in the time series by the calculus of the spectrum width $\Delta\alpha=\alpha_{max}-\alpha_{min}$, and obtained $\Delta\alpha = 0.1515 \slash 0.027 \slash 0.2574$ for the original, shuffled, and surrogate data, respectively (Table \ref{tab1}). This supports our analysis based on the generalized Hurst exponent and the scaling exponent by demonstrating again that there is multifractal behaviour in the time series considered and the temporal correlation is responsible for the observed multifractal signature than the fat-tailed distribution. The left-side truncated spectrum reflects the existence of a multifractal structure sensitive to the local fluctuations with small magnitudes (intermittency).

\subsection{Radio observations with added noise}
A wide $f(\alpha)$ spectrum is a very good indicative of multifractality signature in the time series considered. As we can see from the fluctuation function plot, in Fig. 4a, the one with added noise of highest variance fluctuates more mainly at the lowest scales which is expected for a multifractal data. This is further strengthened by the non-linearity in the scaling exponent function (Fig. 4b) and width of the multifractal spectrum function (Fig. 4c). The scaling function, $\tau(q)$, is more non-linear (different slope for negative and positive $q$) compared to the other two data sets with noise of lowest variance as a result, the calculated width value is highest for the data set with noise of highest variance. Therefore, based on the analysis given here, we roughly can say that widening of the $f(\alpha)$ curve can also happen due to noise contamination \citep{2009Chaos..19d3129H}.
\begin{figure}
\centering
\includegraphics[scale=0.2]{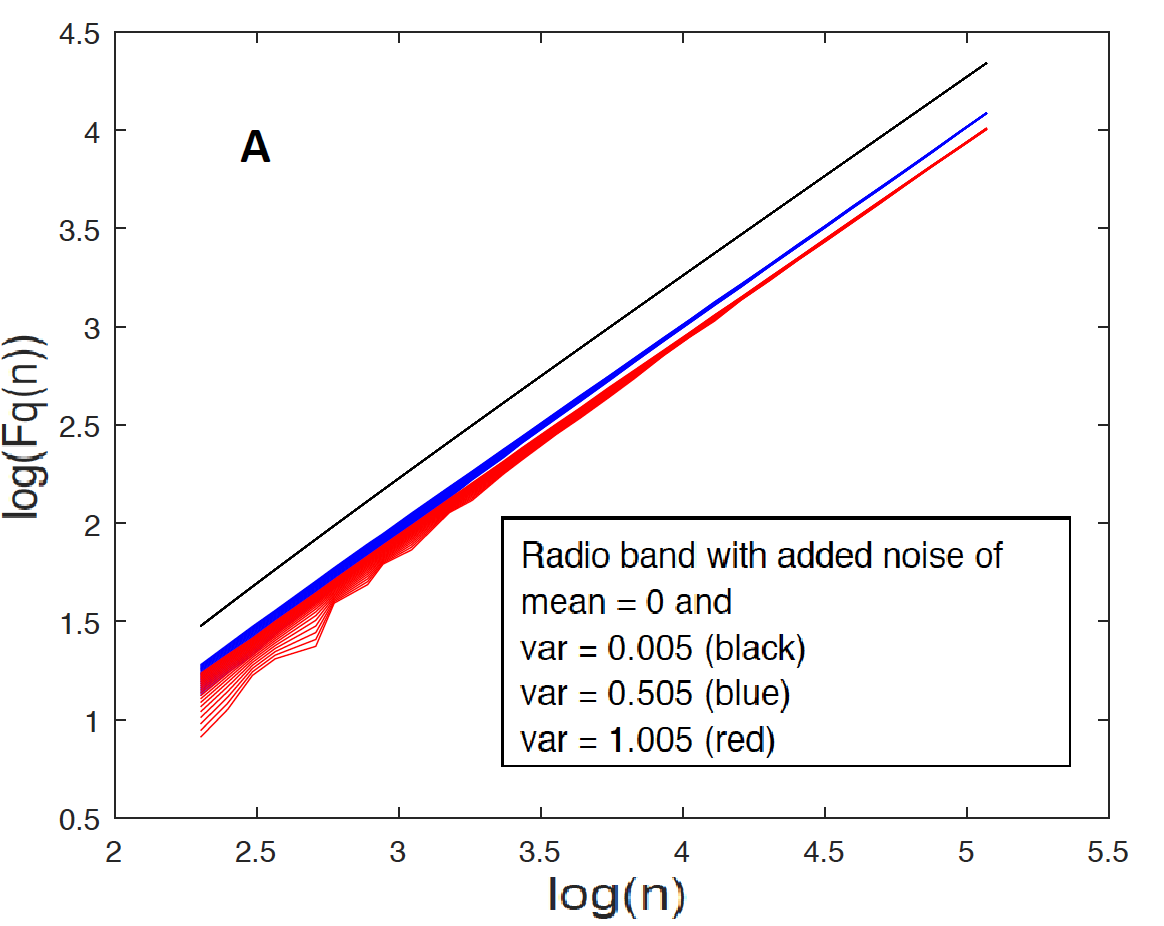}
\includegraphics[scale=0.2]{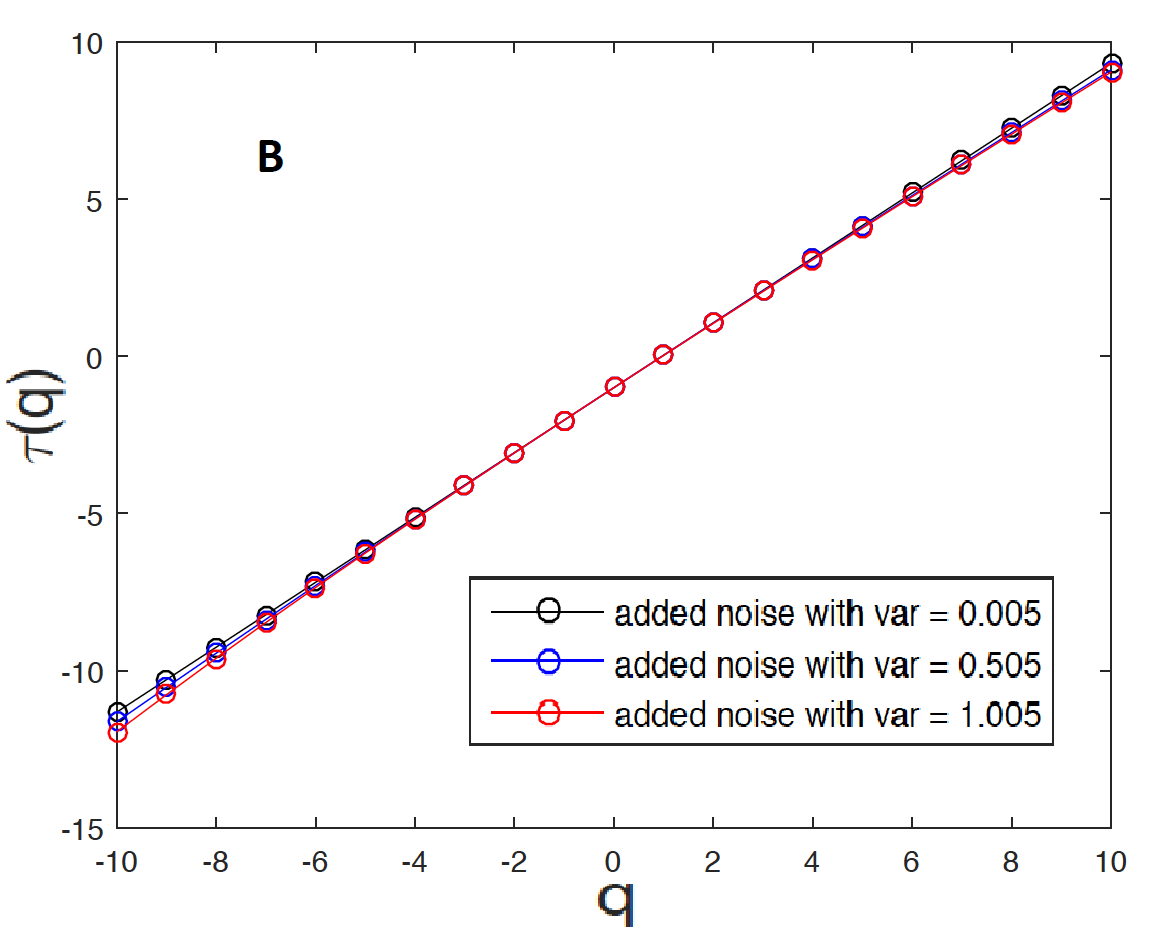}
\includegraphics[scale=0.2]{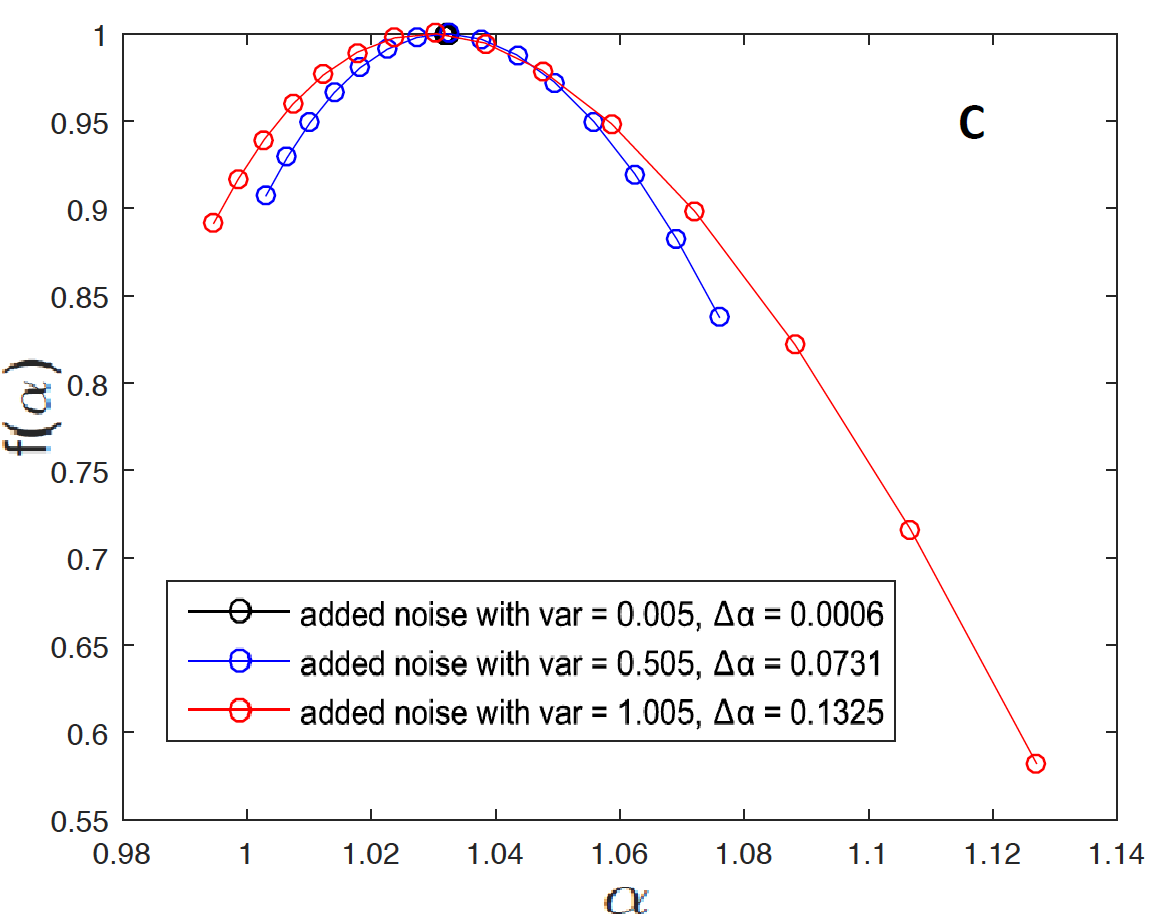}
\caption{A: The logarithm of the fluctuation function, log($F_{q}(n)$), versus the log of the time scale (segment), log($n$), B: The scaling exponent, $\tau(q)$, versus the moment, $q$, and C: The multifractal spectrum $f(\alpha)$.}
\label{fig4}
\end{figure}

\begin{table}
	\centering
	\caption{The best fit parameters; the slopes $h(q)$, and the widths ($\Delta\alpha$).}
	\label{tab:example_table}
	\begin{tabular}{cccc} 
		\hline
		 &&Radio \\ \hline
		 \textit{q} & original & shuffled & surrogate \\
	   -10 & 1.0547$\pm$ 0.0001	& 1.0353$\pm$ 0.0001 & 1.0542$\pm$ 0.0001  \\
        -5 & 1.0477$\pm$0.0001	& 1.0338$\pm$0.0001	&  1.0439$\pm$0.0001\\
         0 & 1.0378$\pm$0.0001	& 1.0323$\pm$0.0001	&  1.0350$\pm$0.0001\\
         5 & 1.0283$\pm$0.0002	& 1.0307$\pm$0.0001	&  1.0267$\pm$0.0002\\
        10 & 1.0223$\pm$0.0002	& 1.0292$\pm$0.0001	&  1.0187$\pm$0.0003\\ 
    $\Delta\alpha$ & 0.0434	            &  0.0085	        &   0.0501\\ \hline
                 &&MM  \\ \hline     
        \textit{q} & original & shuffled & surrogate \\
      -10 & 1.3675$\pm$0.0191	& 1.0914$\pm$0.0005	& 1.5953$\pm$0.0151\\
       -5 & 1.2893$\pm$0.0104	& 1.0732$\pm$0.0003	& 1.5004$\pm$0.0096\\
        0 & 1.0882$\pm$0.0002	& 1.0501$\pm$0.0002	& 1.0786$\pm$0.0005\\
        5 & 0.9916$\pm$0.0002	& 1.0246$\pm$0.0001	& 0.9866$\pm$0.0005\\
       10 & 0.9540$\pm$0.0007	& 1.0023$\pm$0.0001	& 0.9540$\pm$0.0007\\ 
   $\Delta\alpha$ & 0.5234	 & 0.1227	& 0.765\\  \hline
		  &&IR  \\ \hline 
       \textit{q} & original & shuffled & surrogate \\
     -10 & 1.0697$\pm$0.0001   & 1.0592$\pm$0.0001	& 1.0860$\pm$0.0002\\
     -5  & 1.0659$\pm$0.0001   & 1.0580$\pm$0.0001	& 1.0719$\pm$0.0001\\
      0  & 1.0589$\pm$0.0001   & 1.0568$\pm$0.0001	& 1.0606$\pm$0.0001\\
      5  & 1.0464$\pm$0.0001   & 1.0557$\pm$0.0001	& 1.0508$\pm$0.0001\\
     10  & 1.0275$\pm$0.0001   & 1.0546$\pm$0.0001	& 1.0412$\pm$0.0002\\
  $\Delta\alpha$ & 0.0599	& 0.0065	& 0.0634\\ \hline
   &&Optical  \\ \hline 
       \textit{q} & original & shuffled & surrogate \\
	 -10 & 1.0595$\pm$0.0001   & 1.0439$\pm$0.0001	& 1.0590$\pm$0.0001\\
	 -5  & 1.0508$\pm$0.0001    & 1.0430$\pm$0.0001	& 1.0536$\pm$0.0001\\
	  0  & 1.0419$\pm$0.0002   & 1.0420$\pm$0.0001	& 1.0466$\pm$0.0001\\
	  5  & 1.0330$\pm$0.0002   & 1.0411$\pm$0.0001	& 1.0389$\pm$0.0001\\
	 10  & 1.0251$\pm$0.0002   & 1.0401$\pm$0.0001	& 1.0317$\pm$0.0002\\
 $\Delta\alpha$ & 0.0479	& 0.0053	 & 0.0379\\ \hline
    &&UV  \\ \hline 
       \textit{q} & original & shuffled & surrogate \\
     -10 & 1.1313$\pm$0.0025   & 1.0774$\pm$0.0002	& 1.1197$\pm$0.0013\\
     -5  & 1.0963$\pm$0.0005   & 1.0702$\pm$0.0001	& 1.0858$\pm$0.0007\\
      0  & 1.0670$\pm$0.0001   & 1.0636$\pm$0.0001	& 1.0489$\pm$0.0005\\
      5  & 1.0487$\pm$0.0001   & 1.0573$\pm$0.0001	& 1.0214$\pm$0.0007\\
     10  & 1.0321$\pm$0.0002   & 1.0508$\pm$0.0001	& 1.0074$\pm$0.0012\\
  $\Delta\alpha$ & 0.1406	& 0.0375	& 0.1531\\ \hline
	    &&X-ray  \\ \hline 
       \textit{q} & original & shuffled & surrogate \\
     -10 & 1.0997$\pm$0.0001   & 1.0457$\pm$0.0003	& 1.1593$\pm$0.0010\\
     -5  & 1.0743$\pm$0.0001   & 1.0407$\pm$0.0002	& 1.0902$\pm$0.0004\\
      0  & 1.0371$\pm$0.0001   & 1.0359$\pm$0.0001	& 1.0333$\pm$0.0002\\
      5  & 1.0040$\pm$0.0001   & 1.0312$\pm$0.0001	& 0.9993$\pm$0.0002\\
      10 & 0.9862$\pm$0.0001   & 1.0264$\pm$0.0001	& 0.9772$\pm$0.0003\\
   $\Delta\alpha$ & 0.1515	& 0.027	& 0.2574\\ \hline
		\end{tabular}
		\label{tab1}
\end{table}
\section{Summary and Conclusions}\label{concl}

This is the first ever attempt to analyse the multifractal behaviour of the quasar 3C 273 time series using the Multifractal Detrended Moving Average algorithm. In this study, we investigate the multifractal properties of the flux time series of quasar 3C 273 using the backward ($\theta=0$) MFDMA analysis for one-dimensional time series analysis. We first calculate the fluctuation functions from which we estimate the generalized Hurst exponents using least square fitting method. Then we deduce the scaling exponents and the multifractal spectrum of the time series. Our results indicate the presence of a weak multifractal (close to monofractal) signature in some of the time series considered except in the mm, UV, and X-ray bands where the degree of multifractality is the strongest with respect to other bands. Moreover, in order to detect the origin of the observed multifractality, we perform the shuffling and phase-randomization (surrogate) techniques on the original time series. By observing the curves representing the generalized Hurst exponents, the scaling exponents, as well as the multifractal spectrum and calculating the width ($\Delta\alpha$), we conclude that the non-linear temporal correlations are the dominant source of the observed multifractality signature in the time series while the fat-tail probability distribution contributes less in all of the time series considered. Moreover, by adding noise with different variance to the radio band we pointed out that widening of the multifractal spectrum could also happen due to noise contamination. Based on the result obtained, we observed that the quasar 3C 273 is a non-linear, complex and rich dynamic system. Our analysis shows that some of the 3C 273 light-curves have self-similar (intermittent) feature, it is a multifractal time series. The multifractality strength across the EM spectra of the quasar 3C 273 is clearly different, and from which we can conclude that the nature of flux fluctuations across the different bands of the quasar 3C 273 is not the same which in turn provides a good implication as the physical mechanisms that cause the flux fluctuation in the system are not the same for all spectral regions. 

There is much evidence that light-curves of AGN are both non-linear and non-stationary. For example, optical (V, R and I band) studies show the variable flux in the 3C 273 changes with different amplitudes on different timescales revealing the non-linear variability characteristic of blazars \citep{2015MNRAS.451.1356K}. It has been shown by \citet{2015MNRAS.451.1356K} that there is high flux variation in all bands of the quasar 3C 273 in the long term. In our work, we have shown that the quasar 3C 273 time series have a multifractal feature mainly due to temporal correlation which is assumed to be due to flux variation across the different bands of the source. The differences in the degree of multifractality imply that all bands of the source 3C 273 are not only emitting from different regions but also they are driven by different mechanisms. In the work of \citet{2015MNRAS.451.1356K}, it has been presented that there is no relation between X-ray and optical/UV emission of 3C 273 which is more or less in agreement with our work considering the difference in the degree of multifractality observed in these bands. \citet{1999MNRAS.310..571M} has shown that there is no correlation between the X-ray and mm observations of the 3C 273. These all including our results support the presence of high flux variation in the spectra of the source 3C 273 and the variation is not the same in all bands. Multifractal behaviour indicates the presence of non-linearity in a time series which leaves open the possibility that the time series is produced by a chaotic system. However, non-linearity is a necessary but not sufficient condition for a light-curve to be produced by a chaotic system \citep{1992ApJ...391..518V}. Therefore, our work, at least, shows that 3C 273 is a non-linear system. We believe that understanding the fractal behaviour in the light-curves also contributes in terms of providing information from a different point of view about the complexity of the dynamics interior to the system under consideration. Understanding the nature of the scaling in the 3C 273 flux could tell us fundamental things about how the flux varies in each band which in turn provides sufficient information in order to answer basic AGN related questions, such as energy production mechanisms, emission lines problem, and many others. Having clear picture of how the scaling in a time series (the fractal nature) behaves can provide fundamental information for those working on model construction to fully uncover the AGN science.\\ 
\section*{Acknowledgements}

This paper includes data collected by the Integral Science Data Centre (ISDC) linked to the Astronomical Observatory of the University of Geneva. Research activities of the Astronomy Observational Board of the Federal University of Rio Grande do Norte are supported by continuous grants from the Brazilian agencies CNPq and FAPERN. We also warmly thank Prof. Wei-Xing ZHOU, at the East China University of Science and Technology for his continuous comments and communication. A.B.B. acknowledges financial support from the CAPES Brazilian agency. J.P.B. acknowledges a CAPES/PNPD post-doctorate fellowship. Finally, we warmly thank the anonymous Reviewer for providing very helpful comments and suggestions that largely improved this work.
 

\nocite{*}
\bibliographystyle{mnras}
\bibliography{biblio} 

\begin{thebibliography}{}
\makeatletter
\relax
\def\mn@urlcharsother{\let\do\@makeother \do\$\do\&\do\#\do\^\do\_\do\%\do\~}
\def\mn@doi{\begingroup\mn@urlcharsother \@ifnextchar [ {\mn@doi@}
  {\mn@doi@[]}}
\def\mn@doi@[#1]#2{\def\@tempa{#1}\ifx\@tempa\@empty \href
  {http://dx.doi.org/#2} {doi:#2}\else \href {http://dx.doi.org/#2} {#1}\fi
  \endgroup}
\def\mn@eprint#1#2{\mn@eprint@#1:#2::\@nil}
\def\mn@eprint@arXiv#1{\href {http://arxiv.org/abs/#1} {{\tt arXiv:#1}}}
\def\mn@eprint@dblp#1{\href {http://dblp.uni-trier.de/rec/bibtex/#1.xml}
  {dblp:#1}}
\def\mn@eprint@#1:#2:#3:#4\@nil{\def\@tempa {#1}\def\@tempb {#2}\def\@tempc
  {#3}\ifx \@tempc \@empty \let \@tempc \@tempb \let \@tempb \@tempa \fi \ifx
  \@tempb \@empty \def\@tempb {arXiv}\fi \@ifundefined
  {mn@eprint@\@tempb}{\@tempb:\@tempc}{\expandafter \expandafter \csname
  mn@eprint@\@tempb\endcsname \expandafter{\@tempc}}}

\bibitem[\protect\citeauthoryear{{Abdo} et~al.,}{{Abdo}
  et~al.}{2010}]{2010ApJ...714L..73A}
{Abdo} A.~A.,  et~al., 2010, \apjl, 714, L73

\bibitem[\protect\citeauthoryear{{Arneodo}, {Bacry}, {Graves}  \&
  {Muzy}}{{Arneodo} et~al.}{1995}]{1995PhRvL..74.3293A}
{Arneodo} A.,  {Bacry} E.,  {Graves} P.~V.,   {Muzy} J.~F.,  1995, Physical
  Review Letters, 74, 3293

\bibitem[\protect\citeauthoryear{{Ashkenazy}, {Baker}, {Gildor}  \&
  {Havlin}}{{Ashkenazy} et~al.}{2003}]{2003GeoRL..30.2146A}
{Ashkenazy} Y.,  {Baker} D.~R.,  {Gildor} H.,   {Havlin} S.,  2003, \grl, 30,
  2146

\bibitem[\protect\citeauthoryear{{Attridge}, {Wardle}  \& {Homan}}{{Attridge}
  et~al.}{2005}]{2005ApJ...633L..85A}
{Attridge} J.~M.,  {Wardle} J.~F.~C.,   {Homan} D.~C.,  2005, \apjl, 633, L85

\bibitem[\protect\citeauthoryear{{Bacry}, {Delour}  \& {Muzy}}{{Bacry}
  et~al.}{2001}]{2001PhRvE..64b6103B}
{Bacry} E.,  {Delour} J.,   {Muzy} J.~F.,  2001, \pre, 64, 026103

\bibitem[\protect\citeauthoryear{{Barab{\'a}si} \& {Vicsek}}{{Barab{\'a}si} \&
  {Vicsek}}{1991}]{1991PhRvA..44.2730B}
{Barab{\'a}si} A.-L.,  {Vicsek} T.,  1991, \pra, 44, 2730

\bibitem[\protect\citeauthoryear{{Bashan}, {Bartsch}, {Kantelhardt}  \&
  {Havlin}}{{Bashan} et~al.}{2008}]{2008PhyA..387.5080B}
{Bashan} A.,  {Bartsch} R.,  {Kantelhardt} J.~W.,   {Havlin} S.,  2008, Physica
  A Statistical Mechanics and its Applications, 387, 5080

\bibitem[\protect\citeauthoryear{{Carbone}, {Castelli}  \& {Stanley}}{{Carbone}
  et~al.}{2004}]{2004PhRvE..69b6105C}
{Carbone} A.,  {Castelli} G.,   {Stanley} H.~E.,  2004, \pre, 69, 026105

\bibitem[\protect\citeauthoryear{{Chernyakova} et~al.,}{{Chernyakova}
  et~al.}{2007}]{2007A&A...465..147C}
{Chernyakova} M.,  et~al., 2007, \aap, 465, 147

\bibitem[\protect\citeauthoryear{{Collmar} et~al.,}{{Collmar}
  et~al.}{2000}]{2000A&A...354..513C}
{Collmar} W.,  et~al., 2000, \aap, 354, 513

\bibitem[\protect\citeauthoryear{{Courvoisier} \& {T{\"u}rler}}{{Courvoisier}
  \& {T{\"u}rler}}{2005}]{2005A&A...444..417C}
{Courvoisier} T.~J.-L.,  {T{\"u}rler} M.,  2005, \aap, 444, 417

\bibitem[\protect\citeauthoryear{{Courvoisier}, {Robson}, {Blecha}, {Bouchet},
  {Hughes}, {Krisciunas}  \& {Schwarz}}{{Courvoisier}
  et~al.}{1988}]{1988Natur.335..330C}
{Courvoisier} T.~J.-L.,  {Robson} E.~I.,  {Blecha} A.,  {Bouchet} P.,  {Hughes}
  D.~H.,  {Krisciunas} K.,   {Schwarz} H.~E.,  1988, \nat, 335, 330

\bibitem[\protect\citeauthoryear{{Courvoisier} et~al.,}{{Courvoisier}
  et~al.}{1990}]{1990A&A...234...73C}
{Courvoisier} T.~J.~L.,  et~al., 1990, \aap, 234, 73

\bibitem[\protect\citeauthoryear{{Dai} et~al.,}{{Dai}
  et~al.}{2009}]{2009MNRAS.392.1181D}
{Dai} B.~Z.,  et~al., 2009, \mnras, 392, 1181

\bibitem[\protect\citeauthoryear{{\lowercase{D}e Freitas}, Nepomuceno, de
  Souza, Le{\~a}o, Chagas, Costa, Martins  \& Medeiros}{{\lowercase{D}e
  Freitas} et~al.}{2017}]{0004-637X-843-2-103}
{\lowercase{D}e Freitas} D.~B.,  Nepomuceno M. M.~F.,  de Souza M.~G.,
  Le{\~a}o I.~C.,  Chagas M. L.~D.,  Costa A.~D.,  Martins B. L.~C.,   Medeiros
  J. R.~D.,  2017, The Astrophysical Journal, 843, 103

\bibitem[\protect\citeauthoryear{{Fan}, {Peng}, {Tao}, {Qian}  \& {Shen}}{{Fan}
  et~al.}{2009}]{2009AJ....138.1428F}
{Fan} J.~H.,  {Peng} Q.~S.,  {Tao} J.,  {Qian} B.~C.,   {Shen} Z.~Q.,  2009,
  \aj, 138, 1428

\bibitem[\protect\citeauthoryear{{Fan}, {Kurtanidze}, {Liu}, {Richter},
  {Chanishvili}  \& {Yuan}}{{Fan} et~al.}{2014}]{2014ApJS..213...26F}
{Fan} J.~H.,  {Kurtanidze} O.,  {Liu} Y.,  {Richter} G.~M.,  {Chanishvili} R.,
   {Yuan} Y.~H.,  2014, \apjs, 213, 26

\bibitem[\protect\citeauthoryear{{Feder}}{{Feder}}{1988}]{1988Feder}
{Feder} J.,  1988, Fractals.
Springer US

\bibitem[\protect\citeauthoryear{{Frisch}}{{Frisch}}{1995}]{1995tlan.book.....F}
{Frisch} U.,  1995, {Turbulence. The legacy of A. N. Kolmogorov.}.
Cambridge University Press

\bibitem[\protect\citeauthoryear{{Grandi} \& {Palumbo}}{{Grandi} \&
  {Palumbo}}{2004}]{2004Sci...306..998G}
{Grandi} P.,  {Palumbo} G.~G.~C.,  2004, Science, 306, 998

\bibitem[\protect\citeauthoryear{{Greve} et~al.,}{{Greve}
  et~al.}{2002}]{2002A&A...390L..19G}
{Greve} A.,  et~al., 2002, \aap, 390, L19

\bibitem[\protect\citeauthoryear{{Gu} \& {Zhou}}{{Gu} \&
  {Zhou}}{2010}]{2010PhRvE..82a1136G}
{Gu} G.-F.,  {Zhou} W.-X.,  2010, \pre, 82, 011136

\bibitem[\protect\citeauthoryear{{Halsey}, {Jensen}, {Kadanoff}, {Procaccia}
  \& {Shraiman}}{{Halsey} et~al.}{1986}]{1986PhRvA..33.1141H}
{Halsey} T.~C.,  {Jensen} M.~H.,  {Kadanoff} L.~P.,  {Procaccia} I.,
  {Shraiman} B.~I.,  1986, \pra, 33, 1141

\bibitem[\protect\citeauthoryear{{Harikrishnan}, {Misra}, {Ambika}  \&
  {Amritkar}}{{Harikrishnan} et~al.}{2009}]{2009Chaos..19d3129H}
{Harikrishnan} K.~P.,  {Misra} R.,  {Ambika} G.,   {Amritkar} R.~E.,  2009,
  Chaos, 19, 043129

\bibitem[\protect\citeauthoryear{{Heck} \& {Perdang}}{{Heck} \&
  {Perdang}}{1991}]{1991HeckandPerdang}
{Heck} A.,  {Perdang} J.~M.,  1991, {Applying Fractals in Astronomy}.
Springer-Verlag Berlin Heidelberg

\bibitem[\protect\citeauthoryear{{Hu}, {Gao}  \& {Wang}}{{Hu}
  et~al.}{2009}]{2009JSMTE..02..066H}
{Hu} J.,  {Gao} J.,   {Wang} X.,  2009, Journal of Statistical Mechanics:
  Theory and Experiment, 2, 02066

\bibitem[\protect\citeauthoryear{{Ida}, {Hayakawa}, {Adalev}  \& {Gotoh}}{{Ida}
  et~al.}{2005}]{2005NPGeo..12..157I}
{Ida} Y.,  {Hayakawa} M.,  {Adalev} A.,   {Gotoh} K.,  2005, Nonlinear
  Processes in Geophysics, 12, 157

\bibitem[\protect\citeauthoryear{{Ihlen}}{{Ihlen}}{2012}]{2012Ihlen}
{Ihlen} E. A.~F.,  2012, Frontiers in Physiology, 3, 141

\bibitem[\protect\citeauthoryear{{Jester}, {Harris}, {Marshall}  \&
  {Meisenheimer}}{{Jester} et~al.}{2006}]{2006ApJ...648..900J}
{Jester} S.,  {Harris} D.~E.,  {Marshall} H.~L.,   {Meisenheimer} K.,  2006,
  \apj, 648, 900

\bibitem[\protect\citeauthoryear{{Kalita}, {Gupta}, {Wiita}, {Bhagwan}  \&
  {Duorah}}{{Kalita} et~al.}{2015}]{2015MNRAS.451.1356K}
{Kalita} N.,  {Gupta} A.~C.,  {Wiita} P.~J.,  {Bhagwan} J.,   {Duorah} K.,
  2015, \mnras, 451, 1356

\bibitem[\protect\citeauthoryear{{Kantelhardt}, {Koscielny-Bunde}, {Rego},
  {Havlin}  \& {Bunde}}{{Kantelhardt} et~al.}{2001}]{2001PhyA..295..441K}
{Kantelhardt} J.~W.,  {Koscielny-Bunde} E.,  {Rego} H.~H.~A.,  {Havlin} S.,
  {Bunde} A.,  2001, Physica A Statistical Mechanics and its Applications, 295,
  441

\bibitem[\protect\citeauthoryear{{Kantelhardt}, {Zschiegner},
  {Koscielny-Bunde}, {Havlin}, {Bunde}  \& {Stanley}}{{Kantelhardt}
  et~al.}{2002}]{2002PhyA..316...87K}
{Kantelhardt} J.~W.,  {Zschiegner} S.~A.,  {Koscielny-Bunde} E.,  {Havlin} S.,
  {Bunde} A.,   {Stanley} H.~E.,  2002, Physica A Statistical Mechanics and its
  Applications, 316, 87

\bibitem[\protect\citeauthoryear{{Kataoka}, {Tanihata}, {Kawai}, {Takahara},
  {Takahashi}, {Edwards}  \& {Makino}}{{Kataoka}
  et~al.}{2002}]{2002MNRAS.336..932K}
{Kataoka} J.,  {Tanihata} C.,  {Kawai} N.,  {Takahara} F.,  {Takahashi} T.,
  {Edwards} P.~G.,   {Makino} F.,  2002, \mnras, 336, 932

\bibitem[\protect\citeauthoryear{{Longo}, {Vio}, {Paura}, {Provenzale}  \&
  {Rifatto}}{{Longo} et~al.}{1996}]{1996A&A...312..424L}
{Longo} G.,  {Vio} R.,  {Paura} P.,  {Provenzale} A.,   {Rifatto} A.,  1996,
  \aap, 312, 424

\bibitem[\protect\citeauthoryear{Mandelbrot}{Mandelbrot}{1983}]{mandelbrot1983fractal}
Mandelbrot B.,  1983, The Fractal Geometry of Nature.
Henry Holt and Company

\bibitem[\protect\citeauthoryear{{Mangalam} \& {Wiita}}{{Mangalam} \&
  {Wiita}}{1993}]{1993ApJ...406..420M}
{Mangalam} A.~V.,  {Wiita} P.~J.,  1993, \apj, 406, 420

\bibitem[\protect\citeauthoryear{{Mantovani}, {Junor}, {McHardy}  \&
  {Valerio}}{{Mantovani} et~al.}{2000}]{2000A&A...354..497M}
{Mantovani} F.,  {Junor} W.,  {McHardy} I.~M.,   {Valerio} C.,  2000, \aap,
  354, 497

\bibitem[\protect\citeauthoryear{{McHardy}, {Lawson}, {Newsam}, {Marscher},
  {Robson}  \& {Stevens}}{{McHardy} et~al.}{1999}]{1999MNRAS.310..571M}
{McHardy} I.,  {Lawson} A.,  {Newsam} A.,  {Marscher} A.,  {Robson} I.,
  {Stevens} J.,  1999, \mnras, 310, 571

\bibitem[\protect\citeauthoryear{{McHardy}, {Lawson}, {Newsam}, {Marscher},
  {Sokolov}, {Urry}  \& {Wehrle}}{{McHardy} et~al.}{2007}]{2007MNRAS.375.1521M}
{McHardy} I.,  {Lawson} A.,  {Newsam} A.,  {Marscher} A.~P.,  {Sokolov} A.~S.,
  {Urry} C.~M.,   {Wehrle} A.~E.,  2007, \mnras, 375, 1521

\bibitem[\protect\citeauthoryear{Moffatt}{Moffatt}{1994}]{moffatt_1994}
Moffatt H.~K.,  1994, Journal of Fluid Mechanics, 263, 375

\bibitem[\protect\citeauthoryear{{Molchanov} \& {Hayakawa}}{{Molchanov} \&
  {Hayakawa}}{1995}]{1995GeoRL..22.3091M}
{Molchanov} O.~A.,  {Hayakawa} M.,  1995, \grl, 22, 3091

\bibitem[\protect\citeauthoryear{{Monin} \& {Yaglom}}{{Monin} \&
  {Yaglom}}{2007}]{monin2007statistical}
{Monin} A.,  {Yaglom} A.,  2007, Statistical Fluid Mechanics: Mechanics of
  Turbulence.
No.~v. 2, Dover Publications

\bibitem[\protect\citeauthoryear{{Muzy}, {Bacry}  \& {Arneodo}}{{Muzy}
  et~al.}{1991}]{1991PhRvL..67.3515M}
{Muzy} J.~F.,  {Bacry} E.,   {Arneodo} A.,  1991, Physical Review Letters, 67,
  3515

\bibitem[\protect\citeauthoryear{{Norouzzadeh}, {Rahmani}  \&
  {Norouzzadeh}}{{Norouzzadeh} et~al.}{2007}]{2007IJMPC..18.1071N}
{Norouzzadeh} P.,  {Rahmani} B.,   {Norouzzadeh} M.~S.,  2007, International
  Journal of Modern Physics C, 18, 1071

\bibitem[\protect\citeauthoryear{{Ouahabi} \& {Femmam}}{{Ouahabi} \&
  {Femmam}}{2011}]{2011Ouahabi}
{Ouahabi} A.,  {Femmam} S.,  2011, S. Analog Integr Circ Sig Process, 69, 3

\bibitem[\protect\citeauthoryear{{Pacciani} et~al.,}{{Pacciani}
  et~al.}{2009}]{2009A&A...494...49P}
{Pacciani} L.,  et~al., 2009, \aap, 494, 49

\bibitem[\protect\citeauthoryear{{Page}, {Turner}, {Done}, {O'Brien}, {Reeves},
  {Sembay}  \& {Stuhlinger}}{{Page} et~al.}{2004}]{2004MNRAS.349...57P}
{Page} K.~L.,  {Turner} M.~J.~L.,  {Done} C.,  {O'Brien} P.~T.,  {Reeves}
  J.~N.,  {Sembay} S.,   {Stuhlinger} M.,  2004, \mnras, 349, 57

\bibitem[\protect\citeauthoryear{{Peitgen}, {J{\''u}rgens}  \&
  {Saupe}}{{Peitgen} et~al.}{2004}]{2004Peitgen}
{Peitgen} H.~O.,  {J{\''u}rgens} H.,   {Saupe} D.,  2004, New Frontiers of
  Science.
Springer-Verlag New York

\bibitem[\protect\citeauthoryear{Pietronero \& Siebesma}{Pietronero \&
  Siebesma}{1986}]{PhysRevLett.57.1098}
Pietronero L.,  Siebesma A.~P.,  1986, Phys. Rev. Lett., 57, 1098

\bibitem[\protect\citeauthoryear{{Robson} et~al.,}{{Robson}
  et~al.}{1993}]{1993MNRAS.262..249R}
{Robson} E.~I.,  et~al., 1993, \mnras, 262, 249

\bibitem[\protect\citeauthoryear{{Sadegh Movahed}, {Jafari}, {Ghasemi},
  {Rahvar}  \& {Rahimi Tabar}}{{Sadegh Movahed}
  et~al.}{2006}]{2006JSMTE..02..003S}
{Sadegh Movahed} M.,  {Jafari} G.~R.,  {Ghasemi} F.,  {Rahvar} S.,   {Rahimi
  Tabar} M.~R.,  2006, Journal of Statistical Mechanics: Theory and Experiment,
  2, 02003

\bibitem[\protect\citeauthoryear{{Sambruna}, {Urry}, {Tavecchio}, {Maraschi},
  {Scarpa}, {Chartas}  \& {Muxlow}}{{Sambruna}
  et~al.}{2001}]{2001ApJ...549L.161S}
{Sambruna} R.~M.,  {Urry} C.~M.,  {Tavecchio} F.,  {Maraschi} L.,  {Scarpa} R.,
   {Chartas} G.,   {Muxlow} T.,  2001, \apjl, 549, L161

\bibitem[\protect\citeauthoryear{{Savolainen}, {Wiik}, {Valtaoja}  \&
  {Tornikoski}}{{Savolainen} et~al.}{2006}]{2006A&A...446...71S}
{Savolainen} T.,  {Wiik} K.,  {Valtaoja} E.,   {Tornikoski} M.,  2006, \aap,
  446, 71

\bibitem[\protect\citeauthoryear{{Schmidt}}{{Schmidt}}{1963}]{1963Natur.197.1040S}
{Schmidt} M.,  1963, \nat, 197, 1040

\bibitem[\protect\citeauthoryear{{Shao}, {Gu}, {Jiang}, {Zhou}  \&
  {Sornette}}{{Shao} et~al.}{2012}]{2012NatSR...2E.835S}
{Shao} Y.-H.,  {Gu} G.-F.,  {Jiang} Z.-Q.,  {Zhou} W.-X.,   {Sornette} D.,
  2012, Scientific Reports, 2, 835

\bibitem[\protect\citeauthoryear{{Shimizu}, {Thurner}  \&
  {Ehrenberger}}{{Shimizu} et~al.}{2002}]{2002SHIMIZU}
{Shimizu} Y.,  {Thurner} S.,   {Ehrenberger} K.,  2002, Fractals, 10, 103

\bibitem[\protect\citeauthoryear{{Soldi} et~al.,}{{Soldi}
  et~al.}{2008}]{2008A&A...486..411S}
{Soldi} S.,  et~al., 2008, \aap, 486, 411

\bibitem[\protect\citeauthoryear{{Tanna} \& {Pathak}}{{Tanna} \&
  {Pathak}}{2014}]{2014Ap&SS.350...47T}
{Tanna} H.~J.,  {Pathak} K.~N.,  2014, \apss, 350, 47

\bibitem[\protect\citeauthoryear{{Telesca} \& {Lapenna}}{{Telesca} \&
  {Lapenna}}{2006}]{2006Tectp.423..115T}
{Telesca} L.,  {Lapenna} V.,  2006, Tectonophysics, 423, 115

\bibitem[\protect\citeauthoryear{{Telesca}, {Cuomo}, {Lapenna}  \&
  {Macchiato}}{{Telesca} et~al.}{2001}]{2001GeoRL..28.4323T}
{Telesca} L.,  {Cuomo} V.,  {Lapenna} V.,   {Macchiato} M.,  2001, \grl, 28,
  4323

\bibitem[\protect\citeauthoryear{{Telesca}, {Balasco}, {Colangelo}, {Lapenna}
  \& {Macchiato}}{{Telesca} et~al.}{2004}]{2004PCE....29..295T}
{Telesca} L.,  {Balasco} M.,  {Colangelo} G.,  {Lapenna} V.,   {Macchiato} M.,
  2004, Physics and Chemistry of the Earth, 29, 295

\bibitem[\protect\citeauthoryear{{T{\"u}rler} et~al.,}{{T{\"u}rler}
  et~al.}{1999}]{1999A&AS..134...89T}
{T{\"u}rler} M.,  et~al., 1999, \aaps, 134, 89

\bibitem[\protect\citeauthoryear{{T{\"u}rler}, {Courvoisier}  \&
  {Paltani}}{{T{\"u}rler} et~al.}{2000}]{2000A&A...361..850T}
{T{\"u}rler} M.,  {Courvoisier} T.~J.-L.,   {Paltani} S.,  2000, \aap, 361, 850

\bibitem[\protect\citeauthoryear{{T{\"u}rler} et~al.,}{{T{\"u}rler}
  et~al.}{2006}]{2006A&A...451L...1T}
{T{\"u}rler} M.,  et~al., 2006, \aap, 451, L1

\bibitem[\protect\citeauthoryear{{Varotsos}, {Sarlis}  \& {Skordas}}{{Varotsos}
  et~al.}{2002}]{2002PhRvE..66a1902V}
{Varotsos} P.~A.,  {Sarlis} N.~V.,   {Skordas} E.~S.,  2002, \pre, 66, 011902

\bibitem[\protect\citeauthoryear{{Varotsos}, {Sarlis}  \& {Skordas}}{{Varotsos}
  et~al.}{2003}]{2003PhRvE..67b1109V}
{Varotsos} P.~A.,  {Sarlis} N.~V.,   {Skordas} E.~S.,  2003, \pre, 67, 021109

\bibitem[\protect\citeauthoryear{{Vio}, {Cristiani}, {Lessi}  \&
  {Salvadori}}{{Vio} et~al.}{1991}]{1991ApJ...380..351V}
{Vio} R.,  {Cristiani} S.,  {Lessi} O.,   {Salvadori} L.,  1991, \apj, 380, 351

\bibitem[\protect\citeauthoryear{{Vio}, {Cristiani}, {Lessi}  \&
  {Provenzale}}{{Vio} et~al.}{1992}]{1992ApJ...391..518V}
{Vio} R.,  {Cristiani} S.,  {Lessi} O.,   {Provenzale} A.,  1992, \apj, 391,
  518

\bibitem[\protect\citeauthoryear{{Vyushin}, {Zhidkov}, {Havlin}, {Bunde}  \&
  {Brenner}}{{Vyushin} et~al.}{2004}]{2004GeoRL..3110206V}
{Vyushin} D.,  {Zhidkov} I.,  {Havlin} S.,  {Bunde} A.,   {Brenner} S.,  2004,
  \grl, 31, L10206

\bibitem[\protect\citeauthoryear{{Wagner} \& {Witzel}}{{Wagner} \&
  {Witzel}}{1995}]{1995ARA&A..33..163W}
{Wagner} S.~J.,  {Witzel} A.,  1995, \araa, 33, 163

\makeatother
\end{thebibliography}

\label{lastpage}
\end{document}